\documentclass{sigma}

\usepackage{cite}            

\begin{document}

\renewcommand{\PaperNumber}{??}

\FirstPageHeading

\ShortArticleName{Symmetry of Complexity: Unification of Causal QM,
Relativity, and Cosmology}

\ArticleName{Consistent Cosmology, Dynamic Relativity and Causal
Quantum Mechanics as Unified Manifestations of the Symmetry of
Complexity \footnote [0] {Report presented at the Sixth
International Conference ``Symmetry in Nonlinear Mathematical
Physics'' (Kiev, 20-26 June 2005),
\texttt{http://www.imath.kiev.ua/\~{ }appmath/conf.html}.} }

\Author{Andrei P. KIRILYUK}

\AuthorNameForHeading{A.P. Kirilyuk}

\Address{Institute of Metal Physics, 36 Vernadsky Avenue, 03142 Kyiv-142, Ukraine} 
\EmailD{kiril@metfiz.freenet.kiev.ua} 
\URLaddressD{http://myprofile.cos.com/mammoth/} 


\Abstract{The universal symmetry, or conservation, of complexity
underlies any law or principle of system dynamics and describes the
unceasing transformation of dynamic information into dynamic entropy
as the unique way to conserve their sum, the total dynamic
complexity. Here we describe the real world structure emergence and
dynamics as manifestation of the universal symmetry of complexity of
initially homogeneous interaction between two protofields. It
provides the unified complex-dynamic, causally complete origin of
physically real, 3D space, time, elementary particles, their
properties (mass, charge, spin, etc.), quantum, relativistic, and
classical behaviour, as well as fundamental interaction forces,
including naturally quantized gravitation. The old and new
cosmological problems (including ``dark'' mass and energy) are
basically solved for this explicitly emerging, self-tuning world
structure characterised by strictly positive (and large)
energy-complexity. A general relation is obtained between the
numbers of world dimensions and fundamental forces, excluding
plausible existence of hidden dimensions. The unified, causally
explained quantum, classical, and relativistic properties (and types
of behaviour) are generalised to all higher levels of complex world
dynamics. The real world structure, dynamics, and evolution are
exactly reproduced by the probabilistic dynamical fractal, which is
obtained as the truly complete general solution of a problem and the
unique structure of the new mathematics of complexity. We outline
particular, problem-solving applications of always exact, but
irregularly structured symmetry of unreduced dynamic complexity to
microworld dynamics, including particle physics, genuine quantum
chaos, real nanobiotechnology, and reliable genomics.}


\Classification{34K23;47J10;60A99;70H33;70K55;81V25;83D05;85A40} 

%
\section{Universe structure emergence by the symmetry of
complexity}\label{Sec:UnivByComp}
\subsection{Unreduced interaction dynamics and elementary particle structure}\label{Subsec:CompIntDyn}
The \emph{universal symmetry (conservation and transformation) of
complexity} underlies any real interaction process development (any
system dynamics and evolution) and constitutes both the origin and
the result of structure emergence at any level of world dynamics,
providing a large scope of problem-solving applications
\cite{Kir:USciCom,Kir:USymCom,Kir:SelfOrg,Kir:QuMach,Kir:Fractal:1,Kir:Fractal:2,Kir:QFM,Kir:100Quanta,Kir:Cosmo,Kir:QuChaos,Kir:Channel,Kir:Nano,Kir:Conscious,Kir:CommNet,Kir:SustTrans}.
Contrary to usual symmetry expression by \emph{externally imposed},
formal \emph{operators} \cite{Bluman,Fushchich,Elliott}, the
symmetry of complexity expresses \emph{real interaction dynamics} in
the form of unceasing chaotic \emph{transitions} between system
realisations and complexity levels \cite{Kir:USymCom}. In this
report we consider explicit emergence of the lowest complexity
levels of the universe, represented by elementary particles, fields,
all their properties and interactions, as well as global and cosmic
structure (cosmological) features
\cite{Kir:USciCom,Kir:QuMach,Kir:QFM,Kir:100Quanta,Kir:Cosmo,Kir:QuChaos}.
We demonstrate how the symmetry of complexity determines the
properties of \emph{real-world} structures and provides
\emph{solution} to fundamental and practical problems remaining
otherwise unsolved or even growing (e.\,g. in cosmology, quantum and
classical gravity and field theory).
\begin{figure}
\centerline{\includegraphics[width=15cm]{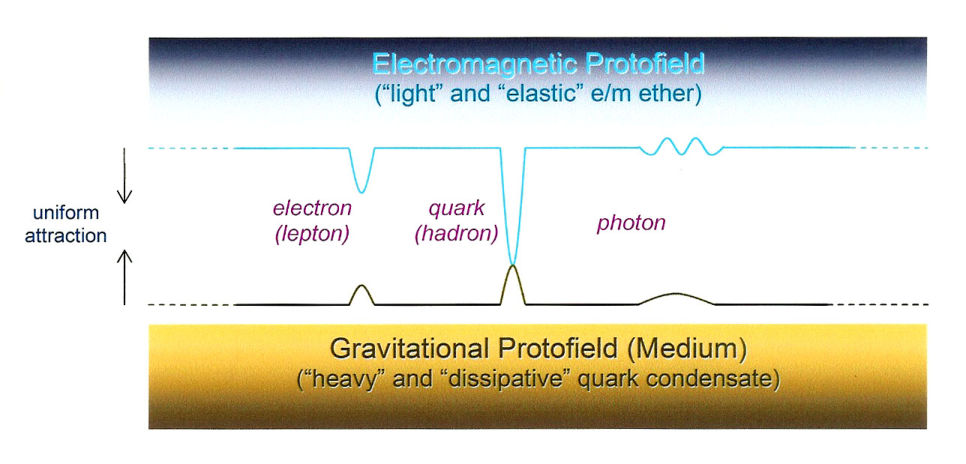}}
\caption{Scheme of protofield interaction configuration and
development, giving rise to \emph{emerging} elementary particles,
fields, their properties and interactions (progressively derived).}
\label{Protofields}
\end{figure}

The ``fundamental'' world structures of lowest complexity emerge
necessarily from the \emph{simplest} possible interaction
configuration, which is uniquely represented, at the universe scale,
by two homogeneous, physically real protofields \emph{uniformly
attracted} to each other (Figure \ref{Protofields}). The dense and
dissipative gravitational protofield, or medium, plays the role of
inert world ``matrix'' and eventually gives rise to (dynamically
emerging) universal gravitation, while light and elastic
electromagnetic (e/m) protofield is the ``swift" system component
that underlies e/m properties. Interaction development in the
protofield system leads first to emergence of most fundamental world
structures, elementary particles (and fields), and we are going to
\emph{explicitly obtain} them as \emph{unreduced solutions} of a
quite general equation, called \emph{existence equation} and
actually only fixing the initial system configuration (it also
generalises various ``model'' equations):
\begin{equation}\label{Eq:ExistProto}
\left[ {h_{\rm g} \left( \xi  \right) + V_{{\rm eg}} \left( {\xi,q }
\right) + h_{\rm e} \left( q \right)} \right]\Psi \left( {\xi,q }
\right) = E\Psi \left( {\xi,q } \right),
\end{equation}
where $\Psi \left( {\xi,q } \right)$ is the system state-function
expressing its state and development (to be found), $h_{\rm g}
\left( \xi \right)$ and $h_{\rm e} \left( q \right)$ are generalised
Hamiltonians for the free (non-interacting) gravitational and e/m
protofields (i. e. measures of dynamic complexity defined below),
$V_{{\rm eg}} \left( {\xi,q } \right)$ is arbitrary (but actually
attractive and binding) interaction potential between the fields of
$\xi$ and $q$, and $E$ is the generalised Hamiltonian eigenvalue
(energy). The Hamiltonian form of existence equation generalises
various, linear and ``nonlinear'', models and is self-consistently
confirmed below (section \ref{Subsec:SymCom}) as \emph{unified}
expression of the \emph{symmetry of complexity} (the latter can
already be traced in equation (\ref{Eq:ExistProto}) as the expressed
permanence of energetic system content). The starting problem
formulation of equation (\ref{Eq:ExistProto}) does not contain
either space or time that will be obtained as \emph{emerging}, real
manifestations of universe structure development (section
\ref{Subsubsec:SpaceTime}).

Expressing $\Psi \left( {q,\xi } \right)$ in terms of the free e/m
protofield eigenfunctions, $\{\phi _n (q)\}$, we get
\begin{equation}\label{Eq:StFuncExpan}
\Psi \left( {\xi,q } \right) = \sum\limits_n {\psi _n \left( \xi
\right)} \phi _n \left( q \right),\qquad h_{\rm e} \left( q
\right)\phi _n \left( q \right) = \varepsilon _n \phi _n \left( q
\right),
\end{equation}
which after substitution into equation (\ref{Eq:ExistProto}) and
standard eigenfunction separation gives a system of equations for
$\psi _n \left( \xi \right)$, equivalent to the starting existence
equation (\ref{Eq:ExistProto})
\cite{Kir:USciCom,Kir:USymCom,Kir:QFM,Kir:100Quanta,Kir:Cosmo}:
\begin{gather}
\left[ {h_{\rm g} \left( \xi  \right) + V_{00} \left( \xi \right)}
\right]\psi _0 \left( \xi  \right) + \sum\limits_n {V_{0n} } \left(
\xi  \right)\psi _n \left( \xi  \right) = \eta \psi _0
\left( \xi  \right),\nonumber\\
\left[ {h_{\rm g} \left( \xi \right) + V_{nn} \left( \xi \right)}
\right]\psi _n \left( \xi \right) + \sum\limits_{n' \ne n} {V_{nn'}
} \left( \xi \right)\psi _{n'} \left( \xi  \right) = \eta _n \psi _n
\left( \xi \right) - V_{n0} \left( \xi \right)\psi _0 \left( \xi
\right),\label{Eq:ExistSystem}
\end{gather}
where $\eta _n  \equiv E - \varepsilon _n$,
\begin{equation*}
V_{nn'} \left( \xi  \right) = \int\limits_{ \Omega _q } {dq} \phi
_n^ *  \left( q \right)V_{{\rm eg}} \left( {\xi,q } \right)\phi
_{n'} \left( q \right),
\end{equation*}
equation with $n = 0$ is separated from others, so that $n \ne 0$
from now on, and $\eta \equiv \eta _0$. Note that one obtains
exactly the same system of equations (\ref{Eq:ExistSystem}) starting
from a general existence equation for arbitrary system configuration
and number of components ($N$)
\cite{Kir:SelfOrg,Kir:QuMach,Kir:Fractal:2,Kir:Nano,Kir:Conscious,Kir:CommNet},
\begin{equation}\label{Eq:ExistGen}
\left\{ {h_0 \left( \xi  \right) + \sum\limits_{k = 1}^N {\left[
{h_k \left( {q_k } \right) + V_{0k} \left( {\xi ,q_k } \right)} +
\sum\limits_{l > k}^N {V_{kl} \left( {q_k ,q_l } \right)} \right]} }
\right\}\Psi \left( {\xi ,Q} \right) = E\Psi \left( {\xi ,Q}
\right),
\end{equation}
where $Q \equiv \left\{ {q_1 ,...,q_N } \right\}$. This fact should
not be surprising, as arbitrary interaction between protofield
elements is implied in (\ref{Eq:ExistProto}). It demonstrates also
the deep underlying \emph{universality} of real world structure
emergence at all levels, properly reflected in our unreduced
description.

The unreduced interaction complexity emerges if instead of
perturbative reduction of ``nonintegrable'' system
(\ref{Eq:ExistSystem}) we try to ``solve'' it by expressing
$\psi_n(\xi)$ through $\psi_0(\xi)$ by the standard Green function
technique and inserting the result into the equation for
$\psi_0(\xi)$, which gives the \emph{effective existence equation}
of the generalised effective (optical) potential method
\cite{Kir:Channel,Dederichs}:
\begin{equation}\label{Eq:ExistEff}
\left[ {h_{\rm g} \left( \xi  \right) + V_{{\rm eff}} \left( {\xi
;\eta } \right)} \right]\psi _0 \left( \xi  \right) = \eta \psi _0
\left( \xi  \right),
\end{equation}
where the \emph{effective potential (EP)}, $V_{{\rm
eff}}(\xi;\eta)$, is given by
\begin{gather}
V_{{\rm eff}} \left( {\xi ;\eta } \right) = V_{00} \left( \xi
\right) + \hat V\left( {\xi ;\eta } \right),\qquad \hat V\left( {\xi
;\eta } \right)\psi _0 \left( \xi  \right) = \int\limits_{ \Omega
_\xi  } {d\xi 'V\left( {\xi ,\xi ';\eta } \right)} \psi _0 \left(
{\xi '} \right),\nonumber\\
V\left( {\xi ,\xi ';\eta } \right) = \sum\limits_{n,i}
{\frac{{V_{0n} \left( \xi  \right)\psi _{ni}^0 \left( \xi
\right)V_{n0} \left( {\xi '} \right)\psi _{ni}^{0*} \left( {\xi '}
\right)}}{{\eta  - \eta _{ni}^0  - \varepsilon _{n0} }}}\ ,\qquad
\varepsilon _{n0}  \equiv \varepsilon _n  - \varepsilon _0\ ,
\label{Eq:EP}
\end{gather}
and $\{\psi_{ni}^0(\xi)\}$, $\{\eta_{ni}^0\}$ are the complete sets
of eigenfunctions and eigenvalues for an auxiliary, truncated system
of equations:
\begin{equation}\label{Eq:AuxSyst}
\left[ {h_{\rm g} \left( \xi  \right) + V_{nn} \left( \xi \right)}
\right]\psi _n \left( \xi  \right) + \sum\limits_{n' \ne n} {V_{nn'}
\left( \xi  \right)} \psi _{n'} \left( \xi  \right) = \eta _n \psi
_n \left( \xi  \right).
\end{equation}

The state function (\ref{Eq:StFuncExpan}) of the initial existence
equation (\ref{Eq:ExistProto}) is then obtained as
\cite{Kir:USciCom,Kir:QuMach,Kir:QuChaos,Kir:Channel}:
\begin{gather}
{ \Psi} \left( {\xi,q } \right) = \sum\limits_i {c_i } \left[ {\phi
_0 \left( q \right) + \sum\limits_n {\phi _n } \left( q \right)\hat
g_{ni} \left( \xi  \right)} \right]\psi _{0i} \left(
\xi  \right),\nonumber\\
\psi _{ni} \left( \xi  \right) = \hat g_{ni} \left( \xi \right)\psi
_{0i} \left( \xi  \right) \equiv \int\limits_{ \Omega _\xi  } {d\xi
'g_{ni} \left( {\xi ,\xi '} \right)\psi _{0i}
\left({\xi '} \right)},\nonumber\\
g_{ni} \left( {\xi , \xi '} \right) = V_{n0} \left( {\xi '}
\right)\sum\limits_{i'} {\frac{{\psi _{ni'}^0 \left( \xi \right)\psi
_{ni'}^{0*} \left( {\xi '} \right)}}{{\eta _i  - \eta _{ni'}^0  -
\varepsilon _{n0} }}}\ ,\label{Eq:StateFunc}
\end{gather}
where $\{\psi_{0i}(\xi)\}$ are eigenfunctions and $\{\eta_i\}$
eigenvalues found from equation (\ref{Eq:ExistEff}), while the
coefficients $c_i$ should be determined by state-function matching
on the boundary where the effective interaction vanishes. The
observed system density, $\rho(\xi,q)$, is given by the squared
modulus of the state-function, $\rho(\xi,q) = \left|{
\Psi}(\xi,q)\right|{}^2$ (for ``quantum" and other ``wave-like"
levels of complexity), or by the state-function itself, $\rho(\xi,q)
= { \Psi}(\xi,q)$ (for ``particle-like" levels) \cite{Kir:USciCom}.
\begin{figure}
\centerline{\includegraphics[width=14cm]{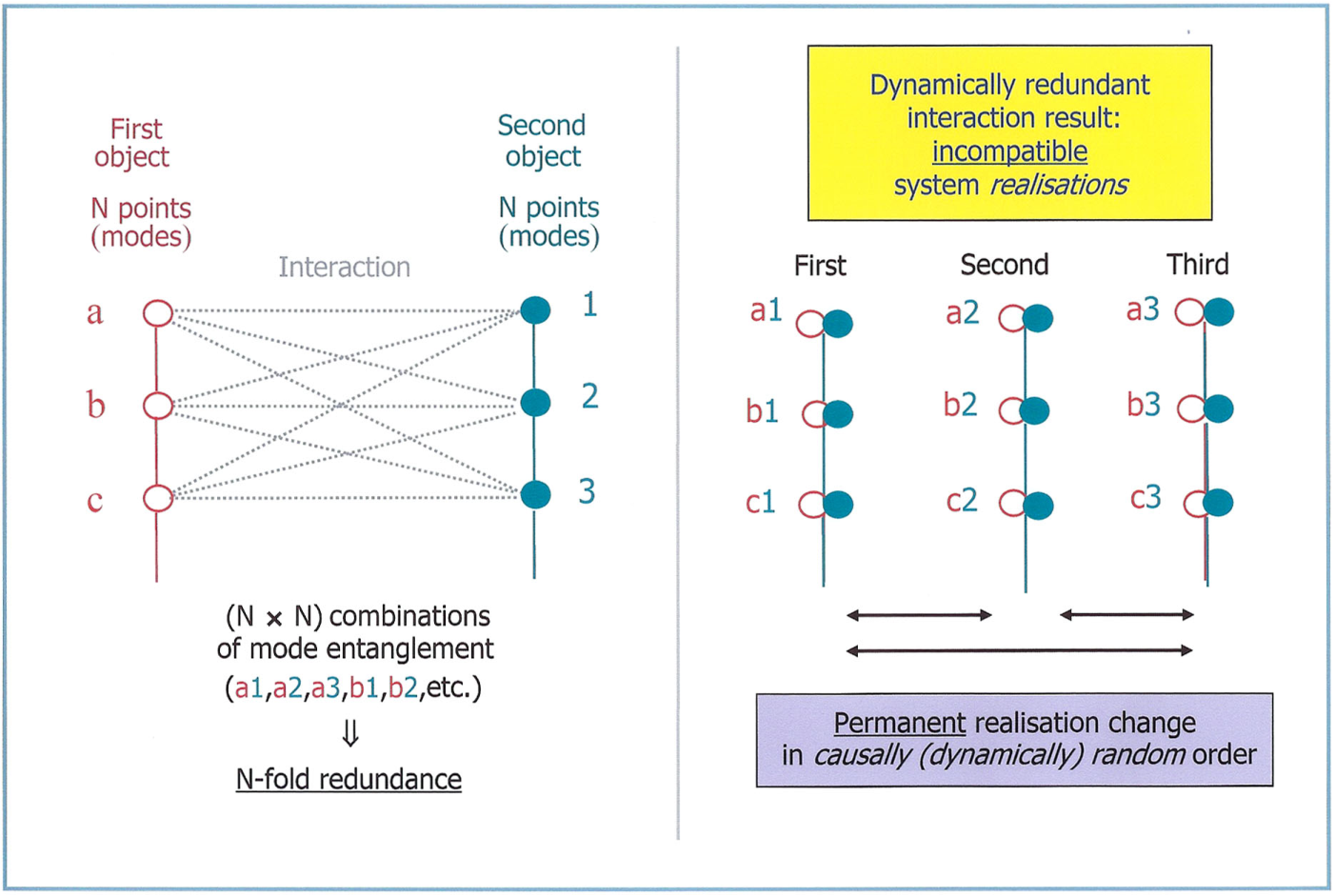}}
\caption{Dynamic multivaluedness emergence in any real interaction
process. Unreduced interaction between two objects (e.\,g.
protofields) with $N$ interacting points or modes each (left)
includes $N^2$ mode combinations, leading to $N$-fold
\emph{redundance} of \emph{incompatible} system realisations thus
formed (in agreement with equations
(\ref{Eq:EP}),~(\ref{Eq:EP-Full})). It is related to the
\emph{symmetry of complexity}, as the ``number of places'' for
interaction components and products, determined by the total system
complexity, \emph{cannot change} in the course of interaction.
Therefore all $N_\Re = N$ equally real system realisations are
forced, by the same driving interaction, to \emph{permanently
replace} each other in a \emph{dynamically random} order thus
defined (right).} \label{Multivalued}
\end{figure}

The unreduced EP problem formulation (\ref{Eq:ExistEff}) reveals the
key property of \emph{dynamic multivaluedness} (or
\emph{redundance}) of \emph{any real} interaction process that
remains hidden in the formally equivalent initial problem expression
(\ref{Eq:ExistProto})--(\ref{Eq:ExistGen}) and is artificially
reduced in usual, perturbative theories (including conventional EP
applications \cite{Dederichs}). It is due to the self-consistent,
\emph{dynamically} nonlinear dependence of the unreduced EP
(\ref{Eq:EP}) on eigen-solutions to be found that leads to dramatic
increase of the maximum eigenvalue power in the characteristic
equation and corresponding growth of the number of unreduced problem
eigen-solutions with respect to their usual, ``unique-solution'' set
\cite{Kir:USciCom,Kir:USymCom,Kir:SelfOrg,Kir:QuMach,Kir:Fractal:1,Kir:Fractal:2,Kir:QFM,Kir:100Quanta,Kir:Cosmo,Kir:QuChaos,Kir:Channel}.
The unique, incomplete solution of perturbative interpretations is
replaced by \emph{many} equally real, physically ``complete'' and
therefore \emph{mutually incompatible} solutions of the unreduced
problem called (system) \emph{realisations}, which are forced to
\emph{permanently replace each other} in a \emph{dynamically
random}, or \emph{chaotic}, order thus defined. In addition to the
above algebraic derivation of dynamic multivaluedness (confirmed by
geometric analysis \cite{Kir:USciCom,Kir:QuChaos}), it can also be
understood in terms of \emph{transparent physical picture}, Figure
\ref{Multivalued}, directly reflecting the key features of the
unreduced EP expression (\ref{Eq:EP}). Measured system density
$\rho(\xi,q) = \left|{ \Psi}(\xi,q)\right|{}^2$ is obtained then as
a \emph{dynamically probabilistic} sum of all realisation densities,
$\{\rho _r(\xi,q)\}$, where each $r$-th realisation is provided by
the \emph{dynamically derived, a priori} value of its emergence
probability $\alpha _r$:
\begin{equation}\label{Eq:ProbabSum}
\rho \left( {\xi,q} \right) = \sum\limits_{r = 1}^{N_\Re  } {^
\oplus  \rho _r \left( {\xi,q} \right)} , \ \ \ \ \rho _r(\xi,q) =
\left|{ \Psi} _r(\xi,q)\right|{}^2 \ ,
\end{equation}
\begin{equation}\label{Eq:RealProbab}
\alpha _r \left( {N_r } \right) = \frac{{N_r }}{{N_\Re }}\quad
\left( {N_r  = 1, \ldots ,N_\Re  ;\ \sum\limits_r {N_r }  = N_\Re }
\right),\qquad \sum\limits_r {\alpha _r }  = 1\ .
\end{equation}
where $N_\Re$ is the total number of elementary realisations (equal
to the number $N$ of interacting protofield modes, $N_\Re = N$, see
Figure \ref{Multivalued}), $N_r$ is the number of elementary
realisations within actually observed, ``compound'' $r$-th
realisation, and the sign $\oplus$ serves to designate the special,
dynamically probabilistic meaning of the sum derived above.

The dynamically probabilistic sum (\ref{Eq:ProbabSum}) over system
realisations provides (together with its fractal extension
(\ref{Eq:FractGenSol})) a universal expression of the \emph{complete
general solution} to a problem, as opposed to its usual version of
\emph{compatible} eigenfunction \emph{superposition} similar to
(\ref{Eq:StFuncExpan}) that does not contain any intrinsic,
dynamically probabilistic change and is obtained as solution to a
perturbative, ``mean-field'' problem approximation of the form
\begin{equation*}
\left[ {h_0 \left( \xi  \right) + V_{nn} \left( \xi  \right) +
\tilde V_n \left( \xi  \right)} \right]\psi _n \left( \xi  \right) =
\eta _n \psi _n \left( \xi  \right)\ ,
\end{equation*}
where $ \left| {V_0 \left( \xi  \right)} \right|  \le \left| {\tilde
V_n \left( \xi  \right)} \right| \le \left| {\sum\limits_{n'}
{V_{nn'} } \left( \xi  \right)} \right| $. We deal here with the
\emph{dynamically single-valued}, or \emph{unitary}, approximation
or model of the \emph{whole} conventional science that retains
\emph{only one}, averaged system realisation of their really
existing multitude. The real complexity of the unreduced problem
solution becomes evident by comparison of this kind of
oversimplified unitary model with the corresponding unreduced EP and
state-function expressions obtained by explicit substitution of the
EP equation eigenvalues (equation (\ref{Eq:ExistEff})) into
equations (\ref{Eq:EP}) and (\ref{Eq:StateFunc}):
\begin{equation}\label{Eq:EP-Full}
V_{{\rm{eff}}} \left( {\xi ;\eta _i^r } \right)\psi _{0i}^r \left(
\xi  \right) = V_{00} \left( \xi  \right)\psi _{0i}^r \left( \xi
\right) + \sum\limits_{n,i'} { \frac{{V_{0n} \left( \xi  \right)\psi
_{ni'}^0 \left( \xi  \right)\int\limits_{\Omega _\xi  } {d\xi '\psi
_{ni'}^{0*} \left( {\xi '} \right)V_{n0} \left( {\xi '} \right)\psi
_{0i}^r \left( {\xi '} \right)} }}{{\eta _i^r  - \eta _{ni'}^0  -
\varepsilon _{n0} }}}\ ,
\end{equation}
\begin{equation}\label{Eq:StFunc-Full}
\Psi _r \left( {\xi,q} \right) = \sum\limits_i {c_i^r \left[ {\psi
_{0i}^r \left( \xi  \right)\phi _0 \left( q \right)\left. { +
\sum\limits_{n,i'} { \frac{{\psi _{ni'}^0 \left( \xi \right)\phi _n
\left( q \right)\int\limits_{\Omega _\xi  } {d\xi '\psi _{ni'}^{0*}
\left( {\xi '} \right)V_{n0} \left( {\xi '} \right)\psi _{0i}^r
\left( {\xi '} \right)} }}{{\eta _i^r  - \eta _{ni'}^0  -
\varepsilon _{n0} }}} } \right]} \right.}\ ,
\end{equation}
where the last state-function expression should be used in the
probabilistic general solution (\ref{Eq:ProbabSum}).

The expressions for $r$-th realisation EP (\ref{Eq:EP-Full}) and
state-function (\ref{Eq:StFunc-Full}) reveal the emerging
realisation configuration characterised by autonomous
\emph{dynamical squeeze} of the protofield system. It is determined
by the resonant denominators in combination with the cutting
numerator integrals of the unreduced EP formalism, leading to $r$-th
realisation ``concentration'' around a particular eigenvalue $\eta
_i^r$, which can be interpreted as \emph{dynamically} emerging
\emph{space point} and elementary particle core
\cite{Kir:USciCom,Kir:QuMach,Kir:QFM,Kir:100Quanta,Kir:Cosmo}. The
eigenvalue separation $\Delta x = \Delta _r \eta _i^r$ for different
$r$ provides the elementary space distance between
``point''-particles appearing to be close to the ``barred'' Compton
wavelength $\mathchar'26\mkern-10mu\lambda _{\rm C} = \lambda
_{\rm{C}} / 2 \pi$, $\Delta x = \Delta _r \eta _i^r =
\mathchar'26\mkern-10mu\lambda _{\rm C}$, while eigenvalue
separation for different $i$ gives the size of the squeezed state
determined, for the electron, by the ``classical electron radius''
$r_e$, $\Delta _i \eta _i^r  = r _e$ (see section
\ref{Subsubsec:ConstPlanckian} for more details). As follows from
equations (\ref{Eq:EP-Full})--(\ref{Eq:StFunc-Full}), the dynamical
squeeze has a self-consistent character demonstrating real
``self-organisation'' (structure formation) process and mechanism,
where the more is state-function localisation in the emerging EP
well, the deeper is that well around the state-function localisation
centre. This rigorously derived property has a clear physical
interpretation: a local density increase of (sufficiently strongly)
attracting protofields will grow until saturation, since the more is
local protofield density the stronger is their local attraction and
vice versa.

Dynamical protofield squeeze saturates at the point where protofield
attraction is compensated by internal repulsion forces between the
protofield elements (underlying final compressibility of every real
medium). After that the self-amplifying system collapse, or
``reduction'', loses its force and the opposite protofield extension
develops due to the same kind of instability related to protofield
interaction at neighbouring locations. The system thus transiently
returns to its initial, quasi-free state before falling into the
next reduction phase involving another, randomly chosen realisation
(physical ``point''). We obtain thus the unceasing process of
\emph{quantum beat}, consisting of repeating cycles of reduction and
extension around different, chaotically changing centres, which is
equivalent to a dynamically random walk of the squeezed, corpuscular
state called \emph{virtual soliton} (as opposed to usual,
\emph{permanently} localised and regular solitons). Quantum beat
process in the coupled protofield system thus obtained constitutes
the \emph{physically real structure} of any (massive) elementary
particle called also \emph{field-particle} in view of its permanent
\emph{dualistic} change between corpuscular (local) and undular
(extended) system states
\cite{Kir:USciCom,Kir:QuMach,Kir:QFM,Kir:100Quanta}. Note that a big
change of configuration involves mainly e/m protofield due to its
much larger compressibility (smaller effective density), whereas the
relatively dense and (almost) incompressible gravitational
medium-matrix shows much smaller external change of properties
(similar to a liquid), which leads to essentially e/m origin of
directly observed structures and properties.

The transient extended state of the field-particle provides the
causal, physically real interpretation of the quantum-mechanical
\emph{wavefunction}, remaining otherwise completely mysterious and
abstract in the unitary theory framework. The realistic wavefunction
thus revealed constitutes a separate, specific system realisation
called \emph{intermediate, or main, realisation} and differing
essentially from all other, localised, ``regular'' realisations. It
is \emph{explicitly obtained} within the unreduced EP formalism as a
special solution with effectively weak, perturbative
(``mean-field'') interaction where essentially nonlinear additions
to $V_{00}$ in the general EP expression (\ref{Eq:EP-Full}) are
self-consistently small (contrary to the case of regular, localised
realisations) \cite{Kir:USciCom,Kir:QuMach,Kir:QFM}. That's why it
is only the main, effectively linear and weak-interaction
realisation that remains in the usual, dynamically single-valued
theory, in quantum mechanics and beyond, whereas all regular,
essentially nonlinear and strong-interaction realisations are
neglected, which leads to the well-known ``mysteries'' and unsolved
problems.

The \emph{dynamically discrete}, or \emph{quantised}, structure of
the field-particle realisation change within the quantum beat
process results from the \emph{holistic}, self-consistent character
of \emph{unreduced} interaction, where any local component
displacement entrains neighbouring component shifts and propagates
thus to the whole system, including the initial perturbation point.
The system can have therefore only a limited number of discrete,
more stable, structure-forming configurations, while the
``continuum'' of other possible configurations plays the role of
quickly changing intermediate phases during system transitions
between those regular realisations (and the main realisation of the
wavefunction). When the system is measured at the lowest,
``quantum'' complexity level, it can only be ``caught'' in one of
its realisations, but not ``between'' them. Because of that quantum
beat pulsation cannot be traced in detail, but it can be registered
as a whole using e.\,g. a resonance effect, and recent electron
channeling experiment \cite{ChannQuantBeat} provides a clear
evidence of that kind, confirming \emph{quantum beat reality} (see
also section \ref{Subsubsec:ParticleProp}).

Thus \emph{dynamically derived} discreteness of the quantum beat
process constitutes \emph{causal, physically real} basis for all
``quantisation''/``uncertainty'' effects and properties (including
Planck's constant origin and universality)
\cite{Kir:USciCom,Kir:QuMach,Kir:QFM,Kir:100Quanta}, which are
postulated as unprovable ``mysteries'' in the usual theory (see also
sections
\ref{Subsubsec:ConstPlanckian},~\ref{Subsubsec:Relativity}). Dynamic
discreteness should be distinguished from mechanistic, non-dynamical
discreteness used in unitary simulation of quantised behaviour: the
``steps'' of dynamically discrete realisation change are
\emph{causally determined} by the \emph{unreduced interaction}
process and \emph{cannot} be replaced by \emph{arbitrary} values. At
the lowest complexity level, dynamic discreteness is represented by
physically real \emph{quantum jumps} of virtual soliton through a
distance of $\Delta x = \Delta _r \eta _i^r =
\mathchar'26\mkern-10mu\lambda _{\rm C}$ for a field-particle
globally at rest.

One obtains also causally specified \emph{events} of protofield
reduction and extension (realisation change) and with them the
\emph{emerging change} and \emph{time}, although none of these were
present in or inserted into initial system configuration and problem
formulation (see equations
(\ref{Eq:ExistProto})--(\ref{Eq:ExistGen})). Once the
reduction-extension events are obtained in the unreduced EP
formalism, time becomes dynamically determined as \emph{intensity}
(represented by \emph{frequency}) of those structure formation
processes. Specifically, quantum beat frequency is directly related
to the above quantum jump length, $\Delta t = {{\Delta x}
\mathord{\left/ {\vphantom {{\Delta x} c}} \right.
\kern-\nulldelimiterspace} c} = {{\mathchar'26\mkern-10mu\lambda
_{\rm C} } \mathord{\left/ {\vphantom
{{\mathchar'26\mkern-10mu\lambda _{\rm C} } c}} \right.
 \kern-\nulldelimiterspace} c} = \tau  = {1 \mathord{\left/
 {\vphantom {1 \nu }} \right. \kern-\nulldelimiterspace} \nu }$,
where $c$ is the velocity of perturbation propagation in a
\emph{physically real} medium of e/m protofield coupled to the
gravitational protofield, or the \emph{speed of light} thus
\emph{causally} introduced, $\Delta t = \tau$ is the quantum beat
period, and $\nu$ is its frequency. Quantum beat processes
\emph{within} each (massive) elementary particle represent thus the
fundamental physical \emph{clock} of the universe \cite{Kir:USciCom}
whose ``mechanism'' is driven by unreduced interaction of two
initially homogeneous protofields (where the interaction magnitude
determines $\Delta x = \Delta _r \eta _i^r$ and thus $\Delta t =
\Delta x / c$ according to equations
(\ref{Eq:ExistEff})--(\ref{Eq:EP})).

Note that we reveal here the fundamental, universal and physically
real \emph{origin of time} as such, constituting a stagnating
problem of unitary science, despite a lot of most ambitious efforts.
Physically real time we obtain has the main property of
\emph{unceasing} and \emph{intrinsically irreversible} flow due to
\emph{permanent}, interaction-driven change of \emph{multiple}
realisations (absent in principle in any unitary theory) and
\emph{dynamically random} order of realisation emergence
respectively. It is dynamically related to naturally quantised space
structure described above. Both quantised space and irreversibly
flowing time (see also section \ref{Subsubsec:SpaceTime}) have
eventually emerging \emph{multi-level} structure following that of
the unreduced dynamic complexity (section
\ref{Subsubsec:Relativity}).

The key property of dynamic multivaluedness of the unreduced
interaction process is completed by equally important \emph{dynamic
entanglement} of interacting components (here protofields) within
each emerging system realisation. It is described mathematically by
dynamically weighted, ``inseparable'' products of functions
depending on interacting degrees of freedom $\xi$ and $q$ in the
total state-function expressions
(\ref{Eq:StateFunc}),~(\ref{Eq:StFunc-Full}). Taking into account
realisation plurality, one obtains \emph{dynamically multivalued
entanglement} as the unreduced interaction content and meaning.
Physically real protofield entanglement (whose magnitude varies for
different particle species) constitutes tangible, ``material''
filling, or ``texture'', of the emerging field-particles determining
their perceived material \emph{quality}, which obtains thus its
\emph{rigorous} expression (contrary to purely \emph{abstract},
``immaterial'' quantities of unitary models).

Dynamically multivalued entanglement is further amplified by the
\emph{dynamically probabilistic fractality} of the unreduced problem
solution
\cite{Kir:USciCom,Kir:QuMach,Kir:Fractal:1,Kir:Fractal:2,Kir:Conscious}
that extends essentially usual, dynamically single-valued fractality
and gives rise to important system properties, such as \emph{dynamic
adaptability}. The unreduced fractality, specifying also the notions
of \emph{nonseparability} and \emph{nonintegrability} of any real
interaction, originates from the truncated system (\ref{Eq:AuxSyst})
whose solutions enter the unreduced EP expressions
(\ref{Eq:EP}),~(\ref{Eq:StateFunc}),~(\ref{Eq:EP-Full},~(\ref{Eq:StFunc-Full})
of the first level. Applying \emph{universal} EP method to the
truncated system (\ref{Eq:AuxSyst}), one obtains its effective,
externally ``separated'' version:
\begin{equation}\label{Eq:AuxEff}
\left[ {h_{\rm g} \left( \xi  \right) + V_{{\rm{eff}}}^n \left( {\xi
;\eta _n } \right)} \right]\psi _n \left( \xi  \right) = \eta _n
\psi _n \left( \xi  \right) \ ,
\end{equation}
where the second-level EP $V_{{\rm{eff}}}^n \left( {\xi ;\eta _n }
\right)$ is similar to its first-level version (\ref{Eq:EP}):
\begin{equation}\label{Eq:AuxEP}
V_{{\rm{eff}}}^n \left( {\xi ;\eta _n } \right)\psi _n \left( \xi
\right) = V_{nn} \left( \xi  \right)\psi _n \left( \xi  \right) +
\sum\limits_{n' \ne n, i} { \frac{{V_{nn'} \left( \xi  \right)\psi
_{n'i}^{0n} \left( \xi  \right)\int\limits_{\Omega _\xi  } {d\xi
'\psi _{n'i}^{0n*} \left( {\xi '} \right)V_{n'n} \left( {\xi '}
\right)\psi _n \left( {\xi '} \right)} }}{{\eta _n  - \eta
_{n'i}^{0n}  + \varepsilon _{n0}  - \varepsilon _{n'0} }}} \ ,
\end{equation}
and $\left\{ {\psi _{n'i}^{0n} \left( \xi  \right),\eta _{n'i}^{0n}
} \right\}$ is the complete eigen-solution set of a second-level
truncated system:
\begin{equation}\label{Eq:AuxSyst2Lev}
h_{\rm{g}} \left( \xi  \right)\psi _{n'} \left( \xi  \right) +
\sum\limits_{n'' \ne n'} {V_{n'n''} \left( \xi  \right)} \psi _{n''}
\left( \xi  \right) = \eta _{n'} \psi _{n'} \left( \xi  \right) \ ,
\ \ \ n' \ne n \ , \ \ n,n' \ne 0 \ .
\end{equation}
Similar to dynamic multivaluedness of the first-level EP, its
second-level version is split into many incompatible realisations
(numbered by index $r'$) due to the self-consistent dependence on
the eigen-solutions to be found, leading to corresponding splitting
of system (\ref{Eq:AuxSyst}) solutions:
\begin{equation}\label{Eq:AuxSplit}
\left\{ {\psi _{ni}^0 \left( \xi  \right),\eta _{ni}^0 } \right\}
\to \left\{ {\psi _{ni}^{0r'} \left( \xi  \right),\eta _{ni}^{0r'} }
\right\} \ .
\end{equation}

Substituting now those dynamically split solutions of truncated
system (\ref{Eq:AuxSyst}) into the first-level EP expressions
(\ref{Eq:EP}),~(\ref{Eq:StateFunc}),~(\ref{Eq:EP-Full},~(\ref{Eq:StFunc-Full}),
one gets a two-level structure with dynamic multivaluedness, and
thus randomness, at \emph{each} level. As the process of finding the
truly complete problem solution continues, one obtains further
splitting of solutions of the second-level truncated system
(\ref{Eq:AuxSyst2Lev}) that gives the third level of emerging
probabilistic fractal and so on, until one gets all $N$ levels of
dynamically probabilistic fractality ($N \gg 1$ is the number of
interacting e/m protofield modes). The \emph{complete general
solution} of the unreduced interaction problem (\ref{Eq:ProbabSum})
can now be further specified in the form of \emph{dynamically
probabilistic fractal}:
\begin{equation}\label{Eq:FractGenSol}
\rho \left( {\xi ,q} \right) = \sum\limits_{r,r',r''...}^{N_\Re}
{{^\oplus } \thinspace \rho _{rr'r''...} \left( {\xi ,q} \right)} \
,
\end{equation}
where indexes $r,r',r'',\dots$ enumerate obtained realisations at
consecutive levels of dynamically probabilistic fractality. The
average, \emph{expectation} value of the dynamically probabilistic
fractal density (valid for long enough observation time) is obtained
as
\begin{equation}\label{Eq:FractExpValue}
\rho \left( {\xi ,q} \right) = \sum\limits_{r,r',r''...}^{N_\Re  }
{\alpha _{rr'r''...} \rho _{rr'r''...} \left( {\xi ,q} \right)} \ ,
\end{equation}
where $\{ \alpha _{rr'r''...} \}$ are \emph{dynamically determined
probabilities} for the respective levels of dynamical fractal (cf.
equation (\ref{Eq:RealProbab})):
\begin{equation}\label{Eq:FractProb}
\alpha _{rr'r''...}  = \frac{{N_{rr'r''...} }}{{N_\Re  }} \ , \ \ \
\sum\limits_{rr'r''...} {\alpha _{rr'r''...} }  = 1 \ .
\end{equation}

The obtained dynamically probabilistic fractal of the unreduced
general solution (\ref{Eq:FractGenSol}) is essentially different
from any unitary ``perturbative expansion series'', since any term
and level of the dynamically probabilistic sum
(\ref{Eq:FractGenSol}) expresses the exact, really existing object
structure. The whole unitary solution approximately corresponds, in
the best case, to a single term of the dynamically probabilistic
sum. Major physical consequence of the obtained multivalued
extension of usual, dynamically single-valued fractality is the
property of \emph{interactive dynamic adaptability} of the unreduced
system structure which can \emph{autonomously} adapt to the changing
interaction configuration and efficiently find its ``way'' for the
most complete interaction process development due to
\emph{permanent} chaotic, ``searching'' motion of multivalued
fractal branches on all scales (giving a ``living arborescence''
kind of structure). The multi-level, multivalued fractal structure
of dynamic entanglement of interacting entities provides a
\emph{physically real} version and true meaning of mathematical
``nonseparability'' of a real (generic) interaction process, while
\emph{transient} component separation (disentanglement) happens
locally all the time, during system transition between realisations
(in the phase of thus \emph{quasi-linear} wavefunction).

Finally, we can now provide the \emph{causally complete} and
\emph{universally applicable} definition of the main physical
quantity of \emph{(dynamic) complexity}, $C$, as a growing function
of the total number, $N_\Re$, of (explicitly obtained) system
realisations, or rate of their change, equal to zero for the
(unrealistic) case of only one system realisation:
\begin{equation}\label{Eq:Complexity}
C = C(N_\Re) \ , \ \ {{dC} \mathord{\left/ {\vphantom {{dC} {dN_\Re
> 0}}} \right. \kern-\nulldelimiterspace} {dN_\Re   > 0}} \ , \ \
C(1) = 0 \ .
\end{equation}
Suitable examples are provided by $C\left( {N_\Re  } \right) = C_0
\ln N_\Re$, $C\left( {N_\Re  } \right) = C_0 \left( {N_\Re - 1}
\right)$, generalised action and entropy, generalised energy/mass
(temporal rate of realisation change), and momentum (spatial rate of
realisation emergence)
\cite{Kir:USciCom,Kir:QuMach,Kir:QFM,Kir:Conscious} (see also
sections \ref{Subsec:SymCom},~\ref{Subsec:Properties}). As any real
system, object, or phenomenon results from an interaction process
with at least few components and interacting modes, it becomes clear
that \emph{any real entity}, starting from (massive) elementary
particle like the electron, has a strictly \emph{positive} dynamic
complexity (and actually a great realisation number, $N_\Re \gg 1$).
Since dynamic multivaluedness ($N_\Re
> 1$) constitutes the basis of genuine, intrinsic \emph{chaoticity}
(dynamic randomness), it is evident that dynamic complexity thus
defined \emph{includes} chaoticity as its major content and aspect
(we shall see in the next section that chaoticity is represented
directly by one of the two complexity forms, \emph{dynamic
entropy}). It is evident also that the whole unitary, dynamically
single-valued science and paradigm ($N_\Re = 1, \ C = 0$), including
its versions of ``complexity'' and ``chaoticity'', consider
exclusively over-simplified, zero-complexity, zero-chaoticity
(regular) models of real world dynamics equivalent to its
effectively zero-dimensional (point-like) projection (which is
sometimes mechanistically extended to one-dimensional projection,
using a formally imposed time variable). Therefore unitary
definitions of e.\,g. ``chaoticity'' by exponential divergence of
close trajectories or infinitely long motion period (let alone the
totally lost case of quantum chaos) describe at best
``sophisticated'', ``chaotically looking'' regularity cases devoid
of any genuine, dynamic randomness and complexity (internal
inconsistency of those unitary definitions using e.\,g. incorrect
extension of perturbative approximation is a separate issue
considered elsewhere \cite{Kir:USciCom}).

\subsection{Universal symmetry and transformation of complexity}\label{Subsec:SymCom}
As the full number of system realisations $N_\Re$ determining its
total complexity $C(N_\Re)$ (see equation (\ref{Eq:Complexity}))
depends only on the initial system configuration (e.\,g. the number
$N$ of interacting protofield modes, $N_\Re = N$, see Figure
\ref{Multivalued}), the \emph{total system complexity remains
unchanged} during interaction development, $C = {\rm const}, \
\Delta C = 0$
\cite{Kir:USciCom,Kir:USymCom,Kir:QuMach,Kir:Fractal:1,Kir:Fractal:2,Kir:QFM,Kir:100Quanta,Kir:Cosmo,Kir:Conscious}.
This \emph{universal complexity conservation law} constitutes both
the result and the origin of unreduced interaction, underlying thus
any real structure emergence and existence. In this sense it is
equivalent to a universal symmetry of nature called the
\emph{symmetry of complexity}: contrary to unitary conservation
laws, in the universal science of complexity, describing
\emph{explicit} (and unceasing) \emph{structure emergence}, there is
no difference between complexity \emph{conservation law} and
resulting structure \emph{symmetry}.

A straightforward, ``horizontal'' manifestation of the universal
symmetry of complexity is the symmetry between all (elementary)
system realisations at a given complexity level: they are equal by
their origin and therefore taken by the system in a \emph{causally
random} order (section \ref{Subsec:CompIntDyn}), so that (true)
randomness \emph{results} from the symmetry of complexity. The
latter is uniquely and completely realised by the system
\emph{motion dynamics} (realisation change process), rather than any
formal ``operators'' transforming one abstract structure into
another (unitary symmetry concept). Always \emph{exact} (unbroken)
symmetry between \emph{irregularly structured} and
\emph{chaotically} changing elementary realisations leads, in
particular, to unequal but \emph{well-defined} probabilities of
compound realisations (\ref{Eq:RealProbab}) containing different
numbers of elementary realisations.

There is a more involved, ``vertical'' manifestation of complexity
symmetry concerning interaction process development with multiple
\emph{levels of complexity}. Indeed, emerging system structures
(grouped realisations) start interacting among them and produce
multivalued structure of the next level, and so on. Every such
\emph{qualitative} change of system configuration (also in each
transition between realisations) corresponds to \emph{complexity
transformation}, or development, or unfolding, from the
\emph{permanently decreasing} potential, latent form of
\emph{dynamic information} $I$ to the \emph{always increasing}
realised, explicit form of \emph{dynamic entropy} $S$, whereas their
sum, the \emph{total dynamic complexity} $C = I + S$, remains
unchanged: $\Delta C = 0, \ \Delta I = - \Delta S < 0$
\cite{Kir:USciCom,Kir:USymCom,Kir:QuMach,Kir:Fractal:1,Kir:Fractal:2,Kir:QFM,Kir:Cosmo,Kir:Conscious}.
This permanent complexity transformation from dynamic information to
entropy underlies any interaction process and therefore complexity
\emph{symmetry (conservation)} can be realised only through a
\emph{qualitative change} of its form, determining system dynamics.

In order to derive a unified expression of this relation, we first
specify a \emph{universal integral measure of dynamic complexity} in
the form of (generalised) \emph{action} $\mathcal A$ as the simplest
function, whose increment $\Delta {\mathcal A}$ is simultaneously
and independently proportional to both emerging elements of space
$\Delta x$ and time $\Delta t$ obtained above (section
\ref{Subsec:CompIntDyn}) as universal manifestations of realisation
change process (= unreduced interaction development): $\Delta
{\mathcal A} = - E \Delta t + p \Delta x$, where $E$ and $p$ are
coefficients immediately recognised, however, as \emph{energy} and
\emph{momentum} by comparison to classical mechanics. Their
generalised, universal definitions in terms of complexity-action are
obtained in their \emph{dynamically discrete} (quantised) form:
\begin{equation}\label{Eq:Energy}
E =  - \frac{{\Delta {\mathcal A}}}{{\Delta t}}\left| {_{x =
{\rm{const}}} } \right. \ ,
\end{equation}
\begin{equation}\label{Eq:Momentum}
p = \frac{{\Delta {\mathcal A}}}{{\Delta x}}\left| {_{t =
{\rm{const}}} } \right. \ ,
\end{equation}
where energy and momentum acquire the new, \emph{universal} meaning
of differential \emph{complexity measures} (energy is the temporal
and momentum spatial rate of complexity transformation from dynamic
information to entropy). Dynamic discreteness of system jumps
between realisations is eventually due to the \emph{holistic}
character of \emph{real}, unreduced interaction process and leads to
the causal (dynamic) and universal version of ``(quantum)
\emph{uncertainty relations}'', if we just rewrite the above energy
and momentum definitions as $p {\Delta x} = | \Delta {\mathcal A} |$
and $E {\Delta t} = | \Delta {\mathcal A} |$ \cite{Kir:USciCom}.

As both dynamic information and complexity-action can only
\emph{decrease} in any interaction development (cf. equation
(\ref{Eq:Energy})), generalised action expresses more directly just
informational, potential form of complexity, $I = \mathcal A$,
whereas its dual form of dynamic entropy is measured in the same
units but expresses the opposite in sign, always positive,
complexity increment:
\begin{equation}\label{Eq:CompCons}
\Delta S = - \Delta I > 0 \ , \ \ \Delta {\mathcal A} = - \Delta S \
.
\end{equation}
The last unified expression of conservation and transformation of
complexity leads to the universal dynamic equation if we divide it
by $\Delta t \left| _{x = \rm{const}} \right.$
\cite{Kir:USciCom,Kir:USymCom,Kir:QuMach,Kir:QFM,Kir:Cosmo,Kir:Conscious}:
\begin{equation}\label{Eq:Ham-Jacob}
\frac{{\Delta {\mathcal A}}}{{\Delta t}}\left| {_{x = {\rm const}} }
\right. + H\left( {x,\frac{{\Delta {\mathcal A}}}{{\Delta x}}\left|
{_{t = {\rm const}} ,t} \right.} \right) = 0 \ ,
\end{equation}
where the \emph{Hamiltonian}, $H = H(x,p,t)$, considered as a
function of emerging space-structure coordinate $x$, momentum $p =
\left( {{{\Delta {\mathcal A}} \mathord{\left/ {\vphantom {{\Delta
{\mathcal A}} {\Delta x}}} \right. \kern-\nulldelimiterspace}
{\Delta x}}} \right)\left| {_{t = {\rm const}} } \right.$ (see
equation (\ref{Eq:Momentum})), and time $t$, expresses the
implemented, entropy-like form of differential complexity, $H =
\left( {{{\Delta S} \mathord{\left/ {\vphantom {{\Delta S} {\Delta
t}}} \right. \kern-\nulldelimiterspace} {\Delta t}}} \right)\left|
{_{x = {\rm const}} } \right.$. The obtained generalised,
\emph{universal Hamilton-Jacobi equation} (\ref{Eq:Ham-Jacob})
realises the desired dynamic expression of the symmetry of
complexity and takes a yet simpler form for conservative systems
where the generalised Hamiltonian does not depend explicitly on
time:
\begin{equation}\label{Eq:Ham-JacobCons}
H\left( {x,\frac{{\Delta {\mathcal A}}}{{\Delta x}}\left| {_{t =
{\rm const}}} \right.} \right) = E \ ,
\end{equation}
with the generalised energy $E$ defined by equation
(\ref{Eq:Energy}). Note that action distribution ${\mathcal A}
\left( {x,t} \right)$ corresponds to the above state-function $\Psi
\left( {x,t} \right)$ (see equations
(\ref{Eq:StateFunc}),~(\ref{Eq:StFunc-Full})) for regular, localised
realisations.

The unified differential expression of the symmetry of complexity by
equations (\ref{Eq:Ham-Jacob})--(\ref{Eq:Ham-JacobCons}) would be
incomplete without explicit expression of the related complexity
\emph{transformation} and its \emph{direction} (from dynamic
information to entropy). Due to \emph{unceasing} realisation
emergence in a \emph{causally random} order, system
information-complexity $I = \mathcal A$ can only \emph{decrease},
which means that not only its partial (discrete) derivative ($- E$),
but also total derivative, or (generalised) \emph{Lagrangian} $L$,
is negative:
\begin{equation}\label{Eq:Lagrangian}
L = \frac{\Delta {\mathcal A}}{\Delta t} = \frac{\Delta {\mathcal
A}}{\Delta t} \left| _{x = {\rm const}} \right. + \frac{\Delta
{\mathcal A}}{\Delta x} \left| _{t = {\rm const}} \right.
\frac{\Delta x}{\Delta t} = pv - H < 0 \ ,
\end{equation}
\begin{equation}\label{Eq:TimeArrow}
E,H\left( {x,\frac{{\Delta \cal A}}{{\Delta x}}\left| {_{t = {\rm
const}} ,t} \right.} \right)
> pv \geq 0 \ ,
\end{equation}
where $v = \Delta x / \Delta t$ is the velocity of global, averaged
system motion (i.e. its motion as a whole). In agreement with the
above dynamic origin of time, this dynamic expression of complexity
transformation (within its symmetry), or \emph{dynamically
generalised second law} (``energy degradation''), equation
(\ref{Eq:TimeArrow}), provides also a rigorous, fundamentally
derived expression of the \emph{arrow of time}
\cite{Kir:USymCom,Kir:QuMach,Kir:QFM,Kir:Cosmo}: since $\Delta
{\mathcal A} < 0$, time advances, $\Delta t
> 0$, in the direction of growing (dynamic) entropy $S$ and
decreasing information $\mathcal A$ (i.e. $L < 0$). We see that our
dynamically based symmetry of complexity includes the origin of time
and causally derived direction of its unceasing flow in the form of
interaction complexity development. In fact, dynamic \emph{time
origin} and irreversible flow, permanent \emph{growth} of unreduced
dynamic complexity-entropy (at the expense of decreasing dynamic
complexity-information), and \emph{conservation} of the total
dynamic complexity are obtained as \emph{closely unified}
manifestations of the \emph{single}, holistic symmetry of
complexity.

The universal Hamilton-Jacobi equation
(\ref{Eq:Ham-Jacob})--(\ref{Eq:Ham-JacobCons}) remains naturally
valid for the case of elementary field-particle dynamics (section
\ref{Subsec:CompIntDyn}), but takes into account its unreduced
complexity (multivaluedness). The latter includes causally explained
\emph{quantum duality}, where the localised, corpuscular states of
quantum beat process alternate with extended, undular protofield
configuration in the phase of \emph{wavefunction} (intermediate
realisation). Correspondingly, the above ``classical'', corpuscular
expression of the Hamilton-Jacobi formalism should have its dual
counterpart in the form of explicit undular equation for the
wavefunction. It can be obtained with the help of \emph{causal
quantisation} procedure that describes just those spatially chaotic
transitions between regular (localised) realisations through the
extended wavefunction realisation and involves dynamic complexity
conservation
\cite{Kir:USciCom,Kir:USymCom,Kir:QuMach,Kir:QFM,Kir:100Quanta,Kir:Cosmo,Kir:Conscious}.
Hierarchical structure of multilevel complexity development implies
that the total complexity of several neighbouring levels is equal to
the product of individual level complexities. Since quantum beat
process can be considered as cyclic transitions between neighbouring
complexity sublevels of localised realisations and wavefunction, its
total complexity $C$ is given by the product of localised
realisation complexity $\mathcal A$ and that of the intermediate
realisation expressed by the wavefunction $\mit \Psi$, $C =
{\mathcal A} {\mit \Psi}$. According to complexity conservation,
$\Delta C = \Delta ({\mathcal A} {\mit \Psi}) = {\mathcal A} \Delta
{\mit \Psi} + {\mit \Psi} \Delta {\mathcal A} = 0$, or
\begin{equation}\label{Eq:CausQuant}
\Delta {\mathcal A} =  - {\mathcal A}_0 \frac{{\Delta {\mit \Psi}
}}{\mit \Psi} = - i \hbar \frac{{\Delta {\mit \Psi} }}{\mit \Psi} \
,
\end{equation}
where ${\mathcal A}_0 = i \hbar$ is a characteristic
complexity-action value that may contain also a numerical constant
reflecting specific features of the two considered complexity
sublevels (imaginary unit $i$ in this case) and $\hbar = h / 2\pi$
is Planck's constant.

Note that the above complexity conservation of the quantum beat
process reflects the physically transparent fact of system return to
the \emph{same} wavefunction state after each beat cycle. Causal
quantisation (\ref{Eq:CausQuant}) expresses thus the detailed
complex-dynamic realisation change, or causally specified ``quantum
jumps'', of the quantum beat process within the elementary
field-particle accounting also for its intrinsic ``quantum
uncertainty'' (the corresponding uncertainty and quantisation
relations are only formally postulated in the conventional quantum
mechanics and its unitary modifications describing physically real
particles by purely abstract ``state vectors''). Using relation
(\ref{Eq:CausQuant}) in the Hamilton-Jacobi equation
(\ref{Eq:Ham-Jacob}), we obtain the \emph{causally derived
Schr\"{o}dinger equation} for the \emph{realistically interpreted}
wavefunction:
\begin{equation}\label{Eq:Schrodinger}
i \hbar \frac{{\partial {\mit \Psi} }}{{\partial t}} = \hat H\left(
{x,\frac{\partial }{{\partial x}},t} \right){\mit \Psi} \left( {x,t}
\right) \ ,
\end{equation}
where the Hamiltonian operator, $\hat H\left( {x,\frac{\partial
}{{\partial x}},t} \right)$, is obtained from the Hamiltonian
function $H = H(x,p,t)$ of equation (\ref{Eq:Ham-Jacob}) by the same
causal quantization (\ref{Eq:CausQuant}) and we have used the
continuous derivative notations for brevity. The famous
Schr\"{o}dinger equation containing, in usual theory, the whole
series of inexplicable ``quantum mysteries'' excluding any realistic
physics is obtained now as a \emph{totally causal} consequence of
the universal symmetry of complexity
\cite{Kir:USciCom,Kir:QuMach,Kir:QFM,Kir:100Quanta}.

The \emph{universal} version of Schr\"{o}dinger equation applicable
at \emph{any} complexity level is obtained by the same causal
quantisation of the Hamilton-Jacobi equation:
\begin{equation}\label{Eq:SchrodinUniv}
{\mathcal A}_0 \frac{{\Delta {\mit \Psi} }}{{\Delta t}} \left| _{x =
{\rm const}} \right. = \hat H\left( {x,\frac{\Delta }{{\Delta x} }
\left| _{t = {\rm const}} \right., t} \right){\mit \Psi} \left(
{x,t} \right) \ ,
\end{equation}
where $x$ designates the corresponding dynamically derived system
configuration (section \ref{Subsec:CompIntDyn}) and the generalised
wavefunction, or distribution function, ${\mit \Psi} \left( {x,t}
\right)$ describes intermediate realisation state. The dynamically
derived Schr\"{o}dinger equation
(\ref{Eq:Schrodinger})--(\ref{Eq:SchrodinUniv}) is accompanied by
the generalised, \emph{causally obtained Born rule} for realisation
probabilities $\{\alpha_r\}$ in terms of the wavefunction,
completing the dynamic origin of probabilities in terms of regular
(localised) realisations (\ref{Eq:RealProbab}):
\begin{equation}\label{Eq:BornRule}
\alpha _r  = \left| {{\mit \Psi} \left( {x_r } \right)} \right|^2\ ,
\end{equation}
where $x_r$ is the $r$-th realisation configuration and one may have
the value of the generalised distribution function itself at the
right-hand side for higher, particle-like complexity levels. The
generalised Born rule, extending the corresponding formal postulate
of usual quantum mechanics, is valid for any interaction dynamics at
any level of complexity and results from the dynamic matching
conditions between regular realisations and intermediate realisation
of the wavefunction, giving the values of coefficients $c_i^r$ in
the state-function expression
(\ref{Eq:StateFunc}),~(\ref{Eq:StFunc-Full})
\cite{Kir:USciCom,Kir:QuMach}. This mathematical procedure has a
transparent physical origin in the quantum beat dynamics and the
underlying symmetry of complexity: as the localised, ``corpuscular''
realisations emerge by a direct, interaction-driven dynamical
squeeze of the extended wavefunction realisation to one of redundant
reduction centres, the probability of centre selection will be
proportional to the \emph{physically real} wavefunction magnitude at
the corresponding location. The symmetry of complexity underlies
here the matching condition itself by the evident demand of
\emph{continuity} of complexity transformation in the realisation
change process.

Note that the \emph{dynamic} rules for realisation probabilities
(\ref{Eq:RealProbab}),~(\ref{Eq:FractProb}),~(\ref{Eq:BornRule})
accompanied by their \emph{dynamically fractal} structure (section
\ref{Subsec:CompIntDyn}) describe their ``spontaneous'', unreduced,
but \emph{interaction-driven, purposeful, ``reasonable''} emergence,
which underlies the important property of \emph{dynamic
(probabilistic) adaptability} of real interaction processes
\cite{Kir:USciCom,Kir:QuMach}: the system ``automatically'' goes
everywhere it can and chooses the best possible way for its
complexity development by a natural ``competition'' of dynamically
produced possibilities. That \emph{dynamically probabilistic}
complexity development from dynamic information to entropy
constitutes thus the rigorously defined system \emph{purpose} and
\emph{teleological} power/property of universal symmetry of
complexity.

Equations (\ref{Eq:Ham-Jacob})--(\ref{Eq:BornRule}) form the basis
of the \emph{universal Hamilton-Schr\"{o}dinger formalism} that
unifies extended versions of \emph{all} particular (correct) dynamic
equations postulated in various fields of unitary theory (whereas
the underlying symmetry of complexity unifies causally extended
versions of all usual, postulated laws and ``principles''
\cite{Kir:USciCom}). It can be demonstrated by Hamiltonian expansion
in a power series of momentum and action, which leads to the
following form of universal Schr\"{o}dinger equation
(\ref{Eq:SchrodinUniv})
\cite{Kir:USciCom,Kir:USymCom,Kir:QuMach,Kir:Conscious}
\begin{equation}\label{Eq:SchrodinExpansion}
\frac{{\Delta {\mit \Psi} }}{{\Delta t}} \left| _{x = {\rm const}}
\right. + \sum\limits_{\scriptstyle m = 0 \hfill \atop \scriptstyle
n = 1 \hfill}^\infty  {h_{mn} \left( {x,t} \right)} \left[ {{\mit
\Psi} \left( {x,t} \right)} \right]^m \frac{\Delta ^n {\mit
\Psi}}{{\Delta x^n} } \left| _{t = {\rm const}} \right. = 0\ ,
\end{equation}
where the expansion coefficients $h_{mn} \left( {x,t} \right)$ can
be arbitrary functions and we have taken into account additional
Hamiltonian dependence on action (or wavefunction) through the
``potential energy'' and more generally due to the dynamically
nonlinear EP dependence on the problem solutions (see equations
(\ref{Eq:EP}),~(\ref{Eq:EP-Full}),~(\ref{Eq:AuxEP}) in section
\ref{Subsec:CompIntDyn}). It is important that all dynamic equations
should be provided, within the universal science of complexity, with
the unreduced, \emph{dynamically multivalued} and probabilistic
general solution (\ref{Eq:FractGenSol})--(\ref{Eq:FractProb}), as
opposed to dynamically single-valued solutions of usual theory. The
causally derived Hamiltonian form of the universal formalism
provides also decisive confirmation of the starting existence
equations (\ref{Eq:ExistProto}),~(\ref{Eq:ExistGen}), thus closing
the underlying self-consistent cycle of the symmetry of complexity.

We see that various linear and ``nonlinear'' models and equations,
which are often just semi-empirically ``guessed'' and postulated in
the unitary theory, are obtained in reality as truncated versions of
a general power series of equation (\ref{Eq:SchrodinExpansion}) (or
a similar expansion for the Hamilton-Jacobi equation
(\ref{Eq:Ham-Jacob})) and can therefore be considered as (reduced)
consequences of the single, \emph{unified} law, the symmetry of
complexity. We can also clearly see the difference between the
imitative unitary ``nonlinearity'' due to formal higher powers of a
truncated expansion series and the genuine, \emph{dynamically}
emerging, \emph{essential nonlinearity} due to the unreduced EP
dependence of the solutions to be found. Contrary to popular
confusion of usual ``science of complexity'', the former, imitative
``nonlinearity'' \emph{cannot} provide any true complexity and
chaoticity by itself, without the proper, unreduced analysis of a
real interaction process revealing the dynamically probabilistic
fractal of the complete general solution. That usual nonlinearity
resembles an artificially, trickily entangled one-dimensional thread
that can, however, be completely disentangled and does not change
its basic properties upon any smooth change of configuration. Since,
on the other hand, essential nonlinearity emerges even for formally
``linear'' initial problem formulation (section
\ref{Subsec:CompIntDyn}), one can assume that \emph{any} usual,
formal ``nonlinearity'' is but a reduced representation of genuine,
dynamic nonlinearity of real interaction process.

As noted above, the symmetry of complexity unifies \emph{causally
extended}, universally applicable versions of various
\emph{separated}, individually \emph{postulated} laws and
``principles'' of usual fundamental science, such as conservation of
energy (or ``first law of thermodynamics''), entropy growth
(``second law of thermodynamics''), all ``quantum'' and
``relativistic'' postulates and principles (see section
\ref{Subsubsec:Relativity}). Many of them are related to the
corresponding unitary, abstract symmetries which, besides being
separated among them, appear to be practically always ``broken'' by
the full-scale, real-world dynamics reducing them to a status of
unrealistic, ``approximate'', and therefore \emph{false} symmetry
that can be ``more or less'' valid only within a limited,
ambiguously defined parameter range. Indeed, the evident
irregularity of real-world structures and dynamics is basically
different from the ``too symmetric'', regular and smooth structures
of the unitary, abstract science paradigm. The universal symmetry of
complexity solves the problems of separation, violation and
excessive regularity of usual symmetries by proposing not only
\emph{intrinsically unified}, but also \emph{always exact, unbroken}
symmetry describing real-world irregularity by its own,
\emph{dynamic randomness} (due to chaotic transitions between
asymmetric realisations). Therefore now \emph{all} real-world
structures (described by essentially random general solution of
probabilistic dynamical fractal
(\ref{Eq:FractGenSol})--(\ref{Eq:FractProb})) are explicitly
obtained as \emph{absolutely symmetric} results of complexity
conservation and development supported by the \emph{totality of
existing observations}.

An important general manifestation of the universal symmetry of
complexity takes the form of \emph{complexity correspondence
principle} that can have various particular formulations, but always
emphasises the fact that any interaction result depends critically
and totally upon relative complexities of interacting entities
\cite{Kir:USciCom,Kir:QuMach,Kir:Conscious}. Specifically,
interaction between several (complex) systems can be ``efficient''
(induce essential changes) only for interacting systems of
comparable complexity. Moreover, the system with higher complexity
tends to ``control'', or ``enslave'', less complex interaction
partners, which gives rise to \emph{complex-dynamic control theory}
that unifies and extends essentially usual, unitary control concepts
by showing, in particular, that any real control result and
mechanism are \emph{basically chaotic} and can never be absolute. If
interacting system complexities are very close to each other, a
strong, ``global'' chaos regime can result.

All particular cases of real (complex) interaction dynamics can be
conveniently classified and unified in a single scheme and criterion
of unreduced interaction results
\cite{Kir:USciCom,Kir:SelfOrg,Kir:QuMach,Kir:Conscious,Kir:QuChaos,Kir:Channel}.
If the key interaction parameters (properly represented by
characteristic frequencies) are close enough to each other, one
obtains the limiting case of \emph{uniform, or global, chaos} with
rapidly changing, essentially different system realisations and
homogeneous distribution of their probabilities. If the
characteristic system parameters are essentially different, one gets
the opposite limiting case of generalised, \emph{dynamically
multivalued self-organisation, or self-organised criticality (SOC)},
that unifies, besides those two concepts, the extended versions of
other cases, such as synchronisation, control of chaos, mode
locking, and fractality (they remain separated in their unitary,
dynamically single-valued versions). It contains a small number of
rarely changing ``compound'' realisations that confine, however, a
multitude of rapidly and chaotically changing but externally similar
``elementary'' realisations within them. The almost total
\emph{external} regularity of ultimate SOC cases passes gradually
(though unevenly) to the maximum irregularity of global chaos with
the corresponding change of characteristic frequency ratio, so that
one can describe and classify, in principle, \emph{all} possible
dynamic regimes in any kind of system.

Specifically, the point of transition to the strong, uniform chaos
is expressed by the \emph{universal criterion of global chaos
onset}:
\begin{equation}\label{Eq:ChaosCrit}
\kappa  \equiv \frac{{\Delta \eta _i }}{{\Delta \eta _n }} =
\frac{{\omega _\xi  }}{{\omega _q }} \cong 1\ ,
\end{equation}
where $\kappa$ is the introduced \emph{chaoticity} parameter, while
$\Delta \eta _i$, $\omega _\xi$ and $\Delta \eta _n \sim \Delta
\varepsilon$, $\omega _q$ are energy-level separations and
frequencies for inter-component and intra-component motions,
respectively. At $\kappa \ll 1$ one has an externally regular
multivalued SOC regime, which degenerates into global chaos as
$\kappa$ grows from 0 to 1, and maximum irregularity at $\kappa
\approx 1$ is again transformed into a SOC kind of structure (but
with a ``reversed" configuration) at $\kappa \gg 1$.

Using this universal chaos criterion, it is easy to see, in
particular, the dynamic origin of \emph{fundamental quantum
randomness}, or ``indeterminacy'', appearing in the form of
\emph{inevitably} strong (global) chaoticity of protofield
interaction process at those \emph{lowest}, ``quantum'' levels of
the world structure complexity. Indeed, the characteristic
frequencies, or eigenvalue separations, at the lowest complexity
sublevels containing only elementary structures (field-particles)
coincide par excellence as other, essentially different system
parameters ``have not yet appeared'' in that \emph{essentially
quantum} reality (their definite appearance marks the emergence of
the next complexity level of elementary \emph{classical, permanently
localised} structures with a much more regular, SOC kind of dynamics
\cite{Kir:USciCom,Kir:QuMach,Kir:QFM,Kir:100Quanta,Kir:Cosmo}, see
also section \ref{Subsubsec:Classicality}). Specifically, the
quantum beat frequency determines both internal field-particle
dynamics and its ``external'' motion and interactions. A higher
sublevel of quantum complexity, that of \emph{(true) quantum chaos}
and \emph{(causal) quantum measurement} (section
\ref{Subsubsec:Classicality})
\cite{Kir:USciCom,Kir:QuMach,Kir:QuChaos}, already contains a
possibility of somewhat more regular, SOC kind of dynamics that
further passes to a yet more regular case of classical behaviour.

\subsection{Universe and particle properties by the symmetry of complexity}\label{Subsec:Properties}
Using \emph{only} the unreduced, universally nonperturbative
analysis of sufficiently strong attractive interaction of two
physically real, initially homogeneous protofields, we have shown
above, in section \ref{Subsec:CompIntDyn}, that the elementary
field-particle will generically emerge from that interaction (for
suitably chosen but non-exotic protofield ``material''), in the form
of \emph{spatially chaotic} process of \emph{quantum beat} that can
be described as \emph{unceasing} cycles of protofield
reduction-extension or, alternatively, as \emph{chaotic wandering}
of the transient corpuscular state of \emph{virtual soliton}. The
resulting, dynamically multivalued, intrinsically \emph{unified} and
totally \emph{causal} (realistic) picture of microworld dynamics is
called \emph{quantum field mechanics}
\cite{Kir:USciCom,Kir:QuMach,Kir:QFM,Kir:100Quanta,Kir:Cosmo}, as
opposed to various irreducibly \emph{separated} branches of
\emph{unrealistic (abstract)} and largely \emph{postulated}
(formally imposed) \emph{unitary}, dynamically single-valued theory,
such as quantum mechanics, field theory, particle (high-energy)
physics, special and general relativity, and cosmology, including
their recent, ``advanced'' versions that always remain, however,
within the same, effectively \emph{zero-dimensional projection} of
reality (e.\,g. ``many-world'', ``histories'', and other abstract
``interpretations'' of quantum mechanics, string and spin-network
schemes of modern field theory, brane-world imitations, innumerable
``cosmological'' tricks with ``hidden'' material species, dimensions
and whole ``multi-verses'', etc.). The \emph{universal symmetry of
complexity} (section \ref{Subsec:SymCom}) totally determines the
unreduced interaction development, and we shall continue to derive
further emerging world structures and their properties,
demonstrating the power of the symmetry of complexity to avoid and
solve the accumulating problems of unitary theory and its
simplified, regular symmetries.

\subsubsection{Dynamic origin of 3D space, time,
and elementary particles: Occam's razor}\label{Subsubsec:SpaceTime}
We begin our analysis of the causal, physically real, explicitly
emerging, and \emph{always exactly symmetric} world structure with
recalling the dynamic origin of the naturally quantised,
\emph{tangible space structure} and \emph{irreversibly flowing but
immaterial time} obtained above (section \ref{Subsec:CompIntDyn})
from the protofield interaction description as, respectively,
\emph{eigenvalue separation}, $\Delta x = \Delta _r \eta _i^r =
\mathchar'26\mkern-10mu\lambda _{\rm C}$, of effective existence
equation (\ref{Eq:ExistEff}) and \emph{intensity} (specified as
\emph{frequency}, $\nu$) of quantum beat realisation
emergence/change, $\Delta t = {{\Delta x} \mathord{\left/ {\vphantom
{{\Delta x} c}} \right. \kern-\nulldelimiterspace} c} =
{{\mathchar'26\mkern-10mu\lambda _{\rm C} } \mathord{\left/
{\vphantom {{\mathchar'26\mkern-10mu\lambda _{\rm C} } c}} \right.
 \kern-\nulldelimiterspace} c} = \tau  = {1 \mathord{\left/
 {\vphantom {1 \nu }} \right. \kern-\nulldelimiterspace} \nu }$. The
space ``coordinate'' $x$ expresses, in general, \emph{configuration}
of \emph{explicitly emerging} system realisation (in the form of
localised virtual soliton), while ``time flow'' (permanently growing
$t$) reflects inevitable \emph{change} of \emph{multiple} and
\emph{incompatible} realisations. Space and time appear thus as
universal, \emph{basic manifestations} of unreduced interaction
complexity and its symmetry/transformation, \emph{together} with the
\emph{system structure and dynamics} itself (represented here by an
elementary field-particle, such as the electron, with the
\emph{dynamically determined size} $\Delta _i \eta _i^r  = r _e$,
performing its \emph{quantum jumps} to the distance $\Delta x =
\Delta _r \eta _i^r = \mathchar'26\mkern-10mu\lambda _{\rm C}$ with
the period of $\Delta t = \Delta x/c$, see sections
\ref{Subsec:CompIntDyn} and \ref{Subsubsec:ConstPlanckian}).

Unitary space-time symmetries are strongly \emph{broken} (and
therefore \emph{illusive}) by \emph{dynamic discreteness}
(quantisation) of space and irreversible, \emph{oriented flow}
(increase) of time variable in a well-defined direction of growing
complexity-entropy (see equation (\ref{Eq:TimeArrow})), whereas the
symmetry of complexity just underlies and \emph{gives rise} to those
properties of space and time, remaining absolutely \emph{exact}
symmetry. That fundamental violation of irreducible ``smoothness''
(regularity) of unitary projections will continue and involve higher
complexity levels and symmetries, e.\,g. those from theories of
relativity, gravity, and cosmology (see sections
\ref{Subsubsec:Self-Tuning},~\ref{Subsubsec:Relativity},~\ref{Sec:DarkMatter}).
In particular, any \emph{direct mixture} between space and time
entities within a single symmetry (constituting the basis of
conventional relativity) is physically \emph{senseless}, already
because of tangible, material structure of real space and immaterial
time origin: real space and time are related \emph{by and only by
the system dynamics}, which is none other than \emph{direct
realisation} of the \emph{symmetry of complexity}.

The same interaction-based origin of physically real space and time
shows that time cannot be ``curved'', or deformed, in any sense at
all, while space emerges as a globally (in average) flat and
homogeneous structure, in agreement with observations and contrary
to the corresponding unitary theories (general relativity and
cosmology). Moreover, any space inhomogeneity emerging at a higher
complexity level is an average density/tension modification of
protofields (see also section \ref{Subsubsec:Relativity}) that can
only formally (and very approximately) be described
``geometrically'', similar to any other long-range interaction
through a (dense enough) continuous medium.

The symmetry of complexity directly determines also the observed
\emph{number (three) of space dimensions} and establishes its
\emph{universal physical origin} and link to the \emph{interactive}
base of \emph{any} real world. Indeed, the initial interaction
configuration includes \emph{three} and only three \emph{global}
entities, the two protofields and their physically real coupling
(interaction) itself (Figure \ref{Protofields}). The symmetry of
complexity tells us that the number of equally global entities
resulting from that interaction should be the same, i.e. equal to
three. But the only resulting entity of the truly \emph{global}
scale is the fundamental \emph{space} structure itself, which
\emph{should} therefore have \emph{three and only three modes, or
``dimensions''}, according to complexity conservation law rigorously
substantiated and supported by the totality of all experimental
observations \cite{Kir:USciCom}. We obtain thus also the genuine,
\emph{physical origin} of those space ``dimensions'' as such,
remaining only empirically and formally defined in usual science. As
the tangible space ``material'' is obtained by dynamically
multivalued \emph{entanglement} of global interaction partners, the
protofields (see section \ref{Subsec:CompIntDyn}), its \emph{global
degrees of freedom}, or ``dimensions'', are none other than
\emph{physically real}, equivalent ``modes'', or realisations, of
that \emph{complex-dynamic mixture} of interacting e/m and
gravitational protofields (including the coupling interaction
itself).

The obtained rule for the number of space dimensions and their
physical origin is valid for any other system, including higher
levels of universe space structure and other possible universes. In
particular, the number of (global) space dimensions of arbitrary
universe is equal to the number of initial interaction components
(including coupling entities). Depending on the driving interaction
details, further split into inhomogeneously structured ``compound''
dimensions is possible, with a three-dimensional space ``unity''
remaining the ``minimal'', most stable combination (because one
\emph{cannot} have less than two interaction components). Although
various complicated cases are possible, the symmetry of complexity
provides a realistic and efficient ordering and understanding
principle, as opposed to arbitrary unitary guesses on the subject
based on the ``demands'' of a purely abstract, \emph{postulated}
formalism that eventually appears to be but an effectively
zero-dimensional projection of any real-world structure. Thus,
according to the symmetry of complexity, higher-dimensional
universes appear from \emph{higher-complexity interactions} as a
sort of ``excited states'' over the exceptionally stable (and
therefore most common, if not unique) ``ground state'' of
three-dimensional world. The latter may have \emph{only one}
irreversibly flowing \emph{time}, which may also be true for any
\emph{unified} higher-dimensional world. But a more complicated
substructure of global space dimensions can give rise to multiple
time flows in a higher-dimensional world that would realise a much
higher, ``excited-state'' complexity of such ``multi-time'' world
\cite{Kir:USciCom}. Despite ``purely theoretical'' character of
those possibilities, one can easily have ``higher-dimensional'' and
``multi-time'' situations with \emph{local} realisation structure at
\emph{higher levels} of complexity, space and time of the present,
\emph{globally} three-dimensional world.

We shall see below (section \ref{Subsubsec:Interactions}) that the
number of fundamental interaction forces (and particle species)
enters into the same physically transparent manifestation of the
symmetry of complexity, leading to considerable reduction of the
(practically unlimited) number of \emph{formal} possibilities of
unitary theory, such as ``hidden'' dimensions and other strangely
``invisible'' entities. It becomes clear that space and its
dimensions have a \emph{physically real} origin in a global
\emph{interaction process} and therefore should \emph{not} be
introduced artificially, by \emph{ad hoc} assumptions in order to
save a contradictory imitation of reality, as it is done in the
unitary theory. The symmetry of complexity provides, in this sense,
the \emph{rigorous} and \emph{practically efficient} extension of
the well-known \emph{Occam's razor, or principle of parsimony}, as
it specifies \emph{how exactly} each real, observed entity emerges
in an \emph{interaction process} from other, equally real entities,
which provides a reliable way of their specification
\cite{Kir:USciCom}. One obtains also a \emph{realistic} extension of
\emph{G\"{o}del incompleteness} theory, where any interaction result
``incompleteness'' is due to its intrinsic uncertainty
(multivaluedness) and (partially unknown) interaction components.

\subsubsection{Universal dynamic origin of particle mass, charge, and spin}\label{Subsubsec:ParticleProp}
Since the first-level world structures, elementary field-particles,
emerge together with physically real space and time (see Figure
\ref{Protofields}), the same complex-dynamic process of quantum beat
should give rise to the ``intrinsic'' particle properties, such as
mass, electric charge, spin, and their observed features. We start
with the major property of \emph{mass} and can state that its key
feature of \emph{inertia} is universally and consistently explained
by the \emph{dynamically chaotic} character of the spatial wandering
of virtual soliton within any (massive) field-particle (quantum beat
process), \emph{rigorously} obtained above (section
\ref{Subsec:CompIntDyn}). Indeed, it is this \emph{already
existing}, never vanishing \emph{internal} motion of the particle
``matter'' that determines its ``resistance'' to any external
\emph{force} (attempt to \emph{change} it) and ensures \emph{finite}
values of acquired \emph{acceleration}. It is evident that anything
different from \emph{purely dynamic, internal} chaoticity
\emph{cannot} solve the problem of \emph{intrinsic} inertia in
principle, including any external influence (e.\,g. of ``zero-point
field'' fluctuations) often arbitrarily assumed in the unitary
theory. Moreover, we show that inertial mass thus \emph{dynamically
emerging} in the unreduced protofield interaction is synonymous, or
``equivalent'' (up to a coefficient or measurement unit), to the
total, ``relativistic'' \emph{energy} and expresses therefore a
differential form of system \emph{complexity} (see equation
(\ref{Eq:Energy}))
\cite{Kir:USciCom,Kir:QuMach,Kir:QFM,Kir:100Quanta,Kir:Cosmo}.
Following universal definitions of complexity-action, energy and
momentum (\ref{Eq:Energy}),~(\ref{Eq:Momentum}) in section
\ref{Subsec:SymCom}, we obtain for the field-particle \emph{at rest}
($p = 0$): $\Delta {\mathcal A} =  - E_0 \Delta t$ and
\begin{equation}\label{Eq:RestMassEnergy}
E_0  =  - \frac{{\Delta {\mathcal A}}}{{\Delta t}} = \frac{h}{{\tau
_0 }} = h\nu _0  = m_0 c^2 \ ,
\end{equation}
where $E_0$ is the particle \emph{rest energy}, $\Delta {\mathcal A}
= - h$ is the \emph{dynamically discrete} complexity-action
increment equal at this \emph{first} complexity level to
\emph{universal} Planck's constant $h$ with the negative sign (since
$E_0,\Delta t > 0$), $\tau _0$ is the quantum beat period and $\nu
_0$ frequency for the field-particle at rest, $m_0$ is the particle
\emph{rest mass} introduced above, and $c^2$ is a coefficient for
the moment, but later rigorously shown to be equal indeed to the
square of light velocity. One also obtains here the explicit
expression of the elementary dynamical \emph{clock of the universe}
within each (massive) particle (section \ref{Subsec:CompIntDyn})
that has a sufficiently high frequency ($\nu _0 \sim 10^{20} \ {\rm
Hz}$ for the electron) and provides the causal, physically real
basis for the famous relation $h\nu _0  = m_0 c^2$ used by Louis de
Broglie in his original derivation of particle (``de Broglie'')
wavelength \cite{deBroglie:These,Kir:75MatWave} and confirmed
recently by an electron channeling experiment \cite{ChannQuantBeat}.
We develop below this unified causal interpretation of mass, energy,
and time to the case of moving particles and obtain the
\emph{dynamically} derived effects of (special) relativity (section
\ref{Subsubsec:Relativity}).

The \emph{multitude} of particle species, reflected by their
observed \emph{mass spectrum}, is obtained as a consequence of
fundamental \emph{dynamic multivaluedness} of the protofield
interaction process, where the light family of \emph{leptons}
represented by the absolutely stable electron is obtained as a
compound realisation with a relatively small quantum beat amplitude,
so that e/m protofield pulsation remains rather ``close'' to the
unperturbed protofield state in Figure \ref{Protofields}. The
opposite case of strongest effective protofield interaction is
obtained for the compound realisation of heavy particles,
\emph{hadrons}, represented by the stable species of proton. Their
composition of explicitly nonseparable quarks corresponds to a
compound structure of quantum beat process that cannot be split,
however, into separate interacting beats for individual quarks. This
involvement of quarks, their unique role in strong interaction
force, and the absence of strong interaction for leptons can be
uniquely explained by the fact that the gravitational protofield, or
medium, is represented by a dense \emph{quark condensate} (probably
with ``quantum'' properties like superfluidity and with unknown
degree of separate quark individuality as localised, corpuscular
states). Recent experimental evidence in favour of \emph{quark-gluon
liquid} \cite{QuarkGluonLiquid} (\emph{rather than} expected
\emph{plasma} of usual theory) confirms this conclusion and the
whole picture of quantum field mechanics.

The compound realisations of leptons and hadrons are further split
into three canonical ``generations'' closely resembling ``excited
states'' of their stable, weakest-interaction species, which
corresponds very well to our interpretation in terms of different,
progressively growing quantum beat (or protofield EP) amplitude. As
for the main \emph{massless} species of \emph{photon}, it is
represented by a basically \emph{regular}, non-chaotic oscillation
process with relatively very small amplitude, which is additionally
stabilised by permanent attraction to the gravitational protofield
and resembles thus \emph{ordinary, regular solitons} (the tiny
remaining dissipation of such photon energy provides just the
necessary features for the consistent explanation for the
\emph{cosmological red shift} phenomenon, \emph{without} the
contradictory Big Bang hypothesis, see \cite{Kir:Cosmo} and section
\ref{Subsec:DarkEnergyBB}). The case of massless photons emphasizes
once more the role of \emph{truly chaotic}, multivalued
\emph{internal} dynamics, and the associated \emph{symmetry of
complexity}, in the emerging property of particle mass. It also
avoids artificial introduction of \emph{additional} entities giving
rise to mass, such as hypothetical but never found Higgs particles
and field from the unitary theory (contradicting the above causally
extended Occam's razor, see section \ref{Subsubsec:SpaceTime}). One
can see here that our interacting protofield construction is indeed
explicitly ``economical'' as it gives rise, within the \emph{same}
quantum beat process, to \emph{both} elementary field-particles and
their intrinsic properties (as well as to all other, dynamic
properties, as we shall see below, in section
\ref{Subsubsec:Relativity}).

Since all massive particles live within the same, \emph{physically
unified} protofield volume (mainly perceptible from the e/m
protofield side), their respective quantum beat processes should be
\emph{synchronised} in time, which is necessary for both
\emph{coherent particle interaction} (especially evident for the
case of \emph{attraction}) and \emph{unified time flow} for the
whole universe. Such complex-dynamic synchronisation is a subject of
separate study, but irrespective of its details one knows the final
result: temporal pulsation phases of all quantum beat processes
\emph{coincide} up to phase inversion (i.e. one may have either the
same or opposite pulsation phases).\footnote{Such synchronisation
provides, in particular, a candidate \emph{dynamic} origin of the
observed \emph{particle-antiparticle asymmetry}, in contradiction
with formal symmetry between particle and antiparticle properties.
The propagating ``wave'' of \emph{inevitable} complex-dynamic
synchronisation of quantum beat processes will automatically leave
only same-phase/antiphase particle species coupled also to their
related spin vorticity (see below in this section). This necessary
dynamic ``ordering'' phase of material universe content implies also
essential modification in the related problems of universe age,
dimension, isotropy, etc. (see also section
\ref{Subsec:GlobalCosmo}).} This important feature leads, in
addition, to dynamic interpretation of the next major intrinsic
property of \emph{electric charge} that emerges now as
\emph{phase-related measure} of the same \emph{quantum beat
complexity}. Indeed, the synchronised field-particles are naturally
subdivided into \emph{two and only two} ``opposite'' species
according to their quantum beat phase, which explains the existence
of two ``opposite'' charges. Like charges represented by quantum
beat processes with the same phases will naturally \emph{repulse}
each other because of their direct, ``mechanical'' competition for
the common e/m protofield material, while unlike charges will
naturally \emph{attract} each other due to a mutual ``help'' of
their reduction-extension processes with opposite phases
\cite{Kir:USciCom}. The famous ``quantisation'' of elementary charge
(its fixed observed value), remaining unexplained in the unitary
theory, is due to the same global \emph{phase synchronisation} of
all quantum beat processes (most probably at the frequency of
\emph{electronic} quantum beat) and thus eventually due to
\emph{quantisation} of their complex dynamics (i.e. dynamically
discrete structure of the symmetry of complexity).

The described direct link between elementary electric charge $e$ and
quantised complexity of the quantum beat process (expressed
according to equation (\ref{Eq:RestMassEnergy}) by the
complexity-action quantum $\hbar = h / 2 \pi$) constitutes the
genuine, causal content of the well-known relation between $e$ and
$\hbar$, $e^2  = \alpha c \hbar$, where $\alpha$ is the fine
structure constant. We shall see below (section
\ref{Subsubsec:ConstPlanckian}) that it leads also to the new
interpretation of the latter (together with Planck's constant
universality). Needless to say, the \emph{electric charge
conservation law}, appearing as a \emph{separate} and
\emph{postulated} (empirical) law in the conventional theory,
obtains now causal and universal extension as a particular case of
\emph{dynamically substantiated} symmetry (conservation) of
complexity.

It is easy to see that dynamic reduction (squeeze) of the
\emph{physically real} e/m protofield within each quantum beat cycle
of an elementary field-particle should involve a strong vortical
\emph{twirl} of the squeezing protofield matter, simply due to its
\emph{finite compressibility}. The phenomenon can be described as a
highly nonlinear (self-amplified) version of a liquid whirlpool
appearing when a liquid is forced, usually by gravitational field,
to pass through a small hole. The \emph{unique} feature of
\emph{unreduced} quantum beat interaction and in particular
self-amplifying \emph{dynamic entanglement} (section
\ref{Subsec:CompIntDyn}) is that it produces, in a \emph{purely
dynamic}, ``spontaneous'' way, a never-ending series of such
``holes'', or protofield reduction centres. The detailed mechanism
of protofield vortex emergence itself is similar to usual
instability against local shift deformation of the liquid/gas flow,
where more rapidly moving parts (closer to the ``hole'') experience
sideways ``twisting'' deviations due to simultaneously emerging
pressure differences in the inhomogeneously moving matter. The
emerging twirl continues in the extension phase, and one obtains in
the whole the \emph{physically real}, dynamic and \emph{essentially
nonlinear} origin of the universal intrinsic property of elementary
particle \emph{spin}. The complex, multivalued dynamics of
protofield interaction provides just a \emph{unique} combination of
properties for this consistent causal interpretation of spin, as
opposed to any unitary ``rotating ball'' models.

Moreover, the quantitative expression of spin, $s$ (and any other
angular momentum $I$), in terms of angular momentum quantum $\hbar$
($s = \hbar / 2$ for the electron) presents it as \emph{another
form} of (naturally quantised) quantum beat \emph{complexity} and
reveals the origin of the deep dynamical connection between
complexity-action quantum $h$ for quantum beat \emph{pulsation} and
angular momentum quantum $\hbar$ for spin \emph{rotation} within
\emph{the same} quantum beat process \cite{Kir:USciCom}. Universal
expression for complexity-action increment will, in general, contain
a ``rotational'' term, $\Delta {\mathcal A} = - E \Delta t + p
\Delta x + I \Delta \phi$, where $\Delta \phi$ is the angle variable
increment, so that the rest energy ($p = 0$) of (for example) the
electron will contain a contribution from the spin rotation energy:
\begin{equation*}
E_0  = \frac{{h\nu _0 }}{2} + s\omega _0  = \frac{{h\nu _0  + \hbar
\omega _0 }}{2} = h\nu _0  = \hbar \omega _0 \ ,
\end{equation*}
where the circular frequency of spin \emph{rotation} $\omega _0$
should coincide with the circular frequency of quantum beat
\emph{pulsation}, $\omega _0 = 2 \pi \nu _0$, as it is \emph{one and
the same process}, so that its energy partition into contributions
from ``pulsation'' and ``spin rotation'' can have only conventional
meaning, as shown above. The ``anomalous'' values of electron spin
and gyromagnetic ratio obtain now a \emph{causal} interpretation in
terms of two-phase structure of the electronic quantum beat process
\cite{Kir:USciCom}. The spin-induced rotation of the e/m protofield
matter can now be seen also as fundamental \emph{physical origin} of
\emph{magnetic field and effects} \cite{Kir:USciCom}. And similar to
the above case of electric charge, all conservation laws involving
angular (spin and orbital) momentum are universally extended now to
causally substantiated, unified symmetry of complexity. We can
clearly, directly see how all the diverse quantities conserved
according to \emph{formally imposed (empirical)} conservation laws
of the unitary science are obtained and conserved as \emph{measures}
of only \emph{externally} different manifestations (or levels) of
the same \emph{dynamic complexity of unreduced interaction process},
which specifies the \emph{physically real, unified origin} of both
conserved quantities and their conservation.

\subsubsection{Dynamically unified fundamental interactions, their number and properties}\label{Subsubsec:Interactions}
The above intrinsic particle properties are related to fundamental
interactions between particles, which naturally emerge in quantum
field mechanics in their \emph{dynamically unified} state and
observed properties
\cite{Kir:USciCom,Kir:QuMach,Kir:QFM,Kir:100Quanta}. The unified
dynamic origin of all particle interactions is the underlying
protofield attraction, in its ``implemented'' form of quantum beat
processes within each field-particle. As every such particle-process
changes the surrounding protofield properties (because of protofield
deformation), it will influence the quantum beat parameters of any
other particle (by certain analogy to ``deformation interaction''
between solid state defects and excitations). As a result, one
obtains two such long-range fundamental interactions through e/m and
gravitational media, the e/m and gravitational interactions, which
explains now the respective protofield names. The e/m interaction
forces are introduced in the previous section, while the dynamic
gravitation mechanism thus obtained provides the causal,
\emph{physically real} basis for \emph{universal gravitation}
(absent in any its unitary description), which possesses all its
observed classical properties, \emph{intrinsically quantum} origin
(due to quantum beat discreteness) avoiding usual quantum gravity
problems (cf. \cite{NoGravitons}), and \emph{relativistic} effects
without any artificial ``geometrisation'' of physically real space
and flowing time (see section \ref{Subsubsec:Relativity} for more
details). The other two, short-range interaction forces, known as
``weak'' and ``strong'' interactions, are simply due to
close-contact forces between constituent elements (remaining
basically unresolved) of e/m and gravitational media respectively.
We obtain thus \emph{exactly the observed number (four)} of
fundamental interactions with their \emph{observed properties}
(including two short-range and two long-range interactions for two
protofields).

Moreover, all the four interactions are \emph{naturally unified}
from the beginning within every quantum beat process (the total
unity is obtained in the maximum reduction phase for the heaviest
hadronic particles), which resolves the notorious ``grand
unification'' problem (also known as the ``theory of everything''),
or even \emph{avoids} any ``big problem'' around such unification
that \emph{seems} a kind of ``magic dream'' and desired
``super-goal'' in the \emph{unitary} theory just because of
\emph{its specific}, effectively zero-dimensional,
\emph{intrinsically split} and postulated, abstract \emph{imitation
(projection)} of reality. We demonstrate below a rigorously
specified expression of dynamic interaction unification that
provides a practically important solution of a series of problems
related to Planckian unit values and interpretation (section
\ref{Subsubsec:ConstPlanckian}). We can also confirm and understand
the \emph{causal}, physical origin of ``partial'' unification of
\emph{e/m and weak interactions} by their common material,
transmitting basis of e/m protofield and can \emph{predict} a
similar (though maybe different in details) unification of
\emph{gravitational and strong interactions} by the \emph{common
gravitational medium} actually represented, as mentioned in the
previous section, by a dense quark condensate. The \emph{forces} of
particle interaction as such emerge as a general consequence of the
symmetry of complexity, in the form of \emph{complexity development
from dynamic information to entropy}, i.e. particles are forced to
move so as to preserve the total system complexity by an optimal
\emph{increase of its dynamic entropy} (through structure creation)
at the expense of dynamic information (or ``generalised potential
energy''). Rigorous expression of this law is provided by the
universal Hamilton-Schr\"{o}dinger formalism
(\ref{Eq:Ham-Jacob})--(\ref{Eq:SchrodinExpansion}), which can be
further specified to reveal unified relativistic, gravitational, and
quantum effects (section \ref{Subsubsec:Relativity}).

The dynamic structure of fundamental particle interactions thus
causally derived can be further specified, including e.\,g.
physically real extension of photon exchange processes for e/m
interactions \cite{Kir:USciCom}, which are described by purely
abstract means as unreal (``virtual'') processes in usual theory.
However, we shall concentrate here on important general relation
between \emph{numbers} of most \emph{fundamental entities}
(dimensions, forces, and particles) valid for \emph{any real}
universe and following from the symmetry of complexity and its
causal manifestations. According to the dynamic origin of real space
dimensions (section \ref{Subsubsec:SpaceTime}), a world emerging
from interaction of $n$ initial entities (protofields) will have $N
_{\rm dim} = n + 1$ \emph{global} space dimensions (and one
irreversibly flowing time), which is a direct consequence of the
symmetry of complexity. As shown above in this section, the same
world will have $N _{\rm F} = 2n$ ``fundamental'' interaction forces
between its (dynamically emerging) particles, physically transmitted
through those $n$ protofields and subdivided into $n$ short-range
and $n$ long-range forces (within protofield configuration). One
obtains thus the following relation between the numbers of (any)
world forces and (space) dimensions:
\begin{equation}\label{Eq:NumForceDim}
N _{\rm F} = 2 \left( N _{\rm dim} - 1 \right) , \ \ N _{\rm dim} =
\frac{N _{\rm F}}{2} + 1 \ ,
\end{equation}
where $N _{\rm dim} - 1$ interaction forces (one half of their total
number) have long-range character, while other $N _{\rm dim} - 1$
forces are short-range, ``contact'' ones. This relation can be more
general than the underlying dependencies of $N _{\rm F}$ and $N
_{\rm dim}$ on the protofield number $n$. Indeed, possible more
complicated, ``non-global'' structure of protofield interactions can
give rise to various ``partial'', (half-) hidden dimensions and
``rare'' forces, but relation (\ref{Eq:NumForceDim}) between numbers
of those \emph{emerging} entities will remain valid due to its
well-specified \emph{physical} basis and related \emph{absolutely
universal} symmetry of complexity. As any starting protofield entity
may have, in principle, more than one short-, middle-, or long-range
kind of excitations and interaction transmission ways, one should
also envisage a yet more general form of relation
(\ref{Eq:NumForceDim}), in the form of \emph{lower limit} to the
number of forces, $N _{\rm F} \ge 2 \left( N _{\rm dim} - 1
\right)$, even though the strict inequality here should be
considered as a more exotic possibility. The latter is hardly
realised in our world, where we have $n = 2$, $N _{\rm dim} = 3$,
and $N _{\rm F} = 4$, according to equation (\ref{Eq:NumForceDim}).
Next higher-dimensional universes should have $\{ n = 3, \ N _{\rm
dim} \ge 4, \ N _{\rm F} \ge 6 \}$, $\{ n = 4, \ N _{\rm dim} \ge 5,
\ N _{\rm F} \ge 8 \}$, and so on. Another relation following from
(\ref{Eq:NumForceDim}) may be its yet more universal consequence: $N
_{\rm F} - N _{\rm dim} = N _{\rm dim} - 2 \ge 1 \left( = n - 1
\right)$, or $N _{\rm F} \ge N _{\rm dim} + 1$, where the universal
inequality follows from the fact that $N _{\rm dim} \ge 3$ (or $n
\ge 2$) for any real, \emph{interaction-based} world and shows that
\emph{any additional, real dimension brings about additional
interaction forces} (and related particles).

The latter statement introduces important application of equation
(\ref{Eq:NumForceDim}) and similar relations. As the number of
interaction forces can be experimentally checked, it \emph{strongly
limits} the number of \emph{any real} (``large'', ``small'', or
``hidden'') dimensions ($N _{\rm dim} \leq \frac{N _{\rm F}}{2} +
1$) and gives rise to additional doubts in various popular
violations, in the usual theory, of Occam's razor principle
(\emph{following} from the symmetry of complexity) by introduction
of additional dimensions \'{a} la carte, according to
\emph{internal}, ``mathematical'' needs of an \emph{abstract}
``model'' (e.\,g. in string theory, quantum gravity, brane world
models, etc.). Those doubts are yet more amplified if we take into
account that each new force implies new (observable) particles (or
excitations), so that the number of particles $N _{\rm part}$ will
in any case be greater than the number of short-range forces (or
protofields), or $N _{\rm part} \ge N _{\rm F} / 2 = N _{\rm dim} -
1$ (which corresponds to unrealistic, absolute minimum of exchange
particles). Thus, in our world with \emph{four} interaction forces
and \emph{three} dimensions we have indeed \emph{two} such ``really
irreducible'' (and \emph{therefore} stable) particles in the form of
electron and proton originating from the respective protofields
(without counting more ephemeral photon and all the unstable and
``excited-state'' species). A world obtained by $n$ protofield
interaction will have at least $N _{\rm F} = 2n$ fundamental
interaction forces, $N _{\rm dim} = n + 1$ space dimensions, and $N
_{\rm part} = n$ ``irreducible'' (i.e. rather stable and
``strongly'' observable) particles. Needless to say, all the
unitary, ``anti-Occamian'', entity-producing theories directly and
strongly violate even least restrictive, unrealistic versions of
those relations between the numbers of dimensions, basic forces and
particles. Instead of incorrect imposition of arbitrary fantasies
from a purely abstract ``reality'', one can use the above
\emph{causally substantiated} consequences of the universal symmetry
of complexity for \emph{deduction} of the total number of (any)
world dimensions from the number of its observed fundamental forces
and ``basic'' elementary particles (as opposed to any ``intuitive''
or formal way of definition of space dimensions and their number).

\subsubsection{Complex-dynamic origin of universal constants and realistic Planckian units}\label{Subsubsec:ConstPlanckian}
Due to the intrinsically creative, structure-forming character of
the symmetry of complexity (section \ref{Subsec:SymCom}), \emph{all}
the fundamental world structures and properties have explicitly
\emph{causal, dynamic} origin in quantum field mechanics, including
such ``intrinsically abstract'' features of usual theory as
universal (physical) constants \cite{Kir:QFM,Kir:Cosmo}. Thus, the
\emph{speed of light} $c$ is introduced as \emph{physically real
velocity} of perturbation propagation in e/m protofield coupled to
gravitational medium (see also section \ref{Subsec:CompIntDyn}),
rather than ``logical'' consequence of the \emph{postulated,
abstract} ``principle of relativity'' of the unitary theory, after
which we \emph{rigorously derive} major ``relativistic'' effects as
causal manifestations of the underlying \emph{complex interaction}
dynamics (section \ref{Subsubsec:Relativity}). Another universal
constant, \emph{Planck's constant} $h$, appears in our approach as a
\emph{dynamically discrete} portion, or ``quantum'', of the
universal complexity measure, complexity-action $\mathcal A$
(sections \ref{Subsec:SymCom},~\ref{Subsubsec:ParticleProp}). Its
universality follows from the \emph{common, physically unified}
structure of the underlying system of coupled protofields and the
fact that complexity-action quantum $h$ appears at the very first,
\emph{least structured} level of world complexity. However, the
truly unlimited, astonishingly large universality of $h$, from
photon energy to nuclear, subnuclear, and even intra-particle
properties, has a more detailed explanation within that general
interpretation, involving also the complex-dynamic origin of the
\emph{fine structure constant} $\alpha$ and \emph{elementary charge}
$e$ (see also section \ref{Subsubsec:ParticleProp}).

The well-known relation between $h$, $e$, and $\alpha$, $e^2  =
\alpha c h / 2 \pi$, can be written as
\begin{equation}\label{Eq:FineStrPlanck}
m_0 c^2  = \frac{ 2 \pi }{\alpha } \thinspace \frac{{e^2
}}{{\lambda _{\rm{C}} }} = N_\Re ^e \thinspace \frac{{e^2
}}{\mathchar'26\mkern-10mu\lambda _{\rm C}} \ , \ N_\Re ^e =
\frac{1}{\alpha} \ , \ \mathchar'26\mkern-10mu\lambda _{\rm C} =
\frac{\lambda _{\rm C}}{2 \pi} \ ,
\end{equation}
where $m_0$ is the electron rest mass and $\lambda _{\rm{C}} = h /
m_0 c$ is the Compton wavelength. As $E _0 = m_0 c^2$ is the
causally defined electron rest energy (see equation
(\ref{Eq:RestMassEnergy})), equation (\ref{Eq:FineStrPlanck})
\emph{means} that $\Delta x = \mathchar'26\mkern-10mu\lambda _{\rm
C}$ can be interpreted as the \emph{length of  virtual soliton jump}
within the quantum beat process (in relation to elementary space
length $\Delta x$ introduced in section \ref{Subsubsec:SpaceTime}),
$N_\Re ^e = 1 / \alpha \approx 137$ as \emph{the electron (quantum
beat) realisation number}, and $\alpha$ as \emph{realisation
emergence probability} (in agreement with the general dynamic
definition of the latter, equation (\ref{Eq:RealProbab})). The
canonical $h$--$e$ relation can now be written also as
\begin{equation}\label{Eq:PlanckConstInterpret}
\hbar = N_\Re ^e \thinspace \frac{{e^2 }}{c} =
\mathchar'26\mkern-10mu\lambda _{\rm C} p_{\rm{0}} \ , \ \
\mathchar'26\mkern-10mu\lambda _{\rm C} = N_\Re ^e r_e \ ,
\end{equation}
where $p_0  = m_0 c = {{E_0 } \mathord{\left/ {\vphantom {{E_0 } c}}
\right. \kern-\nulldelimiterspace} c}$ and $r_e  = {{e^2 }
\mathord{\left/ {\vphantom {{e^2 } {m_0 c^2 }}} \right.
\kern-\nulldelimiterspace} {m_0 c^2 }}$ is the usual ``classical
radius'' of the electron. Equation (\ref{Eq:PlanckConstInterpret})
provides then a transparent physical interpretation of Planck's
constant and its universality: $\hbar$ (or $h$) measures the
``volume'' (in units of action-\emph{complexity}) of the quantum
beat EP well that \emph{remains the same} for \emph{any} particle
species (including massless excitations like photons) and their
coherent-beat combinations, due to both complexity conservation and
permanence of the coupled protofield properties, whereas the
\emph{EP width}, $\mathchar'26\mkern-10mu\lambda _{\rm C}$ ($\lambda
_{\rm{C}}$) or $N_\Re ^e$ (up to $2 \pi$), and \emph{depth}, $p_0$
or $e^2 / c$, \emph{vary} for different species, but with the
permanent value of their complexity-based product, the EP well
volume. Since all equal quantum beat realisations should occupy the
closest two-dimensional vicinity of a current reduction centre, i.e.
a two-dimensional circle with the radius
$\mathchar'26\mkern-10mu\lambda _{\rm C}$, one obtains an estimate
for the virtual soliton size (in the state of maximum dynamical
squeeze), $D_e  = 2{\rm{\pi }}r_e = {\rm{\pi }}d_e$, which
coincides, up to coefficient $\pi$, with the classical electron
diameter, $d_e = 2r_e$ (we have used this result in elementary space
and field-particle size specification, section
\ref{Subsubsec:SpaceTime}).\footnote{Note that similar result for
the virtual soliton size follows directly from equation
(\ref{Eq:FineStrPlanck}), since the total particle energy should be
equal to Coulomb ``self-interaction'' within the squeezed
field-particle state (as an alternative to its equally valid
expression as ``dynamic'' interaction energy through extended
field-particle state during $N _\Re ^e$ jumps of the virtual
soliton). We obtain thus the direct extension of the usual, formal
definition of the classical electron radius, where we specify the
\emph{causal, physical origin} of the ``compressed'' electron state
(as well as its dynamic instability and permanent reappearance). One
should also take into account that the above interpretation will
remain valid if we change $N _\Re ^e$, $\Delta x =
\mathchar'26\mkern-10mu\lambda _{\rm C}$, and $r_e$ correspondingly,
according to their relations
(\ref{Eq:FineStrPlanck})--(\ref{Eq:PlanckConstInterpret}). It means
that the exact number of field-particle realisations and its virtual
soliton size can, in principle, be somewhat different from the above
values. However, the latter should be valid at least approximately,
up to numerical factor like $2 \pi$, as otherwise it would be
difficult to explain the essential difference of the virtual soliton
size from the physically reasonable (and now causally justified)
classical radius (for the electron).} Equation
(\ref{Eq:PlanckConstInterpret}) provides also another,
particle-dependent unit of quantum action-complexity, $\hbar _e =
\hbar / N_\Re ^e = e^2 / c$, that corresponds to \emph{one}
realisation (reduction-extension cycle).

It is essential that the above relations, written formally for the
electron, are directly extendible to arbitrary spatially coherent
(or \emph{quantum}) particles, their systems and excitations. The
electron corresponds to a rather shallow and large EP well,
admitting ``horizontally'' as much as $N_\Re ^e \gg 1$
``corpuscular'' (localised) particle states, as should be expected
for that \emph{light} particle with weak involvement of
gravitational (quark) medium. In the opposite case of heaviest
hadronic species with the effective charge $q$, mass $M_{\rm P}$,
and ``classical'' radius $r_{\rm P}$, one will have a very narrow EP
well with the width $\sim r_{\rm P}$ (or $N_\Re \sim 1$) and depth
$P_{\rm P} = M_{\rm P} c$ (or $q^2 / c$) that corresponds to the
highest protofield deformation/interaction amplitude and $\hbar
_{\rm P} \sim \hbar$. In that way one obtains the causally complete,
complex-dynamic explanation for the remarkable universality (and
physical origin) of Planck's constant that finds its additional
confirmation in the case of \emph{many-particle} quantum system of
atomic nucleus, by the fact that largest \emph{nuclear} masses are
close to the heaviest \emph{elementary particle} mass, $M_{\rm P}
\sim 200 \ {\rm GeV}$ \cite{Kir:QFM,Kir:100Quanta,Kir:Cosmo}.

This heaviest species case brings us to the causally complete
interpretation of the third universal constant, the
\emph{gravitational constant} $\gamma$ from classical Newton's law,
and related \emph{realistic, modified} values of \emph{Planckian
units}. Since any unitary theory does \emph{not} provide the real,
physical mechanism of gravity, the classical gravitational constant
has \emph{purely formal} origin in the usual theory, as a simple
coefficient in Newton's gravity law and its equally formal extension
to relativistic and quantum applications. In quantum field mechanics
gravitational interaction is causally derived as a deformation
influence of one quantum beat process on another, transmitted
through the physically real matter of gravitational protofield
(section \ref{Subsubsec:Interactions}), and the gravitational
constant represents a ``condensed'', resulting expression of that
complex-dynamic (and basically quantised) transmission process
through the gravitational quark condensate. It becomes evident that
this \emph{indirectly transmitted} interaction is driven by, but
remains very \emph{different} from, the underlying \emph{direct
attraction} between the two protofields that gives rise to the
quantum beat processes \emph{within} each particle. It means that
formal combinations of the three universal constants in Planckian
units describe actually the \emph{internal} quantum beat parameters,
i.e. \emph{direct} protofield attraction, and therefore should
contain another, modified value of ``gravitational'' constant,
$\gamma _0$, whose usual value $\gamma$ refers to \emph{much weaker,
indirect} interaction between \emph{different} particles. We obtain
thus the new, modified (or ``renormalised'') values of the Planckian
units of length $L_{\rm{P}}$ ($= r_{\rm P}$), time $T_{\rm{P}}$, and
mass $M_{\rm{P}}$ that just coincide (approximately) with the
\emph{experimentally observed} extreme values of the corresponding
quantities $l_{exp}$, $t_{exp}$, and $m_{exp}$:
\begin{eqnarray}\label{Eq:ModifPlanckUnits}
\nonumber L_{\rm{P}}&=&\left( {\frac{{\gamma _0 \hbar }}{{c^3 }}}
\right)^{\frac{1}{2}}  \approx 10^{ - 17}  - 10^{ - 16}\ {\rm{cm}}
\approx l_{\exp } \ , \\
T_{\rm{P}}&=&\left( {\frac{{\gamma _0 \hbar }}{{c^5 }}}
\right)^{\frac{1}{2}}  \approx 10^{ - 27}  - 10^{ - 26}\ {\rm{s}}
\approx t_{\exp } \ , \\
\nonumber M_{\rm{P}}&=&\left( {\frac{{\hbar c}}{{\gamma _0 }}}
\right)^{\frac{1}{2}}  \approx 10^{ - 22}  - 10^{ - 21}\ {\rm{g}} \
\left( {10^2  - 10^3 \ {\rm{GeV}}} \right) \approx m_{\exp } \ ,
\end{eqnarray}
where the relation between $\gamma _0$ and $\gamma$ can be
specified, for example, using the values of \emph{ordinary}
Planckian unit of length $l_{\rm{P}}$ and measured length $l_{exp}$:
$\gamma _0  = ({{l_{\exp } } \mathord{\left/ {\vphantom {{l_{\exp }
} {l_{\rm{P}} }}} \right. \kern-\nulldelimiterspace} {l_{\rm{P}}
}})^2 \gamma  \approx (10^{33}  - 10^{34} )\gamma$.

Such \emph{essential}, causally derived (i.e. \emph{inevitable})
modification of Planckian units and their new, \emph{realistic}
meaning lead to consistent solution of various stagnating problems.
One of the most remarkable of them is the so-called \emph{hierarchy
problem}, i.e. the problem of huge gap between the values of usual
Planckian values and observed quantities, especially evident for
particle mass spectrum. We see that the hierarchy gap completely
disappears for the modified, \emph{causally substantiated} Planckian
units, which shows that the \emph{whole} particle spectrum is
\emph{already basically covered} by the existing experimental data
and facilities, with evident and important practical implications
for \emph{high-energy physics strategy}
\cite{Kir:QFM,Kir:100Quanta,Kir:Cosmo}. Note the difference of our
\emph{intrinsically parsimonious} solution of the hierarchy problem
from anti-Occamian, unitary imitations of ``brane-world'' models
\cite{HiddenDim:1,HiddenDim:2,HiddenDim:3} arbitrarily postulating
additional (and totally \emph{abstract}), but strangely ``hidden''
dimensions that would inevitably give rise to additional,
experimentally observed forces and particle species (section
\ref{Subsubsec:Interactions}). It is easy to see that such
artificial, unreal entities in this and many other models of unitary
science appear as \emph{unavoidable} replacement for
\emph{incorrectly} rejected \emph{real}, naturally plural entities
and (dynamic) ``dimensions''.

Other applications of the modified Planckian units include major and
\emph{fatal} consequences for standard theory concepts essentially
relying upon usual Planckian units, such as \emph{cosmological
inflation} and \emph{quantum gravity} theories. One obtains also
consistent, physically transparent explanation for relative
\emph{weakness of gravity} (as being due to the small ratio $\gamma
/ \gamma_0$), dynamic \emph{unification of all fundamental forces},
and \emph{causal} theory of ``black holes'' and other dense
``quantum condensates'' \cite{Kir:USciCom} (see also below, section
\ref{Subsubsec:Self-Tuning}). In particular, the above causal
difference between $\gamma$ and $\gamma_0$ effectively disappears in
the \emph{dynamic unification phase} of (hadronic) virtual soliton
(section \ref{Subsubsec:Interactions}) and at corresponding
distances of the order of $L_{\rm{P}}$ ($=r_{\rm P}$), where one
deals with the ultimately dense state of original quark matter of
the gravitational protofield (so that $r_{\rm P} = L_{\rm{P}}$
should be close to the ``quark classical radius''). It becomes clear
also that modified Planckian units and their practical realisation
within quantum beat processes for heavier particle species represent
the real, causally complete version of various ``microscopic/quantum
black holes'' (``Kerr-Newman'' solutions, etc.), often formally
introduced in the unitary theory as particular, exotic possibilities
and models whose zero dynamic complexity is often accompanied by
additional, ``inexplicably plural'' dimensions (see e.\,g.
\cite{Burinskii:1,Burinskii:2} and further references therein).

\subsubsection{Self-tuning, adaptable universe from the creative symmetry of complexity}\label{Subsubsec:Self-Tuning}
We can summarise now those first ``material'', structure-formation
results of complexity symmetry unfolding on the ``cosmological''
scale of the \emph{whole} universe by noting that due to the
\emph{intrinsically creative} character of the unreduced interaction
process and resulting symmetry of complexity, the emerging universe
will automatically have \emph{self-tuning}, internally consistent
structure and properties, as opposed to intrinsically ``anthropic'',
as if very specially ``designed'' properties of any unitary universe
picture. That dynamic consistency of the real, complex-dynamic
universe structure is expressed by general property of \emph{dynamic
adaptability} of unreduced interaction process (section
\ref{Subsec:CompIntDyn}), which is due to the \emph{self-consistent
dependence} of the \emph{unreduced} EP formalism on the solutions to
be found (equations (\ref{Eq:ExistEff})--(\ref{Eq:StFunc-Full}))
amplified by the \emph{probabilistic dynamic fractality} of
interaction-driven structure formation (equations
(\ref{Eq:AuxEff})--(\ref{Eq:FractProb})). It is important that such
``self-tuned'' unfolding of the symmetry of complexity includes even
\emph{most fundamental}, ``intrinsic'' structures and properties
(such as universal constants and internal particle properties),
which are obtained now as \emph{dynamically emerging},
\emph{globally unified} and \emph{physically real} entities
(sections
\ref{Subsubsec:SpaceTime}--\ref{Subsubsec:ConstPlanckian}), contrary
to their unconditionally \emph{imposed}, postulated, and
\emph{abstract} status in any unitary theory version.

The differential form of ``potential'' complexity at the beginning
of interaction process, alias dynamic information, is given by
generalised, \emph{positively} defined \emph{potential energy}
$V_{{\rm{init}}} = - \left( \Delta {\mathcal A} / \Delta t \right)
\left| {_{x = {\rm const}}} \right.$ and enters the initial
existence equation (\ref{Eq:ExistProto}) through the interaction
potential $V_{{\rm eg}} \left( {\xi,q } \right)$. Emergence of
system realisations in the form of spatially chaotic quantum beat
processes within elementary particles transforms potential energy
into the total universe mass-energy-complexity $M_{{\rm{univ}}}
c^2$, with the basic equality between the two due to the symmetry
(conservation and transformation) of complexity:
\begin{equation}\label{Eq:TotUnivMass}
V_{{\rm{init}}}  = M_{{\rm{univ}}} c^2 \ .
\end{equation}
It means that, in accord with the underlying EP formalism (equations
(\ref{Eq:ExistEff})--(\ref{Eq:FractProb})), elementary
field-particle emergence leads to increase of internal e/m
protofield tension until it becomes greater than (sufficient)
attraction between protofields, so that new particles cannot emerge
any more and further complexity development proceeds to its higher
levels driven by interaction between particles (first-level
structures) thus formed (see also Figure \ref{Protofields}). That
multi-level, fractally structured universe complexity development,
always preserving its major self-tuning property, can be
schematically presented as
\begin{equation}\label{Eq:Self-Tuning}
M_{{\rm{univ}}}  \to \sum\limits_{{\rm{part}}} {N_{{\rm{part}}}
m_{{\rm{part}}} }  + \frac{{V_{{\rm{fund}}} }}{{c^2 }} \to
\sum\limits_{{\rm{atom}}} {N_{{\rm{atom}}} m_{{\rm{atom}}} }  +
\frac{{V_{{\rm{chem}}} }}{{c^2 }} \to \dots \ ,
\end{equation}
where``part'' and ``atom'' designate progressively emerging species
of elementary particles (and their interactions $V_{\rm{fund}}$),
atoms (and their interactions $V_{\rm{chem}}$), and so on.

Equations (\ref{Eq:TotUnivMass}),~(\ref{Eq:Self-Tuning}) show that
the more is the initial protofield interaction magnitude, the more
matter will emerge in the universe thus obtained, which is a major
manifestation of the self-tuning property of interaction-driven
universe structure formation. Note that impossibility to satisfy
equations (\ref{Eq:TotUnivMass}),~(\ref{Eq:Self-Tuning}) immediately
at \emph{all} universe locations and for those intrinsically
\emph{chaotic} interaction processes provides a causal explanation
for existence of seemingly ``redundant'' species and generations of
\emph{unstable} elementary particles that can efficiently ``fill in
the (small) gaps'' in the symmetry of complexity, in agreement with
its \emph{dynamically fractal} structure.

The ``anthropic'' universe image of the unitary theory with the
\emph{inexplicably unique} choice of parameters is thus replaced, in
both reality and its causally complete picture of the universal
science of complexity, by the \emph{generically successful} universe
emergence and development, but with naturally, \emph{consistently
variable} quantity and specific features of its material
content.\footnote{Note, however, that a viable universe with
\emph{any} protofield interaction parameters needs certain
``mechanical'' properties of the protofield material and in
particular sufficient e/m protofield ``elasticity''. Such demands do
not seem exotic at that ``subquantum'' level of reality (internal
protofield mechanics) and in any case are fundamentally different
from dynamic or conceptual restrictions of ``anthropic'' origin. The
necessary mechanical properties of the protofield material
constitute the \emph{inevitable minimum} of purely \emph{physical}
and \emph{realistic} ``postulates'' of our theory, as opposed to
\emph{numerous conceptual} and \emph{``mysterious'' (contradictory)}
assumptions of the unitary theory.} Those \emph{generic} cases of
unreduced protofield interaction can be yet better understood by
their \emph{causally} specified \emph{non-generic limits} at the
ultimately strong and weak interaction sides.

The excessively \emph{strong} protofield attraction would create a
\emph{macroscopically large} protofield collapse region (as opposed
to transient \emph{microscopic} collapse within any quantum beat
process). Although such peculiar state differs qualitatively from
any ``ordinary'' matter, it can be causally understood as
\emph{partially coherent, dense condensate} of quantum beat
pulsations with many discrete states (``phases'') of different
density providing causally complete, \emph{physically specified}
versions of such ``contradictory'' stellar objects as black holes
and neutron stars \cite{Kir:USciCom}, which are only
``phenomenologically'' (macroscopically) introduced and
\emph{formally} described in usual theory, leaving too much place
for ambiguity and related (often justified) doubts.

The ultimately \emph{weak} protofield attraction is insufficient for
appearance of a genuine, chaotic quantum beat and can give rise only
to small, quasi-linear protofield fluctuations. This is the
``primordial ether'' state of the coupled protofield system that can
have its modern realisation far enough from massive particles, in
the (physically real) ``vacuum'', where it can account for the
realistic, fundamentally substantiated version of ``microwave
radiation background''. The latter appears thus not as a
``definite'' sign of the past Big Bang event and related hot
universe state, but as a \emph{generic vacuum state} of \emph{any
real} universe, where those photonic ``vacuum fluctuations'' are
driven by \emph{weak protofield interaction} and configured in
detail by their multiple interactions \emph{within} the e/m
protofield, tending to \emph{equilibrium, thermodynamically
determined} state in an old enough universe (like ours) with
basically created massive particle content.\footnote{This
interpretation of microwave background radiation as protofield
fluctuations shows also why much \emph{larger} fluctuations, in the
form of ``virtual'' \emph{massive} particles, are actually
\emph{impossible}, contrary to their formal introduction in the
unitary theory. Such greater, massive fluctuations cannot emerge
already because of direct, mechanical impossibility of sufficient
protofield deformation in a ``mature'' universe, but also because
their existence would contradict to the symmetry of complexity
(contrary to the case of effectively zero-complexity photons). We
obtain thus consistent solution of another group of stagnating
unitary problems (in particle physics and cosmology) related to
improper, diverging energy contributions from such ``strong'' vacuum
fluctuations.} It is interesting to note that \emph{both} these
cases of ultimately strong and weak protofield interaction are
realised also within \emph{each} massive field-particle (quantum
beat process), but remain limited there to very small volumes and
short (permanently alternating) time periods.

Describing cosmological results of dynamic adaptability of the
unreduced interaction process, we should finally mention its
causally specified, \emph{exponentially huge efficiency}
\cite{Kir:QuMach,Kir:Fractal:2,Kir:Nano,Kir:Conscious,Kir:CommNet},
which is due to \emph{autonomous dynamic branching} processes of the
\emph{fractal realisation hierarchy} and leads to extremely
efficient, \emph{intrinsically complete} structure creation by
complex-dynamic search and invasion processes. They give rise to the
observed ``unlimited'' diversity and complexity of structures that
demonstrates \emph{creation efficiency} of the underlying
\emph{symmetry of complexity} and inevitably seems ``miraculous''
(inexplicable) within \emph{any} dynamically single-valued
description.

\subsubsection{Positive energy-complexity of the universe and cosmological time arrow}\label{Subsubsec:PositiveUnivEnergy}
As shown in section \ref{Subsec:SymCom}, the universal symmetry of
complexity of any real interaction \emph{necessarily implies} the
\emph{irreversible} time flow in the direction of growing dynamic
entropy, which is equivalent to strictly negative sign of
generalised Lagrangian $L$ and \emph{positive sign of total energy}
$E$ (see equations (\ref{Eq:Lagrangian})--(\ref{Eq:TimeArrow})), $L
= \Delta {\mathcal A} / \Delta t = pv - H < 0$, $E = H
> pv \geq 0$. Being applied to the whole universe (interacting
protofield system), the last inequality imposes strictly positive
total energy of the universe, in contrast with the dominating
unitary assumption about its zero value, obtained as a result of
compensation between positive ``kinetic'' (motion) energy and
negative energy of gravitational attraction. The latter statement is
widely used, often under the reference of ``Hamiltonian
constraint'', in various cosmological models (such as the famous
Wheeler-DeWitt equation in quantum cosmology) and justifies the
possibility of universe emergence ``from nothing'', by
self-amplified ``tunneling'' starting from a (genuine) vacuum
``fluctuation''.

We can see now that the real basis of the zero-energy assumption of
scholar cosmology is the unitary-science ``approximation'' reducing
the strictly positive (and \emph{large}) dynamic complexity of the
real world to the \emph{zero complexity value} of its
\emph{dynamically single-valued projection}. The mechanism that
ensures impossibility of any zero-complexity (totally
\emph{regular}) universe is the fundamental dynamic multivaluedness
of any real interaction (section \ref{Subsec:CompIntDyn}) implying
that any imaginary zero-complexity world configuration would
immediately change to a positive-complexity case once all its
interactions are turned on in their unreduced, \emph{dynamically
chaotic} (multivalued) version providing \emph{permanently growing}
entropy and \emph{positive} total energy.

It follows that \emph{no} zero-energy ``Hamiltonian constraint'' or
other ``nothingness-based'' model can be valid in principle,
irrespective of details, which provides an important restriction on
acceptable cosmological theories. Moreover, even when a positive
energy value is formally inserted in a unitary theory, it can hardly
lead to a correct description, as the dynamically single-valued
world projection in such theories \emph{cannot} reveal the genuine,
complex-dynamical content and meaning of energy-mass, in
\emph{direct relation} to the increasingly acute, ``unsolvable''
problems of \emph{dark mass} and \emph{dark energy} (see section
\ref{Sec:DarkMatter}). Due to the high degree of randomness in
mass-energy universe content, the positive total energy of the
universe is as big as its total material content (and thus cannot be
a relatively small ``unbalanced residue'').

As noted above, the positive energy-mass (or dynamic complexity)
content of the universe is \emph{equivalent} to the \emph{real time
arrow}: since for the (closed) universe $E = - \Delta {\mathcal A} /
\Delta t$ (global motion velocity $v = 0$) and $\Delta {\mathcal A}
< 0$ (chaos-induced loss of dynamic information), time can
\emph{advance} in a real universe, $\Delta t > 0$ (and thus the
universe \emph{can exist}), if and only if $E > 0$. The obtained
time arrow orientation to \emph{always growing} complexity-entropy
(or decreasing complexity-information) solves also all
entropy-related problems by implying that entropy grows in
\emph{all} kind of processes, including an \emph{externally}
``ordered'' structure formation (in this latter case one deals with
the SOC regime of multivalued dynamics, see section
\ref{Subsec:SymCom}). Thus rigorously specified \emph{asymmetry} of
time flow and entropy growth constitutes, however, an integral part
and \emph{inevitable result} of the global \emph{symmetry} of
complexity (whereas unitary symmetry is \emph{opposed} to its
asymmetric ``breaking'').

It is remarkable that the ``old'' problem of universe time arrow
(and the origin of time) is causally solved \emph{together} with the
energy-mass and entropy-information problems, without play on
``quantum'' or other ``mysteries'' and related ambiguous
speculations, but simply due to the \emph{unreduced} interaction
problem solution, revealing the key, \emph{qualitatively} new
phenomenon of \emph{dynamic multivaluedness} and related universal
\emph{symmetry} of complexity.

Note also the causally derived \emph{direct link} between \emph{the
real time flow} and \emph{genuine dynamic randomness}: the basically
\emph{regular}, though arbitrarily involved, ``Laplacian'' world of
the unitary science cannot exist already because it is devoid of any
real, advancing time flow (that's why this direct and fundamental
relation between time and randomness remains ``hidden'' in the
conventional, dynamically single-valued science framework).

\subsubsection{Unified complex-dynamic origin of relativistic and quantum properties}\label{Subsubsec:Relativity}
We have seen in previous sections
\ref{Subsubsec:SpaceTime}--\ref{Subsubsec:PositiveUnivEnergy} how
the global universe structure and properties (space, time, energy),
its universal constants, elementary field-particles, their intrinsic
properties (mass-energy, charge, spin) and interaction forces
causally emerge as \emph{unified manifestations of the symmetry of
complexity} of the underlying protofield interaction process with
generic parameters and simplest possible initial configuration. We
now continue to follow the natural complexity development of the
\emph{same} interaction towards the basic external, \emph{dynamical}
features of the field-particles thus obtained, in the form of their
\emph{unified relativistic and quantum properties} that will be
derived as \emph{totally causal, realistic} manifestations of the
same unreduced \emph{dynamic complexity} and its symmetry. Moreover,
it is the rough \emph{rejection} of the underlying complex
interaction dynamics in the standard, unitary theory that accounts
for the ``inexplicable mystery'' status of official quantum and
relativistic postulates. All ``relativistic'' and ``quantum''
effects emerge as \emph{inevitable, standard, and totally causal}
manifestations of real, unreduced interaction dynamics and therefore
can be generalised to higher complexity levels
\cite{Kir:USciCom,Kir:USymCom,Kir:QuMach,Kir:Fractal:2,Kir:Conscious}
(see also section \ref{Subsec:SymCom} and below in this section).

We start from the causally derived intrinsic property of inertial
particle \emph{rest mass} $m_0$ defined by equation
(\ref{Eq:RestMassEnergy}) (section \ref{Subsubsec:ParticleProp})
that contains already natural, \emph{dynamic} unification of causal
\emph{quantisation} of the underlying quantum beat process and
complex-dynamic origin of \emph{relativistic ``equivalence''}
between mass/inertia and its energetic content.\footnote{It is not a
coincidence that a heuristically postulated version of this relation
was used by Louis de Broglie in his original, realistically based
derivation of his famous formula for the particle wavelength
\cite{deBroglie:These,Kir:75MatWave}.} If the field-particle is not
isolated and interacts with other particles by the causally emerging
interaction forces (section \ref{Subsubsec:Interactions}), it leads
to (further) \emph{growth} of complexity-entropy appearing as
particle \emph{motion}. In other words, we can now \emph{rigorously}
and \emph{universally} define the system \emph{state of motion}
itself as any state with (generalised) energy-complexity
\emph{exceeding} its \emph{minimum} value in the \emph{state of
rest} (also provided thus with absolutely universal and fundamental
definition). As energy-complexity is a positively defined quantity
(see equation (\ref{Eq:TimeArrow})), such minimum always exists.

Because of the maximum homogeneity of initial (protofield) system
configuration giving rise to the emerging system (particle) at rest,
the latter state corresponds to \emph{maximum homogeneity} of
realisation probability distribution (cf. the generalised Born rule
(\ref{Eq:BornRule})). On the other hand, as realisation number is
fixed, growth of energy-complexity in a state of motion is possible
only due to appearing (or growing) \emph{inhomogeneity} of
realisation probability distribution (and thus moving system
structure). It means that action-complexity function ${\mathcal A}$
of a \emph{moving system} acquires dependence on coordinate $x$
(emerging space configuration, see section
\ref{Subsubsec:SpaceTime}), \emph{in addition} to its dependence on
time $t$ in the state of rest, and ordinary (discrete) time
derivative of action in equation (\ref{Eq:RestMassEnergy}) for the
state of rest should be replaced by the total (discrete) time
derivative of action for a moving particle:
\begin{equation}\label{Eq:MotionEnergy}
\frac{\Delta {\mathcal A}}{\Delta t} = \frac{\Delta {\mathcal
A}}{\Delta t} \left| _{x = {\rm const}} \right. + \frac{\Delta
{\mathcal A}}{\Delta x} \left| _{t = {\rm const}} \right.
\frac{\Delta x}{\Delta t} \ , \ \ E =  - \frac{{\Delta {\mathcal
A}}}{{\Delta t}} + \frac{{\Delta {\mathcal A}}}{\lambda
}\frac{{\Delta x}}{{\Delta t}} = \frac{h}{{\mathcal T}} +
\frac{h}{\lambda }\thinspace v = h{\mathcal N} + pv \thinspace ,
\end{equation}
where
\begin{equation}\label{Eq:ParticleEnergy}
E =  - \frac{{\Delta {\mathcal A}}}{{\Delta t}}\left| {_{x  =
{\rm{const}}} } \right. = \frac{h}{\tau } = h\nu
\end{equation}
is the \emph{total energy} of a moving system (in accord with its
previous definition (\ref{Eq:Energy})) specified for the moving
field-particle,\footnote{Expressions containing momentum-complexity
can generally be understood in the sense of corresponding
\emph{vector} definitions and operations. However, in the considered
simplest case of single particle motion, one can interpret $p$ and
$v$ as respective vector moduli.}
\begin{equation}\label{Eq:ParticleMomentum}
p = \frac{{\Delta {\mathcal A}}}{{\Delta x}}\left| {_{t =
{\rm{const}}} } \right. = \frac{\left| {\Delta {\mathcal A}}
\right|}{\lambda } = \frac{h}{\lambda }
\end{equation}
is the universally defined system momentum (see equation
(\ref{Eq:Momentum})) specified now for the moving field-particle,
\begin{equation*}
v = \frac{{\Delta x}}{{\Delta t}} \equiv \frac{\mit
\Lambda}{{\mathcal T}}
\end{equation*}
is the \emph{global motion velocity}, $\tau  \equiv \left( {\Delta
t} \right)\left| _{x = {\rm{const}}} \right.$ is the quantum beat
period at a fixed space point, $\nu = 1 / \tau$, $\lambda  \equiv
\left( {\Delta x} \right)\left| _{t = {\rm{const}}} \right. =
\lambda _{\rm{B}}  = h / p$ is the space element (inhomogeneity)
related to the global field-particle motion and known as \emph{de
Broglie wavelength} $\lambda _{\rm{B}}$, ${\mathcal T} = \Delta t$
is the ``total'' quantum beat period (${\mathcal T} \neq \tau)$,
${\mathcal N} = 1 / {\mathcal T}$, and ${\mit \Lambda} = \Delta x$.

Equation (\ref{Eq:MotionEnergy}) describes the total energy
partition for the globally moving field-particle (quantum beat
process) reflecting its real, complex-dynamical structure that
remains hidden in the unitary theory. It is easy to see that the
second term, $pv$, accounts for the \emph{global}, averaged, and
therefore \emph{regular} motion of the quantum beat process (virtual
soliton wandering), while the first summand, the negative-sign
Lagrangian $- L = - \Delta {\mathcal A} / \Delta t = h / {\mathcal
T}$, describes contribution to the total energy from the
\emph{purely random} deviations of virtual soliton wandering from
that average, global motion tendency (it is the energy of complex
system dynamics in its \emph{moving reference frame}). We see that
any global motion emerges only as an \emph{average tendency} of
internal \emph{chaotic} (dynamically multivalued) process of
structure formation, where the above dynamically derived de Broglie
wave of a moving particle is the corresponding (regular) space
structure of that global motion tendency. However, every single jump
of the virtual soliton within the quantum beat process is
characterised by the \emph{intrinsic uncertainty} of
\emph{dynamically redundant} choice of the next reduction centre,
and therefore the \emph{whole} content of the total energy $E$
possesses the key property of \emph{inertia}, $E = m c ^2$, where
$m$ is the total (``relativistic'') mass and $c ^2$ is a coefficient
to be rigorously specified below.

According to our \emph{causal} definition of the speed of light $c$
(section \ref{Subsec:CompIntDyn}), every virtual soliton jump within
the globally moving field-particle proceeds with the speed $c$. It
becomes clear now that the global motion velocity $v$ is (usually
essentially) different from $c$ just because of the (usually
dominating) tendency of purely random wandering of the virtual
soliton ``around'' global motion tendency, so that only some
(usually very small) part of chaotic quantum jumps falls within that
global, ``systematic'' tendency that forms \emph{explicitly
observed} structure. Specifically, the field-particle moving as a
whole with the velocity $v$ performs (in average) a quantised
global-tendency jump of $\Delta x = \lambda = \lambda _{\rm{B}}$
during the \emph{same} time period $\tau _v = \lambda / c$ that
includes $n _v = c / v$ purely random jumps around global tendency.
As any such jump duration is $\tau$, we have $\tau _v = n _v \tau$,
or $\lambda = V _{\rm ph} \tau$, where $V _{\rm ph} = c ^2 / v$ is
the fictitious, apparently faster-than-light ``phase velocity'' of
``matter wave'' propagation, appearing if one does \emph{not} take
into account the irregular, ``multivalued'' part of the
field-particle dynamics \cite{deBroglie:These}. Energy and momentum
definitions (\ref{Eq:ParticleEnergy}),~(\ref{Eq:ParticleMomentum})
transform this relation between $\lambda$ and $\tau$ into the famous
\emph{relativistic dispersion relation} (which is now obtained as a
\emph{causal result} of underlying \emph{complex} interaction
dynamics):
\begin{equation}\label{Eq:DispRelation}
p = E \thinspace \frac{v}{{c^2}} = mv \ ,
\end{equation}
where $m = E / c ^2$, now by \emph{rigorously obtained} definition
containing the \emph{physically real} speed of light $c$ (we thus
also justify, of course, the corresponding mass-energy equivalence
for the rest mass, equation (\ref{Eq:RestMassEnergy})). The
\emph{genuine} meaning of the famous equivalence of mass and energy,
$E = m c ^2$, becomes now clear due to its \emph{causal, dynamic
derivation} in quantum field mechanics (whereas it is actually only
postulated in the standard theory): particle energy has the property
of enertia due to its well-specified, \emph{complex-dynamic} quantum
beat \emph{content}.

Combining equations (\ref{Eq:DispRelation}) and
(\ref{Eq:ParticleMomentum}), we obtain the standard expression for
the de Broglie wavelength of a moving particle:
\begin{equation}\label{Eq:DeBroglieWave}
\lambda  = \lambda _{\rm{B}}  = \frac{h}{{mv}} \ .
\end{equation}
Now, however, this famous relation, constituting the basis of the
whole quantum physics, is not formally postulated (as in the
standard, unitary theory), or ``phenomenologically'' explained (as
in the original de Broglie approach \cite{deBroglie:These}), but
\emph{rigorously derived} as a totally \emph{consistent} consequence
of the underlying \emph{complex, multivalued interaction dynamics}
within every massive elementary particle. This result and its
derivation include, in particular, remarkable, \emph{intrinsic
unification} of ``relativistic'' and ``quantum'' particle properties
remaining irreducibly split in the unitary theory but in reality
resulting, as we can clearly see now, from the same complex dynamics
of quantum beat process
\cite{Kir:USciCom,Kir:QuMach,Kir:QFM,Kir:100Quanta}. This
omnipresent unification appears, for example, as otherwise
``strange'' combination of \emph{classical} quantity $v$,
\emph{quantum} Planck's constant $h$, and \emph{relativistic} total
mass $m$ in the same relation (\ref{Eq:DeBroglieWave}), or as the
above complex-dynamical origin of inertial property of the total
energy $E$ due to its internal quantisation.

One can also conclude that the basic dispersion relation
(\ref{Eq:DispRelation}) results from the \emph{symmetry of
complexity} as the latter determines the underlying major
\emph{equivalence} between multiple realisations, including those of
both global-motion tendency and irregular deviations from it. This
very familiar and apparently ``simple'' relation, $p = mv$,
includes, as we have seen, the whole complexity of the unreduced
interaction dynamics and has further fundamental and
\emph{universally} valid consequences. In particular, by taking its
time derivative, one obtains \emph{rigorously derived},
\emph{causally relativistic} and universally extended version of
Newton's laws of ``classical'' dynamics (usually postulated),
without any specially introduced classicality or empirically
determined quantities (mass, energy, momentum, etc.):
\begin{equation*}
\frac{\partial \left( mv \right)}{\partial t} = {\mathcal F} \left(
{x,t} \right) \thinspace , \ \ \ {\mathcal F} \left( {x,t} \right) =
\frac{\partial p}{\partial t} = \frac{\partial {\mathcal
A}}{\partial x \partial t} = - \frac{\partial {\mathcal U}}{\partial
x} \ , \ \ \ {\mathcal U} \left( {x,t} \right) = - \frac{\partial
{\mathcal A}}{\partial t} \ ,
\end{equation*}
where \emph{force} ${\mathcal F}  \left( {x,t} \right)$ and
\emph{potential energy-complexity} ${\mathcal U} \left( {x,t}
\right)$ are thus \emph{causally} defined, and continuous derivative
notations are used for brevity, with the general meaning of
\emph{dynamically discrete} derivatives. Therefore Newton's laws
also \emph{result from the symmetry of complexity} (and underlying
\emph{multivalued} dynamics) if one asks for their \emph{consistent
derivation}. Such causally extended Newton's laws are
\emph{universally} applicable to \emph{any} system and complexity
level, although they may be more suitable and efficient in cases of
homogeneous enough, \emph{pseudo}-unitary dynamics.

Inserting now the obtained relativistic dispersion relation
(\ref{Eq:DispRelation}) into the complex-dynamic particle energy
partition (\ref{Eq:MotionEnergy}) and using energy definition
(\ref{Eq:ParticleEnergy}), we get the explicit expression of
\emph{dynamically} derived \emph{time relativity}:
\begin{equation}\label{Eq:TimeRel:1}
\tau  = {\mathcal T}\left( {1 - \frac{{v^2 }}{{c^2 }}} \right) \ .
\end{equation}
As the period ${\mathcal T}$ provides the real (dynamic) time period
for the \emph{intrinsic} clock of the moving particle (system), we
conclude that time slows down \emph{within} the moving
field-particle (${\mathcal T} > \tau$) because time flow is
explicitly \emph{produced} by the \emph{same}, complex-dynamic
(multivalued) interaction process that gives rise to \emph{global
motion}. Combining equation (\ref{Eq:TimeRel:1}) with a relation
involving the quantum beat frequency $\nu _0$ and period $\tau _0$
at rest \cite{Kir:USciCom,Kir:QFM},
\begin{equation}\label{Eq:RelPeriods}
{\mathcal N} \nu  = \left( {\nu _0} \right)^2 \ \ \ {\rm or} \ \ \
{\mathcal T} \tau  = \left( {\tau _0} \right)^2 \ ,
\end{equation}
we get the canonical expression of time relativity (but now
\emph{causally derived} from the underlying \emph{complex}
dynamics):
\begin{equation}\label{Eq:TimeRel:2}
{\mathcal N} = \nu _0 \sqrt {1 - \frac{v^2}{c^2}} \ \ \ {\rm or} \ \
\ {\mathcal T} = \frac{\tau _0}{\sqrt {\displaystyle {1 - \frac{v^2
}{c^2}}} } \ .
\end{equation}
Note that equation (\ref{Eq:RelPeriods}) also follows from the
symmetry (conservation) of complexity: it means that the system
realisation number filling the rectilinear ${\mathcal N} \times \nu$
area remains unchanged. The complex-dynamic time relativity thus
rigorously derived from the symmetry of complexity is easily
extended to other effects of special relativity.

The obtained \emph{intrinsic unification} of causally derived
versions of \emph{relativistic} and \emph{quantum} dynamics in a
single, \emph{complex-dynamical} quantum beat process for a moving
field-particle can be summarised by insertion of the time relativity
expression (\ref{Eq:TimeRel:2}), dispersion relation
(\ref{Eq:DispRelation}) and de Broglie wavelength formula
(\ref{Eq:DeBroglieWave}) into the total energy partition
(\ref{Eq:MotionEnergy}):
\begin{equation}\label{Eq:EnergyPartition}
E = h\nu _0 \sqrt {1 - \frac{{v^2 }}{{c^2 }}}  + \frac{h}{{\lambda
_{\rm{B}} }} \thinspace v = h\nu _0 \sqrt {1 - \frac{{v^2 }}{{c^2
}}}  + h\nu _{\rm{B}}  = m_0 c^2 \sqrt {1 - \frac{{v^2 }}{{c^2 }}}
+ \frac{{m_0 v^2 }}{{\sqrt {\displaystyle {1 - \frac{{v^2 }}{{c^2
}}}} }} \ ,
\end{equation}
where $h\nu _0  = m_0 c^2$ according to equation
(\ref{Eq:RestMassEnergy}) and \emph{de Broglie frequency} $\nu _{\rm
B}$ is defined as
\begin{equation}\label{Eq:DeBroglieFrequency}
\nu _{\rm B}  = \frac{v}{{\lambda _{\rm B} }}  = \frac{{pv}}{h} =
\frac{{\nu _{{\rm B} 0} }}{{\sqrt {\displaystyle {1 - \frac{{v^2
}}{{c^2 }}}} }} = \nu \thinspace \frac{{v^2 }}{{c^2 }} \ , \ \ \nu
_{{\rm{B}}0} = \frac{{m_0 v^2 }}{h}  = \nu _0 \thinspace
\frac{{v^2 }}{{c^2 }} = \frac{v}{{\lambda _{{\rm{B}}0} }} \ , \ \
\lambda _{{\rm{B}}0}  = \frac{h}{{m_0 v}} \ .
\end{equation}
The physical \emph{reality} of de Broglie wave (emerging as a
\emph{complex-dynamic field-particle structure}) is confirmed now by
the standard relation between its length, frequency, and velocity,
$\lambda _{\rm{B}} \nu _{\rm{B}}  = v$, which describes
\emph{occasional quantum jumps} of the moving particle wave field to
the distance $\lambda _{\rm{B}}$, occurring with the \emph{average}
frequency $\nu _{\rm{B}}$ and accompanied by extended \emph{chaotic
wandering} of the particle reduction centre (virtual soliton) around
global motion, reducing its velocity from $c$ (for any single jump)
to $v$. As the frequencies in equation (\ref{Eq:EnergyPartition})
refer to quantised, \emph{causally random} field-particle jumps, it
follows that the quantities $\alpha _1 = {{v^2 } \mathord{\left/
{\vphantom {{v^2 } {c^2 }}} \right. \kern-\nulldelimiterspace} {c^2
}}$ and $\alpha _2 = 1 - \alpha _1 = 1 - {{v^2 } \mathord{\left/
{\vphantom {{v^2 } {c^2 }}} \right. \kern-\nulldelimiterspace} {c^2
}}$ are none other than \emph{dynamically} obtained \emph{(compound)
realisation probabilities} of, respectively, global (average) and
totally random tendencies of the moving field-particle dynamics, in
agreement with their general definition (\ref{Eq:RealProbab}), which
confirms once again the \emph{intrinsic unity} of ``quantum'' and
``relativistic'' manifestations of the \emph{unreduced interaction
complexity}.

Equation (\ref{Eq:EnergyPartition}) provides also \emph{relativistic
transformation of (total) mass} and thus de Broglie wavelength
(\ref{Eq:DeBroglieWave}), the latter demonstrating dynamic and
``quantum'' origin of \emph{relativistic contraction of length} of a
globally moving body (it can also be derived from relativistic time
retardation):
\begin{equation}\label{Eq:RelMassLength}
m = \frac{E}{c^2} = m_0 \left( { \sqrt {1 - \frac{v^2}{c^2}} +
\frac{\displaystyle \frac{v^2}{c^2} }{\sqrt {\displaystyle {1 -
\frac{v^2}{c^2}}} } } \right) = \frac{m_0}{ \sqrt {\displaystyle {1
- \frac{v^2}{c^2}}} } \ , \ \ \lambda _{\rm{B}}  = \frac{h \sqrt
{\displaystyle {1 - \frac{v^2}{c^2}}} }{m_0 v} = \lambda _{{\rm B}
0} \sqrt {\displaystyle {1 - \frac{v^2}{c^2}}} \ .
\end{equation}

The first term of the final complex-dynamic energy partition
(\ref{Eq:EnergyPartition}) taken with the negative sign provides the
\emph{causally derived} expression for \emph{relativistic particle
Lagrangian}, ${\Delta {\mathcal A}} / {\Delta t} = pv - E \equiv L$
(see also equation (\ref{Eq:Lagrangian}) and above in this and the
previous sections), that remains valid, of course, for any
macroscopic body (agglomerate of particles):
\begin{equation}\label{Eq:RelLagrangian}
L =  - h{\mathcal N} =  - h\nu _0 \sqrt {1 - \frac{{v^2 }}{{c^2 }}}
= - m_0 c^2 \sqrt {1 - \frac{{v^2 }}{{c^2 }}} \ .
\end{equation}
We obtain also the causal, complex-dynamic interpretation of
Lagrangian as the energy of the \emph{totally random} part of a
system (field-particle) dynamics, or its ``(internal) heat energy'',
specifying the corresponding heuristically introduced ideas of Louis
de Broglie about ``hidden thermodynamics'' of a single particle
\cite{deBroglie:HiddenTherm}, as well as his anticipation of
\emph{realistic extension} of usual ``least action principle'',
describing now the \emph{real}, dynamically chaotic system wandering
around the average (global) motion tendency, rather than formal
``variations'' of action functional \cite{Kir:USciCom,Kir:QFM}.
Minimisation of action corresponds in our description to
action-complexity transformation into entropy-complexity, within
\emph{conservation} (symmetry) of the total complexity. Recalling
that relativistic Lagrangian (\ref{Eq:RelLagrangian}) is only
mechanistically guessed and postulated in the standard special
relativity and then used, together with artificially imposed
``principle of relativity'' and other postulates, for ``derivation''
of time relativity and other related effects, we can state now that
the symmetry of complexity provides the \emph{unified causal
extension} of \emph{all} those abstract and separated principles of
the unitary theory, including least action and relativity
principles, quantum postulates (see also below), first and second
laws of thermodynamics.

It is clear that the obtained \emph{dynamic} unity of physically
real space (\emph{structure}) and time (\emph{events} of its
explicit emergence) excludes their mechanistic unification in the
same, ``geometric'' construction. Correspondingly, the symmetry of
complexity underlying real world dynamics is much richer (``less
symmetric'') than unitary symmetries of standard relativity, which
allows for a natural solution of all problems of their ``violation''
(including quantisation, irregularities, etc.), so that the symmetry
of complexity remains always exact and gives the real, somewhat
limited and irregular ``relativity'' that can also be directly
extended to arbitrary levels of physically real space and time (see
below) \cite{Kir:USciCom}. These general conclusions concern also
the naturally emerging, \emph{dynamically} based effects of
\emph{general relativity}.

Indeed, we have seen in section \ref{Subsubsec:Interactions} that
gravitational interaction between any material particles (protofield
perturbations) emerges inevitably and universally due to their
``deformation interaction'' through the gravitational medium coupled
everywhere to the equally omnipresent e/m protofield. This
physically real gravity has therefore intrinsically \emph{dynamic}
and \emph{quantised}, but not ``geometric'' origin (even though a
formal geometric \emph{description} can be applied and give correct
results within its validity domain, similar to other cases of
deformation interaction through a quasi-continuous medium). It is
evident that gravitational medium perturbation and interaction
magnitude will grow with the above causally specified inertial mass
of interacting body, which gives the generalised, \emph{causally
substantiated} ``principle of equivalence'', as opposed to its
formally postulated version of the conventional general relativity.
In a usual case of not very dense interaction configurations,
essentially beyond the (modified) Planckian unit situation (section
\ref{Subsubsec:ConstPlanckian}), the same quantity of
\emph{inertial} mass (temporal rate of action-complexity change,
section \ref{Subsubsec:ParticleProp}) will also play the role of
\emph{gravitational} mass (i.e. ``gravitational charge''). Those
gravitational mass-charges and their interaction through
gravitational medium are produced by the \emph{same,
complex-dynamic} quantum beat processes that give rise to electric
charges and their interaction through the e/m protofield, which is
another manifestation of the universal symmetry of complexity and
its ``naturally broken'' character. The latter is due here to
\emph{different physical properties} of e/m and gravitational
protofields (sections
\ref{Subsec:CompIntDyn},~\ref{Subsubsec:Interactions}) and appears
e.\,g. in the fact that there is no ``sign'' of gravitational
mass-charges and they always attract to each other (beyond the
Planckian-scale situation of unified interactions)
\cite{Kir:USciCom}.

The intrinsically \emph{quantised} dynamic origin of mass
determining the local \emph{flow of time} (see also sections
\ref{Subsubsec:SpaceTime},~\ref{Subsubsec:ParticleProp}) naturally
leads to \emph{causally explained, dynamic} effects of general
relativity, demonstrating once more \emph{inseparable unification of
quantum and relativistic manifestations} of dynamic complexity in
quantum field mechanics \cite{Kir:USciCom,Kir:QFM,Kir:100Quanta}. In
particular, the quantum beat frequency $\nu$ (see equations
(\ref{Eq:RestMassEnergy}),~(\ref{Eq:ParticleEnergy})) directly
depends on the local gravitational protofield tension/density
created by other material objects and described as ``gravitational
(field) potential'':
\begin{equation}\label{Eq:GravTimeRetard}
M \left( x \right) c^2 \equiv h\nu \left( x \right) = m c^2 \sqrt
{g_{00} \left( x \right)} \ ,
\end{equation}
where $\nu \left( x \right)$ is the local quantum beat frequency for
a ``test'' particle, while ``metric'' $g_{00} \left( x \right)$
describes in reality the gravitational protofield tension, $g_{00}
\left( x \right) = 1 + 2{{\phi _{\rm g} \left( x \right)} / {c^2
}}$, $\phi _{\rm g} \left( x \right)$ being the classical
gravitational potential (we use the standard relation for the weak
field case \cite{LandauLifshitz:F}). Since $\nu \left( x \right)$
determines the causally derived time flow and $\phi _{\rm g} \left(
x \right) < 0$ ($g_{00} \left( x \right) < 1$) for \emph{attractive}
gravitational interaction, equation (\ref{Eq:GravTimeRetard})
provides the causal, dynamically derived version of ``relativistic
time retardation'' in the gravitational field.

The unified complex-dynamic origin of both relativistic and quantum
effects becomes yet more complete when we provide the explicit
causal derivation of major quantum mechanical wave equations, as
they are associated most closely with the specific ``quantum''
(undular) kind of behaviour. Such complex-dynamic origin and causal
derivation of the Schr\"{o}dinger equation from the underlying
symmetry of complexity are provided in section \ref{Subsec:SymCom},
together with the related causal solution of the unitary ``quantum
mysteries'' and Schr\"{o}dinger formalism generalisation to any
higher complexity levels (see equations
(\ref{Eq:CompCons})--(\ref{Eq:SchrodinExpansion})).

The key condition of causal quantisation (\ref{Eq:CausQuant})
reflecting quantum beat dynamics gives rise to the ``Dirac
quantisation'' rules, which are now dynamically explained
\cite{Kir:USciCom,Kir:QFM,Kir:100Quanta}, but only formally
postulated in the unitary theory:
\begin{equation}\label{Eq:DiracQuant:Momentum}
p = \frac{\Delta {\mathcal A}}{\Delta x} =  - \frac{1}{\mit \Psi}
\thinspace i \hbar \thinspace \frac{\partial {\mit \Psi}}{\partial
x} \ , \ \ p^2 = - \frac{1}{\mit \Psi} \thinspace \hbar ^2
\frac{\partial ^2 {\mit \Psi} }{\partial x^2} \ ,
\end{equation}
\begin{equation}\label{Eq:DiracQuant:Energy}
E = - \frac{\Delta {\mathcal A}}{\Delta t} =  \frac{1}{\mit \Psi}
\thinspace i \hbar \thinspace \frac{\partial {\mit \Psi}}{\partial
t} \ , \ \ E^2 = - \frac{1}{\mit \Psi} \thinspace \hbar ^2
\frac{\partial ^2 {\mit \Psi} }{\partial t^2} \ .
\end{equation}
Inserting these causally obtained rules into relativistic equations
of the same complex-dynamic origin, we can obtain various
relativistic wave equations. Thus, complex-dynamic energy partition
for a moving particle (\ref{Eq:EnergyPartition}) can be written as
\begin{equation}\label{Eq:EnergyPartition'}
E = m_0 c^2 \sqrt {1 - \frac{{v^2 }}{{c^2 }}}  + \frac{{p^2 }}{m}
\ \ \ \ {\rm or} \ \ \ mE = m_0 c^2  + p^2 \ .
\end{equation}
Combining equations
(\ref{Eq:DiracQuant:Momentum})--(\ref{Eq:EnergyPartition'}), we get
the Klein-Gordon or Dirac equation for a free particle:
\begin{equation*}\label{Eq:Klein-Gordon}
\frac{\partial ^2 {\mit \Psi}}{\partial t^2} -  c^2 \frac{\partial
^2 {\mit \Psi}}{\partial x^2} + \omega _0^2 {\mit \Psi} = 0 \ ,
\end{equation*}
where $\omega _0  \equiv m_0 c^2 / \hbar = 2 {\pi} \nu _0$ is the
``circular'' frequency of quantum beat pulsation at rest (see
equation (\ref{Eq:RestMassEnergy})), which actually accounts for
the spin vorticity twirl (see section \ref{Subsec:Properties}).
More complicated versions of relativistic wave equations for
interacting field-particles can be obtained in the same way as
causal consequences of the underlying symmetry of complexity,
whereas their nonrelativistic limit leads again to the causally
substantiated Schr\"{o}dinger equation \cite{Kir:USciCom}.

A straightforward analysis shows also that the Schr\"{o}dinger
equation (\ref{Eq:Schrodinger}) with the Hamiltonian $H ( x,p,t ) =
p ^2 / 2m + V ( x,t )$, where $V ( x,t )$ is a \emph{binding}
potential well, can be satisfied only for a \emph{discrete} set of
configurations of the wavefunction ${\mit \Psi} ( x,t )$ determined
by \emph{integer} numbers of the \emph{same} action-complexity
quantum, $h$, that describes \emph{quantum-beat cycle}, or ``system
realisation change'' (see section \ref{Subsubsec:ConstPlanckian}),
which explains the famous quantum-mechanical \emph{energy-level
discreteness} (e.\,g. in atoms) by \emph{complex-dynamical
discreteness (or causal quantisation)} of the underlying protofield
interaction process \cite{Kir:USciCom} and shows why the same
Planck's constant appears also at this \emph{higher sublevel} of
quantum dynamics.

Another ``postulated mystery'' of the unitary quantum-mechanics,
\emph{linear superposition} of various \emph{probabilistically}
emerging states, including the particular case of \emph{quantum
entanglement} for a many-body system, reflects the \emph{real} but
\emph{multivalued} dynamics of \emph{underlying interaction}, where
the system performs unceasing series of reduction-extension cycles,
or \emph{real quantum jumps}, between the corresponding
\emph{realisations} with the now \emph{dynamically} determined
probabilities (see equations
(\ref{Eq:ProbabSum})--(\ref{Eq:StFunc-Full}),~(\ref{Eq:BornRule})).
The \emph{quasi}-linearity of wavefunction behaviour is due to the
\emph{transiently} weak, perturbative interaction character
\emph{only} within the intermediate (main) system realisation that
constitutes the \emph{ physically real} version of the wavefunction,
whereas the actually \emph{measured} eigenvalue \emph{emergence}
from the wavefunction realisation, obscured by the ``inexplicable''
conventional postulates, is due to its \emph{essentially nonlinear}
and \emph{physically real} reduction to respective regular,
localised realisations (section \ref{Subsec:CompIntDyn}). The
symmetry of complexity between \emph{all} system realisations
naturally provides thus the necessary \emph{dynamic unification} of
those ``opposite'', linear and nonlinear, undular and corpuscular,
distributed and localised, types of behaviour within the single,
holistic interaction process.

The revealed unified, causal origin of quantum, special-relativistic
and general-relativistic effects in the underlying complex
(multivalued) interaction dynamics finds further confirmation in its
straightforward \emph{generalisation to (arbitrary) higher
complexity levels}, where one can also observe dynamic discreteness
(quantisation) and ``relativistic'' modification of the respective
time flow rates and length scales \cite{Kir:USciCom}. Rigorous
derivation of unified quantum and relativistic behaviour for
arbitrary (many-body) interaction process starts from the existence
equation for such a process, equation (\ref{Eq:ExistGen}), that
generalises all (correct) ``model'' equations (see also section
\ref{Subsec:SymCom}) and leads to the same basic system of equations
(\ref{Eq:ExistSystem}) as the protofield existence equation
(\ref{Eq:ExistProto}) at the first complexity level (section
\ref{Subsec:CompIntDyn})
\cite{Kir:USciCom,Kir:USymCom,Kir:QuMach,Kir:Fractal:2,Kir:Conscious,Kir:CommNet}.
We follow then the standard complexity development analysis by the
unreduced EP method, sections
\ref{Subsec:CompIntDyn}--\ref{Subsec:SymCom}, and obtain
\emph{universal dynamic discreteness (quantisation)} of unreduced,
complex interaction dynamics and its detailed description by the
\emph{unified Hamilton-Schr\"{o}dinger formalism}
(\ref{Eq:Ham-Jacob})--(\ref{Eq:SchrodinExpansion}).

The universal physical origin of \emph{discrete} structure of
\emph{unreduced} interaction dynamics accounts for the basic
phenomenon of \emph{dynamic multivaluedness} itself and takes the
form of omnipresent \emph{dynamic instability} by interaction
feedback loops (section \ref{Subsec:CompIntDyn}), where the
self-consistent EP dependence on the solution to be found (equations
(\ref{Eq:ExistEff})--(\ref{Eq:EP})) makes impossible an
evolutionary, smooth change of system configuration that follows
instead a series of highly \emph{inhomogeneous, ``quantum'' jumps}
between its incompatible realisations. Major expression of unreduced
interaction quantisation at any level of complexity is provided by
quantised elements of emerging space structure $\Delta x$ (distance
between neighbouring realisation configurations) and related time
increments $\Delta t$ (duration of transitions between realisation,
or realisation change events), which are determined according to
universal energy-complexity and momentum-complexity definitions
(\ref{Eq:Energy}),~(\ref{Eq:Momentum}) (see also equations
(\ref{Eq:ParticleEnergy}),~(\ref{Eq:ParticleMomentum}) for the
quantum complexity level):
\begin{equation*}\label{Eq:SpaceTimeQuanta}
\Delta x = \frac{{\mathcal A} _0}{p} \ , \ \ \ \Delta t =
\frac{{\mathcal A} _0}{E} \ ,
\end{equation*}
where ${\mathcal A} _0 \gg h$ is a characteristic action value,
which is not as unique/universal, however, as its value $h$ at the
lowest, quantum complexity level. The dynamically determined time
increment $\Delta t = \tau$ is a period of ``generalised quantum
beat'', and for the system globally at rest, it is directly related
to the generalised inertial (rest) mass $m _0$ and energy $E _0$
(cf. equation (\ref{Eq:RestMassEnergy})):
\begin{equation}\label{Eq:GenRestMassEnergy}
E_0  = m_0 v_0^2  =  - \frac{\Delta {\mathcal A}}{\Delta t} =
\frac{{\mathcal A} _0}{\tau _0} = {\mathcal A} _0 \nu _0 \ ,
\end{equation}
where $v_0$ is the perturbation propagation speed in a lower-level
structure (analogous to the speed of light at the first complexity
level), and $\nu _0 = 1 / \tau _0$ is the generalised quantum beat
(realisation change) frequency determining the corresponding
\emph{level} of \emph{causal, irreversible time flow}.

\emph{Generalised (special) relativity} of this \emph{higher-level
time} (and space) for a \emph{globally moving system} follows from
the universal symmetry of complexity in the same way as the
corresponding relativistic effects at the first complexity level
(see above in this section). Universal definitions of the
\emph{states of rest and motion} by, respectively, minimum and
greater than minimum values of differential complexity (energy)
remain directly applicable at any complexity level. The related
\emph{chaotic wandering} of the globally moving system around its
average motion tendency leads to \emph{essential difference} between
its \emph{total} differential complexity (total energy $E$) and
\emph{pseudo-regular, averaged} motion part (determined by momentum
$p$) expressed by the (generalised) ``relativistic dispersion
relation'' between $E$ and $p$ (cf. equation
(\ref{Eq:DispRelation})):
\begin{equation}\label{Eq:GenDispRelation}
E  = pV\left( v \right) \ ,
\end{equation}
where $V\left( v \right) > v, v_0$ is the generalised,
``faster-than-light'' (and actually fictitious) ``phase velocity''
function; for example, in the simplest case of homogeneous chaotic
wandering one has $V\left( v \right) = (v_0) ^2 / v$ (this is the
case of dynamic relativity at the first complexity level with $v_0 =
c$).

Using now the generalised dispersion relation
(\ref{Eq:GenDispRelation}) in combination with universally
applicable equations
(\ref{Eq:MotionEnergy}),~(\ref{Eq:ParticleEnergy}), and
(\ref{Eq:RelPeriods}), one obtains the \emph{universal, dynamically
derived time relativity} (retardation) for a globally moving system
of \emph{any} complexity (and time) level (cf. equations
(\ref{Eq:TimeRel:1}),~(\ref{Eq:TimeRel:2}) for the first complexity
level):
\begin{equation*}\label{Eq:TimeRel:Gen}
\tau  = {\mathcal T}\left( {1 - \frac{v}{V\left( v \right)}} \right)
\ ,  \ \ {\mathcal T} = \frac{\tau _0}{\sqrt {\displaystyle {1 -
\frac{\vphantom{V} v}{V\left( v \right)}}} } \ , \ \ {\mathcal N} =
\nu _0 \sqrt { 1 - \frac{\vphantom{\bar v} v}{V\left( v \right)}} \
.
\end{equation*}
As ${\mathcal N} < \nu _0$, objective, real time \emph{goes
relatively slower} within a \emph{globally moving}, or in general
\emph{developing}, system due to investment of a larger part of the
whole energy-complexity to that global motion tendency and
corresponding decrease of the ``time-producing'' energy of purely
random wandering around that average tendency. Similar to the
fundamental, first-level relativity, the key point here is the
\emph{complex-dynamic origin} of \emph{real physical time} itself.

System interaction with omnipresent environment, or (generalised)
``field'', gives rise to universal effects of general relativity.
Using generalised mass-energy definition
(\ref{Eq:GenRestMassEnergy}) we can directly extend the first-level
expression of complex-dynamical time retardation in the field of
gravity (\ref{Eq:GravTimeRetard}) to arbitrary complexity level:
\begin{equation}\label{Eq:GenGravTimeRetard}
M \left( x \right) v_0^2 \equiv {\mathcal A} _0 \nu \left( x \right)
= m v_0^2 \sqrt {1 + \vphantom{\bar \Phi} {\mit \Phi} \left( x
\right)} \ ,
\end{equation}
where $x$ is the generalised coordinate of the ``test'' system, $\nu
\left( x \right)$ is its generalised quantum beat (realisation
change) frequency, determining flow rate of the corresponding level
of its internal, physically real time, and ${\mit \Phi} \left( x
\right)$ is the (dimensionless) potential of environmental field
($\left| {{\mit \Phi} \left( x \right)} \right| < 1$). Contrary to
attractive gravity field, one may have both ${\mit \Phi} < 0$ and
${\mit \Phi} > 0$ at arbitrary complexity levels, and therefore
``internal'' system time can either slow down or accelerate
depending on the properties of the environment and its interaction
with the ``test'' system.

Universal relativistic modifications of length (spatial dimension)
and generalised mass are obtained in a straightforward way together
with respective time relativity. By analogy to equation
(\ref{Eq:RelMassLength}), the ``generalised de Broglie wavelength''
${\mit \Lambda} _{\rm B}$, or characteristic size of a globally
moving system (at arbitrary complexity level), and its generalised
mass transform as:
\begin{equation*}\label{Eq:UnivRelMassLength:1}
m = \frac{m_0}{ \sqrt {\displaystyle {1 - \frac{\vphantom{V}
v}{V\left( v \right)}}} } \ , \ \ {\mit \Lambda} _{\rm B} =
\frac{{\mathcal A} _0}{mv} = \frac{{\mathcal A} _0}{m_0 v}  \sqrt
{\displaystyle {1 - \frac{\vphantom{\bar v} v}{V\left( v \right)}}}
\ .
\end{equation*}
The universal general-relativistic mass and length transformations
follow from equation (\ref{Eq:GenGravTimeRetard}):
\begin{equation*}\label{Eq:UnivRelMassLength:2}
M \left( x \right) = m \sqrt {1 + \vphantom{\bar \Phi} {\mit \Phi}
\left( x \right)} = m _0 \sqrt { \frac{1 + \vphantom{\bar \Phi}
{\mit \Phi} \left( x \right)}{\displaystyle {1 - \frac{\vphantom{V}
v}{V\left( v \right)}}} } \ ,\ \ {\mit \Lambda} _{\rm B} \left( x
\right) = \frac{{\mathcal A} _0}{mv \sqrt {1 + \vphantom{\bar \Phi}
{\mit \Phi} \left( x \right)} } = \frac{{\mathcal A} _0}{m_0 v}
\sqrt { \frac{\displaystyle {1 - \frac{\vphantom{\bar v} v}{V\left(
v \right)}}}{1 + \vphantom{\vec \Phi} {\mit \Phi} \left( x \right)}
}\ .
\end{equation*}

The unified quantum and relativistic manifestations of the symmetry
of complexity at the first and higher complexity levels provide the
causally complete, realistic and demystified understanding of the
respective types of behaviour that look irreducibly ``weird'' and
only formally postulated in the traditional, unitary description.
Thus, one can now provide an exact, \emph{physically realistic}
answer to the question why a moving clock mechanism goes slower with
respect to the one at rest: it happens because a growing proportion
of total moving system dynamics (measured by its energy-complexity)
goes to this global motion tendency from the internal motion that
just determines the ``proper'' time flow rate, for \emph{any} its
measurement mechanism and \emph{in the same way} at \emph{any level}
of that mechanism. Moreover, the obtained extension of causal
relativity effects to \emph{any} system dynamics provides a
\emph{rigorously specified} explanation even for such traditionally
``subjective'' effects as personal, ``psychological'' time flow
change with the environment (``general relativity'') and internal
development (``special relativity'') of a conscious subject
\cite{Kir:USciCom}. The practically unlimited \emph{power} of the
universal symmetry of complexity to \emph{solve real-world problems}
is thus convincingly demonstrated, in addition to various other
examples described in this paper.

\subsubsection{Genuine quantum chaos, causal quantum measurement,
and complex-dynamic classicality emergence in closed systems}\label{Subsubsec:Classicality}
We have rigorously derived, in previous sections
\ref{Subsubsec:SpaceTime}--\ref{Subsubsec:Relativity}, the lowest,
``quantum'' sublevels of world structure and complexity, in the form
of physically real space and time, elementary particles, their
intrinsic and dynamical properties, and fundamental interaction
forces, all of them emerging as a result of the unified symmetry
(conservation and transformation) of complexity of the underlying
interaction between two initially homogeneous protofields. The next
sublevels of world complexity naturally emerge by the same kind of
unreduced interaction between those elementary structures, appearing
thus as further, dynamically continuous development of the same,
unified protofield interaction (next sublevel of its
\emph{dynamically fractal} hierarchy). They contain the elements of
\emph{both quantum} (undular, nonlocal) \emph{and emerging
classical} (corpuscular, localised) behaviour and can take the form
of \emph{(genuine) quantum chaos} for \emph{essentially
non-dissipative} (Hamiltonian) interaction cases, or \emph{causal
quantum measurement} for \emph{slightly dissipative} systems, or
complex-dynamic \emph{classicality emergence} in elementary (closed)
\emph{bound systems} (like atoms)
\cite{Kir:USciCom,Kir:QuMach,Kir:QFM,Kir:100Quanta,Kir:QuChaos,Kir:Channel,Kir:QuMeasurement}.

The situation of \emph{quantum chaos}
\cite{Kir:USciCom,Kir:QuMach,Kir:QuChaos,Kir:Channel} is described
by a particular case of general existence equation
(\ref{Eq:ExistGen}), the Schr\"{o}dinger equation (now causally
derived) for many (in general) particles interacting among them and
with external, time-dependent field(s), for example:
\begin{equation*}\label{Eq:QuChaos:Schrod}
i \hbar \frac{\partial {\mit \Psi}}{\partial t} = \left[
\sum\limits_{i,j = 1 \atop i \neq j}^N {- \frac{\hbar ^2}{2m _i}
\frac{\partial ^2}{\partial x _i ^2} + V _{ij} (x_i,x_j) +
U_i(x_i,t)} \right] {\mit \Psi} (X,t) \ ,
\end{equation*}
where $X = (x_1,x_2,\ldots,x_N)$ is the vector of all particle
coordinates ($x_i$ are also vectors, in general), $U_i(x_i,t)$ is
the time-dependent external field potential acting on $i$-th
particle with the mass $m_i$, $V _{ij} (x_i,x_j)$ are potentials of
interaction between $i$-th and $j$-th particles, and $N$ is the
number of particles. Time-periodic external fields $U_i(x_i,t)$ are
of special practical and fundamental interest for Hamiltonian chaos
problem (where periodic dependence on time is generally equivalent
to periodic dependence of external interaction on one of space
coordinates). In that ``canonical'' case external field can be
presented as a Fourier series:
\begin{equation*}
U_i(x_i,t) = \sum\limits_{n =  - \infty }^{n = \infty } {U_{in}
(x_i)\exp(i\omega _\pi  nt)}  = U_{i0} (x_i) + \sum\limits_{n \ne 0}
{U_{in} (x_i)\exp(i\omega _\pi  nt)} \ ,
\end{equation*}
where $\omega _\pi$ is the perturbation frequency, $n$ takes only
integer values, and we can consider, without limitation of
generality, that $U_{i0} (x_i)$ constitute integrable, binding
potentials of ``free'' particle motion (i.e. their motion in the
absence of essential, chaos-bringing interaction).

Our general analysis (section \ref{Subsec:CompIntDyn}) shows that
both inter-particle interactions and their interaction with the
external field will lead to dynamic multivaluedness and related
intrinsic randomness in a quantum system with interaction. However,
Hamiltonian chaos emerging due to integrable system interaction with
time- or space-periodic field constitutes a major, most transparent
case, especially for the quantum chaos problem. Application of the
unreduced EP analysis and results to that situation reveals indeed
the \emph{genuine, dynamic randomness} in a \emph{purely quantum}
system (that can be \emph{far} from the semi-classical limit), in
the same, universal form of multiple, incompatible realisations
forced to permanently replace each other in a causally random
(probabilistic) order thus defined
\cite{Kir:USciCom,Kir:QuMach,Kir:QuChaos,Kir:Channel}. The problem
of \emph{genuine} quantum chaoticity, persisting in the usual
theory, acquires thus the direct, universal and transparent
solution.

The universal criterion of global chaos onset (\ref{Eq:ChaosCrit})
remains valid for quantum chaos, but the characteristic frequency
$\omega _q$ and energy-level separation $\Delta \eta _n$ of
intra-component motion are replaced, respectively, by the
perturbation frequency $\omega _\pi$ and ``quantum energy'' $\hbar
\omega _\pi$:
\begin{equation}\label{Eq:QuChaos:Crit}
\kappa  \equiv \frac{\Delta \varepsilon}{\hbar \omega _\pi} =
\frac{\omega _0}{\omega _\pi} \cong 1\ ,
\end{equation}
where $\Delta \varepsilon$ is energy-level separation in the
non-perturbed Hamiltonian system (with the above integrable
potential $U_{i0} (x_i)$) and $\omega _0 = \Delta \varepsilon /
\hbar$ is its characteristic frequency. It is important, in
particular, that the global Hamiltonian chaos criterion
(\ref{Eq:QuChaos:Crit}) obtained by purely \emph{quantum-mechanical}
analysis has a classical form (frequency ratio) that coincides (in
the limit $\hbar \to 0$) with the respective chaos criterion
obtained within \emph{classical mechanics}
\cite{Kir:USciCom,Kir:QuChaos,Kir:Channel} and thus confirms the
usual \emph{correspondence principle} for \emph{real, chaotic}
systems, which constitutes the well-known unsolved problem of the
unitary quantum chaos description. We can conclude that the symmetry
of complexity (here between \emph{all} realisations of a Hamiltonian
quantum system) provides solution to a practically important and
otherwise ``unsolvable'' problem.

The problem of quantum measurement is different from the Hamiltonian
quantum chaos situation by a local, small but finite energy
\emph{dissipation} towards the measurement device that has \emph{no}
special ``classical'' or ``macroscopic'' character in our analysis,
but needs that initial dissipation as a source of \emph{real change}
of its state. The unreduced EP analysis and results remain basically
the same, but local dissipation violates equality between system
realisations and creates a transient compound, SOC-type realisation
(section \ref{Subsec:SymCom}) accompanied by spatial system
reduction (dynamical squeeze) towards the centre of dissipation (cf.
section \ref{Subsec:CompIntDyn}) that explains all quantum
measurement properties by \emph{causal, but complex (multivalued)
interaction dynamics} \cite{Kir:USciCom,Kir:QuMeasurement}. It is
important that \emph{before} (as well as after) the
\emph{dynamically random} emergence of quantum measurement event,
the measured quantum system performs \emph{unceasing transitions},
i.e. \emph{physically real ``quantum jumps''}, between \emph{all its
eigenstates} (with the corresponding, \emph{now dynamically
determined} probabilities), which provides \emph{causal, dynamic}
explanation for the \emph{formal} quantum postulates about ``linear
superposition'' of eigenstates (see also section
\ref{Subsubsec:Relativity}). Self-amplifying complex-dynamic
transformation of \emph{externally} ``linear'' combination into
``classical'', incoherent sum of probabilities provides consistent
solution to the famous ``Schr\"{o}dinger cat'' paradox
\cite{Kir:QuMach}.

\emph{Classical}, permanently localised kind of dynamics emerges
\emph{dynamically} as a natural ``advanced'' case of quantum
measurement, where transient SOC state during measurement event
becomes \emph{permanent}, actually giving rise to the \emph{next,
higher level of dynamic complexity}. Specifically, such elementary
classical states emerge as \emph{bound states of elementary
particles} (such as atoms), which have a classical behaviour
tendency in a totally \emph{closed} system configuration,
\emph{without} any ``environmental decoherence'' effects necessarily
evoked in the standard quantum mechanics and its unitary
modifications. The role of unreduced interaction complexity is
essential in understanding of that \emph{qualitative} transition
(``generalised phase transition'' \cite{Kir:USciCom}) to a higher
complexity level: it is the \emph{dynamically random}, ``quantum''
wandering processes of virtual solitons of \emph{each} of the
\emph{bound} system components that determine \emph{vanishingly
small probability} of longer-distance jump series of \emph{all}
components in the \emph{same} direction (which would determine
``quantum'', delocalised system behaviour)
\cite{Kir:USciCom,Kir:QuMach,Kir:QFM,Kir:100Quanta}. The same
\emph{internal dynamic complexity} of a classical system (in the
form of a SOC kind of state) explains the ``asymptotic'', fractal
boundary between quantum and classical behaviour and occasional
\emph{dynamic} revivals of quantum behaviour for classical,
sometimes macroscopic systems under the influence of their suitable
interactions (in direct contradiction to all ``decoherence''
theories).

In terms of our rigorous criterion of chaoticity
(\ref{Eq:ChaosCrit}), a classical, bound-SOC state is described by
the chaoticity parameter $\kappa = \omega _\xi / \omega _q \approx U
_\xi / m _q c^2$, where $\omega _\xi$ is the bound motion frequency,
$\omega _q$ is the component quantum beat frequency, $U _\xi = \hbar
\omega _\xi$ is the binding energy, and $m _q c^2 = \hbar \omega _q$
is the total component mass-energy. In all ``usual'' bound systems
with well-defined components, including atoms, binding energy is
much smaller than mass-energy, $U _\xi \ll m _q c^2$, or $\omega
_\xi \ll \omega _q$, which determines the \emph{complex-dynamic
origin} of the ``classical'', \emph{localised} and \emph{externally}
quasi-regular, SOC type of system configuration, $\kappa \ll 1$
(section \ref{Subsec:SymCom}) \cite{Kir:QuMach}. It is interesting
to note that in ``ultra-relativistic'' elementary systems where
binding energy can attain the rest energy, $U _\xi \sim m _q c^2$
(so that individual component structure cannot be ensured), the
chaoticity parameter is not small, $\kappa \sim 1$, and thus
classicality does \emph{not} appear, which provides a nontrivial
explanation for globally \emph{quantum} behaviour of hadrons as
``ultra-relativistic'' bound systems of quarks.

In the simplest case of hydrogen atom $\omega _\xi$ coincides with
the Bohr frequency and $U _\xi = \hbar \omega _\xi$ with the atomic
energy unit $\varepsilon _0 = m _e e^4 / \hbar ^2$, while $m _q = m
_e$ is the electron mass, and we have $\kappa = U _\xi / m _q c^2 =
\varepsilon _0 / m _e c^2 = e^4 / \hbar ^2 c^2 = \alpha ^2 \ll 1$,
where $\alpha = e^2 / \hbar c \approx 1/137$ is the fine structure
constant. In that way we confirm the complex-dynamic origin of
hydrogen atom classicality and develop the above interpretation of
fine structure constant in terms of electron realisation number
$N_\Re ^e \thinspace$, $\alpha = 1 / N_\Re ^e$ (section
\ref{Subsubsec:ConstPlanckian}). Indeed, if the electron quantum
beat frequency is the synchronised frequency of virtual soliton
wandering for both electron and proton in the hydrogen atom (cf.
section \ref{Subsubsec:ParticleProp}), then the probability of their
\emph{correlated} quantum jump in the \emph{same direction} will be
of the order of $(N_\Re ^e) ^{-2} = \alpha ^2 = \kappa$, thus
confirming the above classicality interpretation in terms of
\emph{multivalued} SOC dynamics. The probability $\alpha (x)$ of
correlated quantum wandering of two virtual solitons in a bound
system to a distance $x$ from their ``equilibrium'', global-motion
separation is determined by $(N_\Re) ^{-2x/\Delta x}$, where $\Delta
x$ is the quantum jump length ($\Delta x \simeq
\mathchar'26\mkern-10mu\lambda _{\rm C}$ for the electron, see
section \ref{Subsubsec:ConstPlanckian}), so that $\alpha (x)$ drops
exponentially with $x$. The pronounced classical, localised
behaviour of a bound system is obtained if $N_\Re \gg 1$ and
interaction is not so strong as to destroy component individuality
(these two conditions should largely coincide for our unified world
construction, see Figure \ref{Protofields}).

We obtain here a causal, realistic explanation for the ``fuzzy''
atom structure, with ``electron clouds'', etc. that can have only
inexact, figurative meaning in usual theory. In reality, all the
regular electron ``orbits'' (Schr\"{o}dinger wavefunction
configurations) represent but the \emph{average, global-motion} (and
relatively weak) tendency of permanent \emph{chaotic wandering} of a
corpuscular electron state, or virtual soliton (cf. section
\ref{Subsubsec:Relativity}). As we have seen above, larger
deviations from a global motion ``orbit'' are exponentially
suppressed, which explains orbit reality and well defined shape
(especially for the ground state), but the relative \emph{number} of
(small) deviations is \emph{large}. The above expression for the
bound system chaoticity $\kappa$ defines it also as a measure of
global motion ``relativity'', and a comparison with the
complex-dynamic interpretation of relativistic factor $v^2 / c^2$ in
section \ref{Subsubsec:Relativity} shows that $\alpha = 1 / N_\Re$
is also the \emph{probability (proportion)} of quantum jumps
\emph{within} the global motion tendency (which should be expected
in view of multivalued dynamics structure). It is easy to verify
that for the electron $\Delta x = \mathchar'26\mkern-10mu\lambda
_{\rm C} = \alpha a _{\rm B}$, or $a _{\rm B} = N_\Re ^e
\mathchar'26\mkern-10mu\lambda _{\rm C}$, where $a_{\rm B} = \hbar
^2 / m_e e^2$ is the Bohr radius and ``average'' radius of the
ground-state orbit of the hydrogen atom. This well-known relation
acquires now a new meaning as it shows that the size of the main
electron orbit is intimately adjusted to the complex-dynamical
``cycle'' of $N_\Re ^e$ (chaotic) quantum jumps around it. The whole
internal dynamics of an atom appears now as a chaotic,
\emph{complex-dynamical engine} causally driven by the underlying
\emph{protofield interaction}, instead of fixed, abstract
``state-vector'' configurations, related formal ``exact solutions'',
and underlying \emph{irreducible} quantum mysteries of unitary
atomic physics.

\section{Complex-dynamic solution of major cosmological problems}\label{Sec:DarkMatter}
\subsection{Dark mass effects: Unitary projection of multivalued
dynamics}\label{Subsec:DarkMass}
In previous sections we have specified the first, most fundamental
levels of explicit universe structure emergence in the process of
complex-dynamic, unreduced interaction between two protofields
governed by the universal symmetry of complexity. We have shown, in
particular, that this unified symmetry determines self-tuning,
dynamically adaptable universe structure creation without
``anthropic'' problems (section \ref{Subsubsec:Self-Tuning}) and
ensures strict positivity (and large value) of the total universe
energy determining also the physically real, dynamically
irreversible time flow (section \ref{Subsubsec:PositiveUnivEnergy}).
We shall continue now to study \emph{cosmological manifestations} of
the symmetry of complexity at its higher, \emph{macroscopic} levels
confirming its status of the unified Order of the World. In this
section we show that the \emph{same} unreduced dynamic complexity
that determines non-zero material content of the universe (its
positive mass-energy) provides also a natural and universal solution
to multiple problems of apparently strongly excessive, hidden, or
``dark'' mass content of major cosmological objects (galaxies,
clusters, etc.).

The dark mass problem involves various observations showing that
universe structure dynamics, mostly on the scale of galaxies and
related structures, would need larger, and often much larger,
quantities of massive matter, than those that can actually be
perceived (see e.\,g.
\cite{DarkMatter:1,DarkMatter:2,DarkMatter:3,DarkMatter:4}). Great
\emph{variability} of the missing mass effect is a serious
additional complication of a problem. We show that these
difficulties of the \emph{unitary interpretation} are actually
spurious and originate from the same incorrect \emph{neglect} of the
\emph{main, chaotic} part of \emph{real} system dynamics, now
occurring at the level of cosmic object dynamics. If one considers
the unreduced, \emph{dynamically multivalued} system behaviour, the
problem will not even appear and the truly chaotic dynamics of real
objects will account for observed dynamical features with the
``visible'', normal mass values. It is important that one should
take into account the \emph{genuine}, dynamically multivalued chaos,
rather than one of its unitary imitations by ``involved'' but
basically regular behaviour.

The main idea is straightforward: because of \emph{artificial} cut
of all system realisations but one in the unitary theory (this is an
\emph{exponentially big} reduction for a many-body system, see
section \ref{Subsubsec:Self-Tuning}), one inevitably obtains a
\emph{missing motion} problem, which is interpreted as inexplicably
``missing mass''. One can specify this result in various ways, and
we start with a demonstration of incompleteness of the standard
virial theorem application to the real, multivalued dynamics of a
many-body system, since it shows how the key balance between
potential and kinetic energy can be modified by the true chaos.

If system components move under the influence of gravitational
attraction, e.\,g. in a galaxy, then the ordinary virial theorem
gives the following relation between the time-averaged values of
kinetic $\bar T$ and potential $\bar U$ energy of a system or any
its subsystem (see e.\,g. \cite{LandauLifshitz:M}):
\begin{equation}\label{Eq:DarkM:T-U}
2\bar T =  - \bar U \ ,
\end{equation}
whereas in reality this \emph{regular}-motion kinetic energy, $\bar
T = \bar T_{\rm reg}$, is a \emph{small} part of its true,
\emph{chaotic} content $\bar T_{\rm real}$:
\begin{equation*}\label{Eq:DarkM:Treal}
\bar T_{\rm real}  = \bar T_{\rm reg} N_\Re \ ,
\end{equation*}
where $N_\Re$ is an \emph{effective} realisation number for a given
kind of observation and averaging (usually $N_\Re \gg 1$, while
$N_\Re = 1$ for unitary models of the standard theory).

The \emph{observed} potential energy,  $\bar U_{\rm obs}$, gives
\emph{real} kinetic energy:
\begin{equation}\label{Eq:DarkM:T-Ureal}
2\bar T_{\rm real}  =  - \bar U_{\rm obs} \ .
\end{equation}
However, if observations are interpreted within a unitary,
\emph{deficient} version of dynamics (\ref{Eq:DarkM:T-U}) implying
that
\begin{equation}\label{Eq:DarkM:T-Uunit}
2\bar T_{\rm reg}  =  - \bar U_{\rm obs} \ ,
\end{equation}
one obtains a ``mysterious'' \emph{discrepancy}, $\delta$, between
(\ref{Eq:DarkM:T-Ureal}) and (\ref{Eq:DarkM:T-Uunit}):
\begin{equation*}\label{Eq:DarkM:Delta}
\delta  = \frac{\bar T_{\rm real}}{\bar T_{\rm reg}} = N_\Re \ .
\end{equation*}
It is explained \emph{within the unitary model} as being due to
``invisible'', but actually present, or ``dark'' mass, $M_{\rm dark}
= M_{\rm real}  - M_{\rm reg}$, whose relative value can be
estimated as
\begin{equation*}\label{Eq:DarkM:RelValue}
\frac{M_{\rm real}}{M_{\rm reg}} = \frac{\bar T_{\rm real}}{\bar
T_{\rm reg}} = \delta = N_\Re \ .
\end{equation*}
According to the unreduced, \emph{complex-dynamic interpretation},
the observed discrepancy $\delta$ can be used for estimation of
effective $N_\Re$ values. Since $\bar T \propto \overline {Mv^2}$,
one can say that in reality there is \emph{too much motion}, or
(deviating) \emph{velocity}, in a system with respect to unitary
expectations, so that one has rather a ``dark velocity (or kinetic
energy)'' effect:
\begin{equation*}\label{Eq:DarkM:DarkVel}
(\overline {v^2 }) _{\rm real}  = N_\Re (\overline {v^2 }) _{\rm
reg} \ .
\end{equation*}

One can easily refine this result for a \emph{distance-dependent
case}, $N_\Re   = N_\Re ( r )$ (where $r$ is a coordinate within the
system), in terms of velocity-distance dependence curves, or
``rotation curves'', for galaxies. In that case an ``anomalous'' $v
(r)$ dependence is \emph{not} due to anomalies of mass distribution,
$M (r)$ (attributed to ``dark matter halos''), but due to
``unexpected'' (in the \emph{unitary} model) contribution to average
velocity from \emph{chaotic} motion parts, so that $v (r)$ is
proportional not to $\sqrt {M_{\rm reg} (r) + M_{\rm dark} (r)}$,
but to $\sqrt {N_\Re (r)}$. In a general case,
\begin{equation}\label{Eq:DarkM:V-r}
v\left( r \right) = \sqrt {\frac{{\gamma N_\Re  \left( r
\right)M_{{\rm{obs}}} \left( r \right)}}{r}} \ \ \ {\rm{or}} \ \ \
N_\Re \left( r \right) = \frac{{rv^2 \left( r \right)}}{{\gamma
M_{{\rm{obs}}} \left( r \right)}} \ ,
\end{equation}
where $M_{\rm obs} (r) = M_{\rm real} (r)$ is the \emph{ordinary},
``visible'' mass within radius $r$, and one can \emph{derive} the
features of \emph{chaotic} system dynamics, $N_\Re (r)$, from the
observed $v (r)$ and $M_{\rm obs} (r)$ dependences for perceivable,
``normal'' object components.

As should be expected, $N_\Re (r)$, and thus chaoticity, will
typically have a wide, often irregular maximum in ``looser'' system
parts, such as galactic halos and central, inter-component regions
of a cluster. This result correlates with the \emph{empirically
based} MOND hypothesis interpreting ``unusual'' motion in those
weak-interaction regions in terms of modification of Newtonian
gravitational attraction itself (see e.\,g.
\cite{DarkMatter:3,DarkMatter:4,MOND:1,MOND:2,MOND:3}). There is
even a deeper link between MOND hypothesis and our unreduced EP
approach: in a real many-body system one always deals with an
\emph{effective}, rather than direct, interaction that bears the
self-consistent influence of \emph{all} system components,
\emph{differs} essentially from the direct interaction, and
possesses \emph{many} contributing, chaotically changing
realisations. By contrast, if one takes any MOND-like assumption
\emph{without} reference to the underlying complex dynamics of the
system in question, then any its explanation should still inevitably
rely upon additional ``dissipation'' of unknown origin.

The observed \emph{big variations} of dark mass effects for
different objects represent a ``heavy'' difficulty for any
explanation in terms of additional, ``invisible'' entities, but are,
on the contrary, \emph{inevitable} for the above unified explanation
in terms of the true (multivalued) chaos effects. Such ``unlimited''
variability and  visible ``asymmetry'' are just \emph{unique}
properties of the symmetry of complexity (section
\ref{Subsec:SymCom}) appearing at all, but especially higher
complexity levels. Moreover, one can trace a definite qualitative
correlation between the expected object chaoticity (degree of
irregularity), its spatial dependence, and the observed magnitude of
``missing mass'' effects (further extended verification is certainly
necessary). It seems also to be much more consistent to explain an
observed, variable system property by a \emph{fundamental} property
of its \emph{dynamics}, rather than by a new, strangely escaping,
and inevitably \emph{fixed} entity (this situation is quite similar
to interpretation of the \emph{origin of mass} at the \emph{first}
level of complexity, see section \ref{Subsubsec:ParticleProp}). One
should also take into account the spatial dependence of chaotic mass
distribution effects (or ``structural'' chaos) that tend to
accumulate just outside of the main mass and interaction
concentration in the system, in agreement with data interpretation
using equation (\ref{Eq:DarkM:V-r}).

Finally, we emphasize once more the discovered \emph{unified
solution}, within the \emph{symmetry of complexity}, of the missing
mass problems at different levels of world dynamics, including
elementary particle mass (section \ref{Subsubsec:ParticleProp}), the
(total) mass-energy of the universe (section
\ref{Subsubsec:PositiveUnivEnergy}), and ``dark mass'' effects at
the level of galactic structures (this section), all of them related
to consistent solution of the \emph{unreduced interaction problem}
(sections \ref{Subsec:CompIntDyn}--\ref{Subsec:SymCom}).

\subsection{Complex-dynamic solution of dark energy and Big Bang problems}\label{Subsec:DarkEnergyBB}
The origin of \emph{globally} missing, ``distributed'' universe
energy, or ``dark energy''
\cite{DarkMatter:1,DarkMatter:2,DarkMatter:3}, is directly related
to the vicious circle of the \emph{unitary} cosmology scheme centred
on the \emph{zero-energy universe assumption} and related \emph{Big
Bang hypothesis}, or ``exploding vacuum'' solution. Indeed, the
latter starts from \emph{postulated}, artificially imposed
nothingness of the universe mass-energy content (see section
\ref{Subsubsec:PositiveUnivEnergy}), in the form of dynamically
single-valued, \emph{zero-complexity} reduction of universe dynamics
(irrespective of particular ``model'' details and including
occasional models with formally positive energy, but always zero
dynamic complexity). Because of the \emph{intrinsic instability} of
that fundamentally \emph{fixed}, static construction, one is forced
to impose a mechanistic ``general expansion'' (or the reverse
squeeze) of the universe as a single possible mode of its (totally
illusive) ``development''. The choice for expansion, or Big Bang, is
justified by a \emph{particular interpretation} of the observed
``red shift'' effect (involving a number of \emph{serious
contradictions} in itself). However, the conceptual instability of
\emph{any} unitary model (absence of evolving, adaptable degrees of
freedom, as opposed to abstract ``parameters'') persists in the form
of multiple problems of the Big Bang model whose proposed
``solutions'' only transform them to other formulations or
artificially introduced entities. The dark energy problem represents
only the latest in the list, though scandalously big and long
hidden, rupture in the \emph{basically frustrated} construction: the
discovered \emph{slightly} uneven red-shift dependence on distance
leads to a \emph{huge} deficiency in the source of uneven expansion,
supposed to be a distributed stock of mysterious, invisible energy
that should take very exotic, normally \emph{impossible} forms.

That \emph{final impasse} of missing energy (and mass) content of
the universe (see also the previous section) simply takes us back to
the beginning of the unitary vicious circle, where such emptiness of
the universe content has been \emph{explicitly imposed} by the
unitary paradigm itself. In fact we deal here with another, though
unrealistically simplified case of the \emph{symmetry (conservation)
of complexity}, astonishing in its long-lasting reduction, $0 = 0$,
applied here to the \emph{whole universe content}. In other words,
the symmetry of complexity provides the rigorous and properly
universal substantiation of the fact that \emph{all} the
artificially reduced, \emph{dynamically single-valued} universe
models with zero value of genuine dynamic complexity will
\emph{inevitably} and \emph{essentially} fail in description of
\emph{real, dynamically multivalued} universe structure
characterised by \emph{positive (and high)} value of unreduced
dynamic complexity
\cite{Kir:USciCom,Kir:USymCom,Kir:Cosmo}.\footnote{Note that any
usual, zero-complexity cosmology necessarily implies, due to its
dynamic single-valuedness, total basic \emph{regularity} and thus
zero entropy of the universe and any its quasi-closed subsystem, in
contradiction to entropy growth principle. Any observed or described
``chaoticity'' or randomness of such universe content, on any scale,
is inevitably reduced to mere ``entangled regularity'', in agreement
with the old Laplacian vision of totally mechanistic,
\emph{basically predictable}, but maybe \emph{practically}
noncomputable world.}

By contrast, the unreduced, dynamically multivalued and
probabilistically fractal structure of real interaction dynamics
leads to \emph{globally stable} concept of universe structure
development, just because it is based on the omnipresent and
massively adaptable \emph{local, dynamic instability} of
\emph{explicit structure creation} (see also section
\ref{Subsubsec:Self-Tuning}). The explicit universe structure
emergence in the \emph{initially homogeneous} system of interacting
protofields, starting from the physically real space, time, and
elementary particles, intrinsically unified with their fundamental
properties and interactions (section \ref{Subsec:Properties}), can
be described as a distributed \emph{implosion} of ubiquitous,
fractally structured \emph{creation}, as opposed to mechanistic and
intrinsically \emph{destructive explosion} of the unitary Big Bang
(and ``inflation'') schemes.

Therefore the ``dark energy'' problem \emph{does not even appear} in
the complex-dynamic, intrinsically creative cosmology, quite similar
to all ``anthropic'' kind of problems (section
\ref{Subsubsec:Self-Tuning}). The self-tuning universe structure,
liberated from artificial unitary instabilities and related
``anthropic'' speculations, emerges naturally and self-consistently,
simply due to the unreduced, \emph{truly exact} picture of the
\emph{underlying interaction} processes.

As for the origin of the observed \emph{red shift effect} in
radiation spectra of distant objects, it finds its consistent
explanation in terms of \emph{intrinsically nonlinear} radiation
propagation properties in the system of coupled protofields (see
section \ref{Subsubsec:ParticleProp} and Figure \ref{Protofields}),
where some (relatively weak) loss of energy by soliton-like photons,
propagating in the e/m protofield medium, is \emph{inevitable}
because of their weak, but finite coupling to the gravitational
medium. Note the essential difference of this nonlinear energy
dissipation from linear scattering effects in any ordinary model.
The soliton-like photon, remaining stabilised by interaction with
the gravitational protofield, can \emph{slowly} give its
\emph{energy} to the gravitational degrees of freedom (most probably
quarks) \emph{without} any noticeable change of its direction of
propagation (i.e. without any ``blur'' effects in the distant object
images). Characteristic ``transpiercing'' and ``circumventing''
modes of soliton interaction with small enough obstacles can explain
anomalously small loss and vanishing angular deviation effects for
photons and very high-energy particles (see below).

One should also take into account possible contribution from
modified protofield parameters around big mass concentration or
various ``special'' objects, as well as ``older'' photon propagation
at earlier stages of universe structure development. Detailed
calculations of the effect will inevitably involve many unknown
parameters of the system, but \emph{qualitative} properties and
\emph{consistency of the whole picture} provide convincing evidence
in favour of this kind of \emph{fundamentally new} explanation for
the red shift effect (within a broader scope of ``tired light''
approach) and its expected refinement, including the necessary
clarification of the \emph{detailed physical origin of photon}
(missing persistently in the unitary theory framework).

The nonlinear red shift dependence on distance that gives rise to
\emph{catastrophic} consequences in the unitary cosmology can only
be natural in the complex-dynamic, \emph{essentially nonlinear}
picture (section \ref{Subsec:CompIntDyn}). The nonlinear energy-loss
mechanism of soliton-like photons explains why this loss grows more
slowly with distance, than any usual mechanism of diffuse scattering
would imply (cf. the above note on soliton scattering dynamics).
Similar dynamics could solve, by the way, the persisting puzzle of
GZK effect for the ultra-relativistic particles, since at those
super-high energies the motion of a massive particle approaches that
of (a group of) photons, according to the results of quantum field
mechanics \cite{Kir:USciCom,Kir:QFM,Kir:100Quanta}. Another, though
maybe less specific, feature of red-shift data correlating with our
explanation is (increased) growth of average scatter of data points
with distance.

\subsection{Complex-dynamic cosmology: Global universe structure development}\label{Subsec:GlobalCosmo}
Returning to the general picture of emerging universe (section
\ref{Sec:UnivByComp}), note once more that according to the
underlying \emph{symmetry of complexity}, it \emph{cannot} contain
``motion-on-circles'' dynamics, on \emph{any} scale of structure
creation, so that the initial, \emph{positive} amount of
\emph{dynamic information}, in the form of protofield interaction,
gives rise to generalised, \emph{complex-dynamical system birth},
followed by its uneven, \emph{irreversible}, and global
\emph{transformation} into \emph{dynamic entropy} (developed
structure) within thus \emph{universally} defined, \emph{finite}
system \emph{life}, which ends up in the state of \emph{generalised
death, or equilibrium}, around the total transformation of the
initial dynamic information into entropy (unless additional dynamic
information is introduced into the system) \cite{Kir:USciCom}.

The generalised ``potential energy" of interacting protofields can
be introduced e.\,g. by their explicit separation from the
pre-existing state of ``totally unified'' (mixed) protofields that
could have the form of a generally inert quark-gluon condensate in
its ``absolute'' ground state. Although these ``prehistoric''
assumptions are subject to inevitable and increased uncertainty,
they can be estimated rather definitely by general consistency and
parsimony principles, now \emph{rigorously specified} by the
universal symmetry of complexity (see section
\ref{Subsubsec:SpaceTime}). What appears to be much more certain,
however, is that one does need an initial form of ``potential''
interaction energy, positively defined and specified here as
``dynamic information'', since the birth of a structured, real
universe from absolute ``nothingness'', without genuine interaction
development (which is the preferred dogma of the conventional
unitarity), contradicts the fundamentally substantiated and
\emph{universally} confirmed symmetry (conservation) of complexity
(section \ref{Subsec:SymCom}).

We can add here other perspectives of our complex-dynamical universe
description, whose consistent development within the standard,
unitary cosmology paradigm seems much less probable (cf. e.\,g.
\cite{Corredoira}). The highly uneven, long-distance concentration
of various anomalous, super-intense sources of energy, as well as
their ``peculiar'' red-shift tendency, point to a (probably moving)
``shape of the world'', which looks quite natural in our interacting
protofield logic, while it would need additional, ``unnatural''
assumptions in the Big Bang logic of ``exploding emptiness''.
Growing problems with the \emph{universe age} can be naturally
solved in our complex-dynamic cosmology as it traces
\emph{explicitly} the \emph{real life-cycle dynamics} of emerging
structures, while the unitary theory encounters here another series
of its \emph{inbred} ``instabilities'' (due to the rigidly fixed,
imposed ``models'' and mechanistic data fit). The same refers to
structural difficulties of the omnipresent expansion and natural
elimination in our approach of this and other ``old'' difficulties
of the unitary cosmology, such as average \emph{space flatness} and
\emph{homogeneity} (section \ref{Subsubsec:SpaceTime}),
``anthropic'' problems (section \ref{Subsubsec:Self-Tuning}), causal
origin of high-density states, real Planckian units and microwave
background radiation (section \ref{Subsubsec:ConstPlanckian}).
Intrinsic inclusion of \emph{realistic, unified} solution of
stagnating problems of quantum mechanics, field theory, and
relativity (sections
\ref{Subsec:CompIntDyn},~\ref{Subsec:Properties}) constitutes the
\emph{unique feature} of our theory that, being highly desirable,
cannot be even expected for any unitary approach. Finally,
\emph{irreducibly complex} dynamics of detailed formation and
evolution of galaxies, stars, and planetary systems is among further
applications of the present theory that will similarly profit from
the universal problem-solving power of the symmetry of complexity
demonstrated above.

\section{New mathematics of complexity and emergence}\label{Sec:NewMath}
We have demonstrated, in previous sections, how the \emph{universal
symmetry of complexity}, including \emph{conservation} and unceasing
\emph{development} of unreduced dynamic complexity describes the
\emph{explicit emergence} and \emph{properties} of \emph{real}
universe structures, starting from elementary particles, their
properties and interactions, and provides thus \emph{consistent} and
\emph{unified} solutions to many stagnating problems of usual,
zero-complexity models. This problem-solving power of the symmetry
of complexity centered on the obtained property of explicit
structure emergence necessarily involves a qualitatively new,
extended application of familiar mathematical tools and ideas
\cite{Kir:USciCom,Kir:USymCom,Kir:Fractal:2}. In this section we
summarise the main features of the \emph{new mathematics of
emergence} thus obtained (it can also be called \emph{new
mathematics of complexity}), with the reference to previous sections
presenting its more detailed framework (sections
\ref{Subsec:CompIntDyn}--\ref{Subsec:SymCom}) and applications to
fundamental world structures and properties (sections
\ref{Subsec:Properties},~\ref{Subsec:DarkMass}--\ref{Subsec:GlobalCosmo}).

The most important, embracing feature of the \emph{new mathematics
of emergence and complexity} is that it is represented by the
\emph{unified, single structure} of \emph{dynamically probabilistic
fractal} obtained as explicit, \emph{causally complete solution} of
\emph{real, unreduced interaction problem}
(Sect.~\ref{Subsec:CompIntDyn}). All its properties, describing the
\emph{exact} world structure and dynamics as it is, are unified
within the single,  \emph{absolutely exact} (never broken)
\emph{symmetry, or conservation, of complexity} including its
\emph{unceasing transformation} from complexity-information to
complexity-entropy (Sect.~\ref{Subsec:SymCom}). It means, in
particular, that \emph{all real-world structures}, and thus the
\emph{world/universe as a whole}, are \emph{absolutely symmetric}
(and \emph{dynamically complex}) and in this sense \emph{represent}
the symmetry of complexity as such, the latter \emph{explicitly
producing}, in particular, all the observed \emph{irregularities}.
By contrast, omnipresent violations of usual, unitary symmetries
result \emph{inevitably} from their \emph{artificially reduced},
dynamically single-valued basis, including all imitative models of
usual, unitary ``science of complexity'' (cf.
\cite{Perplexity,EndScience}).

One can emphasize several \emph{specific, but universally appearing
features} of this unified structure and law of the new mathematics,
distinguishing it essentially from the unitary framework
\cite{Kir:USciCom,Kir:Fractal:2}:
\begin{list}{(\roman{enumi})}{\usecounter{enumi}}

\item
\emph{Non}uniqueness of any \emph{real, unreduced} (interaction)
problem solution, in the form of its \emph{dynamic multivaluedness
(redundance)}; exclusively \emph{complex-dynamic} (multivalued,
internally \emph{chaotic}) existence of any real system (cf. usual
``existence and uniqueness" theorems).

\item
Omnipresent, explicit \emph{emergence} of \emph{qualitatively new}
structure and \emph{dynamic origin of time} (change) and
\emph{events}: $\mathbf {A \ne A}$ for \emph{any} structure/element
$\mathbf A$ in the new mathematics \emph{and} reality, while
$\mathbf {A = A}$ (self-identity postulate) in the \emph{whole}
usual mathematics, which thus excludes any \emph{real} change in
principle.

\item
Fractally structured \emph{dynamic entanglement} of unreduced
problem solution (interaction-driven, \emph{physically real}
intertwining between system components \emph{within any
realisation}): it is a \emph{rigorous} expression of \emph{material
quality} of a real structure in mathematics (as opposed to
``immaterial'', qualitatively ``neutral'', ``dead'' structures of
usual mathematics).

\item
Basic deficiency of perturbation theory and ``exact solution"
paradigm: the unreduced problem solution is \emph{dynamically
random} (permanently, chaotically changing), \emph{dynamically
entangled} (internally textured and ``living") and \emph{fractal}
(hierarchically structured). One obtains \emph{unified dynamic
origin} and \emph{causally specified} meaning of such basic
properties of unreduced problems and underlying real systems as
\emph{nonintegrability}, \emph{nonseparability},
\emph{noncomputability}, \emph{(genuine) randomness},
\emph{uncertainty (indeterminacy)}, \emph{undecidability},
\emph{``broken symmetry''}, etc. Real interaction problem is
nonintegrable and nonseparable \emph{but} solvable. Realistic
mathematics of complexity is \emph{well defined} (\emph{certain},
\emph{unified} and \emph{complete}, cf. \cite{Kline}), but its
structures are intrinsically ``fuzzy'' (dynamically
\emph{indeterminate} and really \emph{fluctuating}) and properly
\emph{diverse} (\emph{not} reduced to numbers or geometry).

\item
\emph{Dynamic discreteness (causal quantisation)} of unreduced
interaction products (realisations) resulting simply from the
\emph{holistic} character of every unreduced interaction process. It
appears as \emph{qualitative} inhomogeneity, or \emph{nonunitarity},
of any system structure and evolution and provides universal
\emph{dynamic} origin of (fractally structured) \emph{space}. It
demonstrates \emph{qualitative deficiency} of usual unitarity,
continuity \emph{and} discontinuity, calculus, and \emph{all} major
structures (evolution operators, symmetry operators, \emph{any}
unitary operators, Lyapunov exponents, path integrals, etc.).

\end{list}

\section{Conclusion: Real problem solution by real-world symmetry}\label{Sec:Conclusion}
The rigorously derived concept of \emph{universal dynamic
complexity} and related \emph{symmetry of complexity} involve
qualitatively extended and intrinsically unified properties that
allow for the \emph{causally complete}, totally realistic and
consistent, description of world structure behaviour at any level of
complexity in terms of \emph{unreduced interaction problem solution}
(sections
\ref{Subsec:CompIntDyn}--\ref{Subsec:SymCom},~\ref{Sec:NewMath}).
However, this rigorously based consistency of the unified symmetry
of complexity should also be \emph{confirmed} by various
\emph{applications} to particular systems and levels of complexity.

A part of this applied aspect comes already from the \emph{extended,
causally complete interpretation} of the well-known (but often
unexplained) observation results and related explicit
\emph{unification} of traditionally separated phenomena and levels
of world dynamics. In that way one obtains, for example, not only
\emph{causal, dynamically based} explanation for major \emph{quantum
and relativistic effects}, but also their \emph{intrinsic
unification} by the symmetry of complexity and \emph{extension} to
\emph{any} level of world dynamics (section
\ref{Subsubsec:Relativity}). All the canonical ``mysteries'' and
``inexplicable'', formally imposed ``postulates'' and ``principles''
naturally appear now as \emph{inevitable}, totally realistic
manifestations of the \emph{genuine, complex-dynamic} (multivalued)
content of \emph{any} structure and dynamics. As this content always
obeys the exact symmetry of complexity, it turns out that the
\emph{whole} real world content, including all changes and structure
creation processes, is \emph{absolutely} and \emph{exactly
symmetric}, i.e. it is a unified, but properly diverse manifestation
of the universal and never broken underlying symmetry, the symmetry
of complexity.

The \emph{problem-solving power} of the universal symmetry of
complexity is further confirmed by a growing number of its
\emph{successful applications} to various particular systems
covering the whole hierarchy of world's complexity and involving
explicit solutions of both ``old'' and new, sometimes urgent
problems emerging for both old and new kind of systems (and
remaining ``increasingly'' unsolved within the unitary science
paradigm)
\cite{Kir:USciCom,Kir:Fractal:1,Kir:Fractal:2,Kir:QFM,Kir:100Quanta,Kir:Cosmo,Kir:QuChaos,Kir:Channel,Kir:Nano,Kir:Conscious,Kir:CommNet,Kir:SustTrans}.
One can briefly summarise such applications to systems from both
lowest complexity levels (considered in this paper) and higher
complexity levels (considered elsewhere) in the following way:
\begin{list}{(\arabic{enumi})}{\usecounter{enumi}}

\item
In \emph{particle and quantum physics} one obtains \emph{causal,
unified} origin and structure of \emph{elementary particles},
\emph{all} their \emph{properties} (``intrinsic'', quantum,
relativistic) and interactions (section \ref{Subsec:Properties})
\cite{Kir:USciCom,Kir:QuMach,Kir:QFM,Kir:100Quanta,Kir:Cosmo,Kir:75MatWave}.
\emph{Complex-dynamic origin of mass} (section
\ref{Subsubsec:ParticleProp}) avoids any additional, abstract
entities (Higgs bosons, zero-point field, extra dimensions, etc.).
\emph{Renormalised Planckian units} provide consistent \emph{mass
spectrum} and other stagnating problem solution, including causally
complete explanation for the \emph{physical} origin of
\emph{universal constants} and their \emph{universality} (section
\ref{Subsubsec:ConstPlanckian}). \emph{Complex-dynamic cosmology}
(including higher complexity levels) resolves the dark mass and
energy problems without ``invisible'' entities (sections
\ref{Subsec:DarkMass},~\ref{Subsec:DarkEnergyBB}), together with
other old and new problems of unitary cosmology (sections
\ref{Subsubsec:Self-Tuning},~\ref{Subsubsec:PositiveUnivEnergy},~\ref{Subsec:GlobalCosmo}).
The established \emph{fundamental link between the numbers of
(realistically specified) space dimensions and interaction forces}
(section \ref{Subsubsec:Interactions}) leaves no place for arbitrary
insertion of ``additional'' entities (e.\,g. ``hidden dimensions'').

\item
At a higher complexity sublevel of \emph{interacting particles}
\cite{Kir:USciCom,Kir:QuMach,Kir:QFM,Kir:100Quanta,Kir:QuChaos,Kir:Channel,Kir:QuMeasurement}
(section \ref{Subsubsec:Classicality}) one obtains \emph{genuine,
purely dynamic quantum chaos} for Hamiltonian (nondissipative)
dynamics and \emph{correct correspondence principle} for (real)
\emph{chaotic systems} (natural transition from quantum to classical
behaviour as $\hbar \to 0$). A slightly dissipative interaction
dynamics leads to the \emph{causally complete} understanding of
\emph{quantum measurement} in terms of (causal) \emph{quantum}
dynamics \emph{alone}. \emph{Intrinsic classically} emerges as a
\emph{higher complexity level} in a \emph{closed}, bound system,
like atom, without any ambiguous ``decoherence by environment''.

\item
\emph{Realistic, causally complete} foundation of
\emph{nanobiotechnology} is provided by \emph{rigorous} description
of \emph{arbitrary} nanoscale interaction, revealing the
\emph{irreducible} role of \emph{genuine chaoticity} just on that
smallest scale \cite{Kir:QuMach,Kir:Nano}. \emph{Exponentially huge
power} of unreduced, complex nanobiosystem dynamics explains the
\emph{essential properties of life} and has direct relation to
complex information and communication system development (see item
(\ref{Bio}) below).

\item
\emph{Causally complete} description of \emph{unreduced genome
interactions} leads to \emph{reliable, rigorously substantiated
genetics} and consistent understanding of related \emph{evolutionary
processes} \cite{Kir:Fractal:2}.

\item\label{Bio}
Higher-complexity applications include \emph{general many-body
problem solution} and related description of ``difficult'' cases in
\emph{solid-state physics}, unreduced dynamics and evolution of
\emph{living organisms} (causally complete understanding of the
state of \emph{life} as a high enough level of unreduced dynamic
complexity), \emph{integral (causally complete) medicine}, emergent
(genuine) \emph{intelligence} and \emph{consciousness},
\emph{complex information and communication system dynamics},
\emph{creative ecology} and practically efficient \emph{sustainable
development concept}, \emph{rigorously} specified \emph{ethics} and
\emph{aesthetics}
\cite{Kir:USciCom,Kir:QuMach,Kir:Fractal:1,Kir:Fractal:2,Kir:Conscious,Kir:CommNet,Kir:SustTrans}.

\end{list}
These results explicitly demonstrate the expected advantages of
applying the (exact) \emph{real-world symmetry} to \emph{real
problem solution} and outline practically unlimited development
perspectives of the universal symmetry complexity and its
applications.

\LastPageEnding

\end{document}